\documentclass{mn2e}
\usepackage [nolists,noheads,nomarkers]{endfloat}
\usepackage{graphicx}
\usepackage{epsfig}
\usepackage{verbatim}
\usepackage{subfigure}
\usepackage{appendix}
\usepackage{amssymb}
\usepackage{array}
\usepackage{amsmath}
\usepackage{longtable}

\newcommand{\bq}{\begin{equation}}
\newcommand{\eq}{\end{equation}}

\def\gtsim{\lower.5ex\hbox{$\buSildrel > \over\sim$}}
\def\ltsim{\lower.5ex\hbox{$\buildrel < \over\sim$}}
\def\arcsec{^{\prime\prime}}

\def\farcs{\hbox{$.\!\!^{\prime\prime}$}}

\def\apjl{ApJL}
\def\apj{ApJ}
\def\apjs{ApJS}
\def\mnras{MNRAS}
\def\araa{ARAA}
\def\aj{AJ}
\def\aap{A\&A}
\def\aaps{A\&A Suppl.}
\def\nat{Nature}
\def\pasp{PASP}

\begin{document}
\title[The $\textit{HST}$/ACS Coma Cluster Survey - VII.]{The $\textit{HST}$/ACS Coma Cluster Survey - VII. Structure and Assembly of Massive Galaxies in the Center of the Coma Cluster}
\author[T. Weinzirl et al.]{Tim Weinzirl,$^1$ Shardha Jogee,$^1$ Eyal Neistein,$^2$ Sadegh Khochfar,$^{2,3}$ 
John Kormendy,$^1$ 
\newauthor
Irina Marinova,$^4$ 
Carlos Hoyos,$^{5}$ 
Marc Balcells,$^{6,7,8}$ 
Mark den Brok,$^{9}$ 
Derek Hammer,$^{10,11}$ 
\newauthor
Reynier F. Peletier,$^{12}$ 
Gijs Verdoes Kleijn,$^{12}$ 
David Carter,$^{13}$ 
Paul Goudfrooij,$^{14}$ 
\newauthor
John R. Lucey,$^{15}$ 
Bahram Mobasher,$^{16}$ 
Neil Trentham,$^{17}$ 
Peter Erwin,$^{2,18}$ 
Thomas Puzia$^{19}$\\ 
$^1$Department of Astronomy, 2515 Speedway, Stop C1400, Austin, TX 78712-1205\\
$^2$Max Planck Institut f\"ur extraterrestrische Physik, P.O Box 1312, D-85478 Garching, Germany\\
$^3$Institute for Astronomy, University of Edinburgh, Royal Observatory, Edinburgh EH9 3HJ, UK\\
$^4$Southwestern University, Department of Physics, 1001 E. University Avenue, Georgetown, TX 78626 \\
$^5$Instituto de Astronomia, Geofísica e Ci\^{e}ncias Atmosf\'{e}ricas - Rua do Mat\~{a}o, 1226 - Cidade Universitária S\~{a}o Paulo-SP - Brazil\\
$^6$Isaac Newton Group of Telescopes, Apartado 321, 38700 Santa Cruz de La Palma, Canary Islands, Spain\\
$^7$Instituto de Astrof\'{\i}sica de Canarias, 38200 La Laguna, Islas Canarias, Spain\\
$^8$Departamento de Astrof\'{\i}sica, Universidad de La Laguna, 38206 La Laguna, Islas Canarias, Spain\\
$^{9}$Department of Physics and Astronomy, University of Utah, Salt Lake City, UT 84112, USA\\
$^{10}$Department of Physics and Astronomy, Johns Hopkins, Baltimore, MD 21218, USA\\
$^{11}$NASA Goddard Space Flight Center, Code 662, Greenbelt, MD 20771, USA\\
$^{12}$Kapteyn Astronomical Institute, University of Groningen, Groningen, The Netherlands\\
$^{13}$Astrophysics Research Institute, Liverpool John Moores University, IC2 Liverpool Science Park, 146 Brownlow Hill, Liverpool, L3 5RF, UK\\
$^{14}$Space Telescope Science Institute, 3700 San Martin Drive, Baltimore, MD 21218, USA\\
$^{15}$ Department of Physics, Durham University, South Road, Durham DH1 3LE, UK\\
$^{16}$Department of Physics and Astronomy, University of California, Riverside, CA 92521, USA\\
$^{17}$Institute of Astronomy, Madingley Road, Cambridge, CB3 0HA, UK\\
$^{18}$Universit\"ats-Sternwarte M\"unchen, Scheinerstrasse 1, 81679 M\"unchen, Germany\\
$^{19}$Department of Astronomy and Astrophysics, Pontificia Universidad Cat\'olica de Chile, Avenida Vicu\~na Mackenna 4860, Macul, Santiago, Chile \\ 
}


\maketitle
\clearpage
\begin{abstract}
We constrain the assembly history of galaxies in the projected central 0.5~Mpc 
of the Coma cluster by performing structural decomposition on 69 massive 
($M_\star\geq10^9$~$M_\odot$) galaxies using high-resolution F814W images from 
the $\textit{HST}$ Treasury Survey of Coma. Each galaxy is modeled with up to 
three S\'ersic components having a free S\'ersic index $n$. After excluding 
the two cDs in the projected central 0.5 Mpc of Coma, 57\% of the galactic 
stellar mass in the projected central 0.5~Mpc of Coma resides in classical 
bulges/ellipticals while 43\% resides in cold disk-dominated structures. Most 
of the stellar mass in Coma may have been 
assembled through major (and possibly minor) mergers. Hubble types are assigned 
based on the decompositions, and we find a strong morphology-density relation; 
the ratio of (E+S0):spirals is (91.0\%):9.0\%. In agreement with earlier 
work, the size of outer disks in Coma S0s/spirals is smaller compared with 
lower-density environments captured with SDSS (Data Release 2). Among 
similar-mass clusters from a hierarchical semi-analytic model, no single 
cluster can simultaneously match all the global properties of the Coma cluster. 
The model strongly overpredicts the mass of cold gas and underpredicts the 
mean fraction of stellar mass locked in hot components over a wide range of 
galaxy masses. We suggest that these disagreements with the model result from 
missing cluster physics (e.g., ram-pressure stripping), and certain bulge 
assembly modes (e.g., mergers of clumps). Overall, our study of Coma 
underscores that galaxy evolution is not solely a function of stellar mass, 
but also of environment.
\end{abstract}
\nokeywords

\section{Introduction}\label{intro}

How galaxies form and evolve is one of the primary outstanding problems
in extragalactic astronomy. The initial conditions 
led to the collapse of dark matter halos which clustered hierarchically 
into progressively larger 
structures. In the halo interiors, gas formed rotating disks 
which underwent star formation (SF) to produce
stellar disks (Cole 2000; Steinmetz \& Navarro 2002).  The subsequent
growth of galaxies is thought to have proceeded through a combination
of major mergers,
(e.g., Toomre 1977; Barnes 1988; Khochfar \& Silk 2006, 2009), 
minor mergers (e.g., Oser et al. 2012, Hilz et al. 2013), cold-mode gas
accretion (Birnboim \& Dekel 2003; Kere{\v s} et al. 2005, 2009; 
Dekel \& Birnboim 2006; Brooks et al. 2009; Ceverino et al. 2010; 
Dekel et al. 2009a, b), and secular processes (Kormendy \& Kennicutt 2004).

In early simulations focusing on gas-poor mergers, the major merger of two 
spiral galaxies with mass ratio $M_1/M_2 \geq 1/4$ would inevitably destroy the 
pre-existing stellar disks by violent relaxation, producing a remnant bulge or 
elliptical having a puffed-up distribution of stars with a low ratio of 
ordered-to-random motion ($V/\sigma$) and a steep de Vaucouleurs  $r^{1/4}$ 
surface brightness profile\footnote{A de  Vaucouleurs  $r^{1/4}$ profile
corresponds to a S\'ersic (1968) profile with index $n=4$.} (Toomre 1977).
Improved simulations (Robertson  et al. 2006; Naab et al. 2006;  
Governato et al. 2007; Hopkins et al. 2009a, b) significantly revised this 
picture. In unequal-mass major mergers, violent relaxation of stellar disks is 
not complete. Furthermore, for major mergers where the progenitors have 
moderate-to-high gas fractions, 
gas-dissipative processes build disks on small and large scales 
(Hernquist \& Mihos 1995; Robertson et al. 2006; Hopkins et al. 2009a, b; 
Kormendy et al. 2009). The overall single S\'ersic index $n$ of such 
remnants are typically $2\lesssim n \lesssim 4$ (Naab et al. 2006; 
Naab \& Trujillo 2006; Hopkins et al. 2009a). 
The subsequent accretion of gas from the halo, cold streams, and minor mergers 
can further build large-scale stellar disks, whose size depends on the specific 
angular momentum of the accreted gas (Steinmetz \& Navarro 2000; 
Birnboim \& Dekel 2003; Kere{\v s} et al. 2005, 2009; Dekel \& Birnboim 2006; 
Robertson et al. 2006; Dekel et al. 2009a, b; Brooks et al. 2009; 
Hopkins et al. 2009b; Ceverino et al. 2010). Additionally, 
Bournaud, Elmegreen, \& Elmegreen (2007) and Elmegreen et al. (2009) discuss
bulge formation via the merging of clumps forming within very gas-rich, 
turbulent disk in high-redshift galaxies. These bulges can have a range of 
S\'ersic indices, ranging from $n<2$ to $n=4$.

As far as the structure of galaxies is concerned, we are still actively 
studying and debating the epoch and formation pathway for the main stellar 
components of galaxies, namely flattened, dynamically cold, disk-dominated 
components (including outer disks, circumnuclear disks, and pseudobulges) 
versus puffy, dynamically hot spheroidal or triaxial bulges/ellipticals. 
Getting a census of dynamically hot bulges/ellipticals and dynamically cold, 
flattened disk-dominated components on large and small scales in galaxies 
provides a powerful way of evaluating the importance of violent bulge-building 
processes, such as violent relaxation, versus gas-dissipative disk-building
processes.

We adopt throughout this paper the widely used definition of a bulge as the 
excess light above an outer disk in an S0 or spiral galaxy (e.g., 
Laurikainen et al. 2007, 2009, 2010; Fisher \& Drory 2008; Gadotti 2009; 
Weinzirl et al. 2009). The central bulge falls in three main categories called 
classical bulges, disky pseudobulges (Kormendy 1993; 
Kormendy \& Kennicutt 2004; Jogee, Scoville, \& Kenney 2005; Athanassoula 2005; 
Kormendy \& Fisher 2005; Fisher \& Drory 2008), and boxy pseudobulges 
(Combes \& Sanders 1981; Combes et al. 1990; 
Pfenniger \& Norman 1990;  Bureau \& Athanassoula 2005; Athanassoula 2005; 
Martinez-Valpuesta et al. 2006). Some bulges are composite mixtures 
of the first two classes (Kormendy \& Barentine 2010; 
Barentine \& Kormendy 2012). For remainder of the paper we refer to classical 
bulges simply as ``bulges'' when the context is unambiguous. 

Numerous observational efforts have been undertaken to derive
such a census among galaxies in the field environment. Photometric
studies (e.g., Kormendy 1993; Graham 2001; Balcells et al. 2003, 2007b; 
Laurikainen et al. 2007; Graham \& Worley 2008;  Fisher \& Drory 2008;
Weinzirl et al. 2009; Gadotti 2009; Kormendy et al. 2010)
have dissected field galaxies into  outer stellar disks
and different types of central bulges (classical, disky/boxy pseudobulges) 
associated with different S\'ersic index,
and compiled the stellar bulge-to-total light or mass ratio ($B/T$)
of spirals and S0s. It is found that low-$B/T$ and bulgeless galaxies are 
common in the field  at low redshifts, both among low-mass
or late-type galaxies (B{\"o}ker et al. 2002; Kautsch et al. 2006;
Barazza et al. 2007, 2008) and among high-mass spirals or early-type spirals
(Kormendy 1993; Balcells et al. 2003, 2007b; Laurikainen et al. 2007; Graham \& Worley 2008; Weinzirl et al. 2009; Gadotti 2009; Kormendy et al. 2010).
Balcells et al. (2003) highlighted the paucity of $r^{1/4}$ profiles in the 
bulges of early-type disk galaxies. Working on a bigger sample,
Weinzirl et al. (2009) report that the majority ($66.4 \pm 4.4\%$)
of massive ($M_\star \geq 10^{10}$~$M_\odot$) field spirals have
low $B/T$ ($\leq0.2$) and bulges with low S\'ersic index ($n \leq 2$).

These empirical results can be used to test models of the
assembly history of field galaxies. For instance,
Weinzirl et al. (2009) find that the results reported above 
are consistent with hierarchical semi-analytic models of galaxy evolution from
Khochfar \& Silk (2006) and Hopkins et al. (2009a), which predict that
most ($\sim 85\%$) massive field spirals have had no major merger since $z=2$.
While this work reduces the tension between theory and observations for field 
galaxies, one should note that hydrodynamical models still face challenges in 
producing purely bulgeless massive galaxies in  different environments. 

It is important to extend such studies from the field environment to rich 
clusters. Hierarchical models predict differences in galaxy merger history as 
a function of galaxy mass, environment, and redshift (Cole et al. 2000; 
Khochfar \& Burkert 2001). Furthermore, cluster-specific physical processes, 
such as  ram-pressure stripping (Gunn \& Gott 1972; Fujita \& Nagashima 1999), 
galaxy harassment (Barnes \& Hernquist 1991; Moore et al. 1996, 1998, 1999;
Hashimoto et al. 1998; Gnedin 2003), and strangulation (Larson et al.
1980), can alter SF history and galaxy stellar components (disks, bulges, bars).

Efforts to establish accurate demographics of galaxy components in clusters 
are ongoing. In the nearby Virgo cluster, Kormendy et al. (2009) find more 
than 2/3 of the stellar mass is in classical bulges/ellipticals, including the 
stellar mass contribution from M87\footnote{M87 is considered as a giant ellipticals by 
some authors and as a cD by others. The detection of intra-cluster light around 
M87 (Mihos et al. 2005, 2009) strongly supports the view that it is a cD 
galaxy. In this paper (e.g.,
Table~\ref{tabmdr}) we consider M87 as a cD when making comparisons (e.g., 
Section~\ref{census}) to Virgo.}. Furthermore, there is clear 
evidence for ongoing environmental effects in Virgo; 
see Kormendy \& Bender (2012) for a comprehensive review. 

Yet Virgo is not very rich compared with more typical clusters 
(Heiderman et al. 2009). The Coma cluster at $z=0.024$ ($D=100$~Mpc) 
has a central number density 10,000~Mpc$^{-3}$ (The \& White 1986) and is the
densest cluster in the local universe. However, ground-based data do not 
provide high enough resolution 
($1^{\prime\prime} -2^{\prime\prime} =500-1000$~pc) for accurate structural 
decomposition, an obstacle to earlier work. 

In this paper we make use of data from the $Hubble$ $Space$ $Telescope$
($\textit{HST}$) Treasury Survey (Carter et al. 2008) of Coma which provides 
high-resolution (50~pc) imaging from the Advanced Camera for Surveys (ACS). 
Our goal is to derive the demographics of galaxy components, in particular 
classical bulges/ellipticals and flattened disk-dominated components 
(including both large-scale disks and disky pseudobulges), in the Coma cluster, 
and to compare the results with lower-density environments and to theoretical 
models, to constrain the assembly history of galaxies.

In Section~\ref{data0} we present our mass-complete sample of cluster galaxies
with stellar mass $M_\star \geq 10^9$~$M_\odot$. In  Section~\ref{method} 
we describe our structural decomposition strategy.
Section~\ref{approach} describes our working assumption in this paper of 
using S\'ersic index as a proxy for tracing the disk-dominated structures 
and classical bulges/ellipticals. Section~\ref{special} outlines our procedure 
for structural decomposition, and we refer the reader to Appendix~\ref{app} for 
a more detailed description.
Section~\ref{scheme} overviews the scheme we use to assign morphological
types to galaxies. In Section~\ref{mdr}, we quantitatively assign galaxy types 
based on the structural decompositions. We also make a census 
(Section~\ref{census}) of structures built by dissipation versus violent 
stellar processes, explore how stellar mass is distributed in different galaxy 
components (Section~\ref{discussmass}), and consider galaxy scaling relations 
(Section~\ref{discussScaling}). In Section~\ref{discuss3}, we evaluate and 
discuss the effect of cluster environmental processes. In Section~\ref{theory} 
we compare our empirical results with theoretical models, 
after first identifying Coma-like environments in the simulations. Readers not
interested in the complete details about the theoretical model can
skip to Sections~\ref{globalprop} and \ref{datavmodel}.
We summarize our results in Section~\ref{summary}.

We adopt a flat $\Lambda$CDM 
cosmology with $\Omega_\Lambda = 0.7$ and $H_0 = 73$~km~s$^{-1}$~Mpc$^{-1}$.
We use AB magnitudes throughout the paper, except where indicated otherwise.

\section{Data and Sample Selection}\label{data0}
This study is based on the data products from the $\textit{HST}$/ACS Coma 
Cluster Treasury Survey (Carter et al. 2008), which provides ACS Wide Field 
Camera images for 25 pointings  spanning 274 arcmin$^2$ in the F475W and F814W 
filters.  The total ACS exposure time per pointing is typically 2677 seconds in 
F475W and 1400 seconds in F814W. Most (19/25) pointings are located within 
0.5~Mpc from the central cD galaxy NGC~4874, and the other 6/25 pointings are 
between 0.9 and 1.75~Mpc southwest of the cluster center.  
The FWHM of the ACS point-spread function (PSF) is $\sim 0\farcs1$ (Hoyos et al. 2011), corresponding to $\sim 50$~pc at the 100~Mpc distance of the Coma 
cluster (Carter et al. 2008). Note the 19 pointings cover only 
$19.7\%$ by area of the projected central 0.5~Mpc of Coma. This limited 
spatial coverage of ACS in the projected central 0.5~Mpc of Coma may introduce 
a possible bias in the sample due to cosmic variance. We quantify this effect 
in Appendix~\ref{cosmicvar} and discuss the implications throughout the paper.

Hammer et al. (2010) discuss the images and \texttt{SExtractor} source 
catalogs for Data Release 2.1 (DR2.1). The F814W $5\sigma$ limit for point 
sources is 26.8 mag (Hammer et al. 2010), and we estimate the $5\sigma$ F814W 
surface brightness limit for extended sources within a $0\farcs7$ diameter 
aperture to be 25.6 mag/arcsec$^2$. Several of the ACS images in DR2.1 suffer 
from bias offsets on the 
inter-chip and/or inter-quadrant scale that cause difficulty in removing the sky
background. We use the updated ACS images reprocessed to reduce the impact of 
this issue. The DR2.1 images are used where this issue is not present.

\subsection{Selection of Bright Cluster Members}\label{datas1}

We select our sample based on the eyeball catalog of N. Trentham et al. (in
preparation), with updates from Marinova et al. (2012). This catalog provides
visually determined morphologies and cluster membership status for galaxies
with an apparent magnitude F814W $\leq 24$ mag. 
Morphology classifications in this catalog come from a combination of 
RC3 (de Vaucouleurs et al. 1991) and visual inspection. In Section~\ref{mdr}
we assign Hubble types based only on our own multi-component decompositions.

Cluster membership is ranked from 0 to 4 following the method of Trentham \& 
Tully (2002). Membership class 0 means the galaxy is a spectroscopically 
confirmed cluster member.  The subset of spectroscopically confirmed cluster 
members was identified based on published redshifts (Colless \& Dunn 1996; 
Adelman-McCarthy et al. 2008; Mobasher et al. 2001; Chiboucas et al. 2010) and
is approximately complete in surface brightness at the galaxy half-light radius 
($\mu_\mathrm{e, F814W}$) to $\sim 22.5$ mag/arcsec$^2$ (den Brok et al. 2011). 
The remaining galaxies without spectroscopic confirmation are assigned a rating 
of 1 (very probable cluster member), 2 (likely cluster member), 3 (plausible 
cluster member), or  4 (likely background object) based on a visual estimation 
that also considers surface brightness and morphology.

From this catalog, we define a sample S1 of 446 cluster members having F814W 
$\leq 24$ mag and membership rating 0-3 located within the projected central 
0.5~Mpc of Coma, which is the projected radius probed by the central ACS 
pointings. To S1 we add the second central cD galaxy NGC~4889, which is not 
observed by the ACS data. The majority (179) of S1 galaxies have member class 
0, and 30, 131, and 106 have member 
class 1, 2, and 3, respectively. 

\subsection{Calculation of Stellar Masses}\label{datas2}

Stellar masses are a thorny issue. Uncertainties in the mass-to-light 
ratios of stellar populations ($M/L$) arise from a poorly known initial mass 
function (IMF) as well as degeneracies between age and 
metallicity. We calculate stellar masses based on the $HST$ F475W and 
F814W-band photometry. First, we convert the $HST$ (AB) photometry to the 
Cousins-Johnson (Vega) system using
\begin{equation}
\rm{ I=F814W -0.38}
\end{equation}
from the WFPC2 Photometry Cookbook and 
\begin{equation}
\rm{ B-I=1.287\left(F475W - F814W\right) + 0.538 }
\end{equation}
from Price et al. (2009).

Next, we calculate $I$-band $M/L$ from the calibrations of Into \& Portinari 
(2013) for a Kroupa et al. (1993) initial mass function (IMF) with
\begin{equation}
\rm{ I_{lum}=10^{\left( -0.4\left( I-35-4.08\right) \right)} }
\end{equation}
and
\begin{equation}
\rm{ M_\star=I_{lum} \times 10^{ \left( 0.641\left(B-I\right) -0.997 \right) }},
\end{equation}
where $I$ corresponds to the apparent \texttt{MAG\_AUTO} \texttt{SExtractor} 
magnitude\footnote{For galaxies COMAi125935.698p275733.36 $=$ NGC~4874 and COMAi125931.103p275718.12, 
\texttt{SExtractor} vastly underestimates the total F814W luminosity, and the
calculation is instead made with the total luminosity derived from structural
decomposition (Section~\ref{special}).}, 35 is the distance modulus to Coma, and 4.08 is the solar absolute 
magnitude in $I$-band.

We use the above method to calculate stellar masses for all galaxies in S1
except NGC~4889, which does not have ACS data. 
For NGC~4889, we use $gr$ Petrosian magnitudes from SDSS DR10 
(Ahn et al. 2013). Stellar masses are determined using the 
relations of Bell et al. (2003) and assuming a Kroupa IMF, namely
\begin{equation}
\rm{ g_{lum}=10^{\left( -0.4\left( g-35-5.10\right) \right)} }
\end{equation}
and
\begin{equation}
\rm{ M_\star=g_{lum} \times 10^{ \left( -0.499+1.519\left( g-r\right) -0.1 \right) }},
\end{equation}
where $g$ and $r$ are apparent SDSS magnitudes, 35 is the distance modulus to Coma, and
5.10 is the solar absolute magnitude in $g$-band\footnote{The Kroupa IMF offset term reported as -0.15 in 
Bell et al. 2003 was calculated assuming unrealistic conditions (Bell, E., private 
communication). The correct value is -0.1 and is used in Borch et al. (2006).}.

It is hard to derive the stellar mass of cD galaxies for several reasons.
The stellar $M/L$ ratio of cDs is believed to be high ($M_{\rm dyn}/L_B>100$; 
Schneider 2006),  but is very uncertain as most of the light
of a cD is in an outer envelope made of intra-cluster light 
and galaxy debris. Another problem is that even if one knew the correct 
stellar $M/L$ ratio, it is likely that the available photometry from ACS and
SDSS is missing light from the extended low surface brightness envelope.
Given all these factors, it is likely that the above equations, which are 
typically used to convert  color to $M_\star$ for normal representative 
galaxies, are underestimating the $M/L$ ratios and stellar masses of the cDs, 
so that the adopted stellar masses for the cDs 
($M_\star \sim6-8\times 10^{11}$~$M_\odot$) are lower limits. Due to the 
uncertain stellar masses of the cDs, we present many of our results without 
them, and we take care to consider them separately from the less massive galaxy 
population of E, S0, and spiral galaxies.

\subsection{Selection of Final Sample of Massive Galaxies}\label{datas3}

The left panels of Figure~\ref{lummasstype} show the distributions of F814W 
magnitudes (upper panel) and stellar masses (lower panels) for sample S1, 
while in the right panels of the same figure the correlations of stellar 
masses with F814W magnitudes (upper panel) and $g-r$ colors (lower panel) 
are shown.

In this paper, we focus on massive ($M_\star \geq 10^9$~$M_\odot$) galaxies. 
Our rationale is that we are specifically interested in 
understanding the evolution of the most massive cluster galaxies through 
comparisons with model clusters (Section~\ref{theory}) which show mass 
incompleteness at galaxy stellar masses $M_\star < 10^9$~$M_\odot$.
We found for sample S1 that imposing the mass cut \
$M_\star \geq 10^9$~$M_\odot$ 
effectively removes most galaxies identified in the Trentham et al. catalog as 
dwarf/irregular and very low surface brightness galaxies. With this cut,
we are left with 75 galaxies that consist primarily of E, S0, and spiral 
galaxies, two cDs, and only six dwarfs. Three out of 75 galaxies are 
significantly cutoff the ACS detector, and we ignore these sources.  
Of the remaining 72 galaxies, 69/72 have spectroscopic redshifts. The 3/72
galaxies without spectroscopic redshifts appear too red to be in Coma
(Figure~\ref{lummasstype}d), and the estimated SDSS DR10 photometric redshifts 
are much larger than the redshift of Coma (0.024). We also neglect these three 
sources as they are unlikely to be Coma members. Our final working sample S2 
consists of the 69 galaxies inside the projected central 0.5~Mpc with stellar 
mass $M_\star \geq 10^9$~$M_\odot$ and spectroscopic redshifts.
Table~\ref{tab:cross_ids} cross references our sample with other datasets.

\section{Method and Analysis}\label{method}

\subsection{Using S\'ersic Index as a Proxy For Tracing Disk-Dominated Structures 
and Classical Bulges/Ellipticals}\label{approach}
As outlined in Section~\ref{intro}, galaxy bulges and stellar disks hold 
information on galaxy assembly history. The overall goal in this work is to 
separate galaxy components into groups of classical bulges/ellipticals versus 
disk-dominated structures.

It is common practice (e.g., Laurikainen et al. 2007; Gadotti 2009; 
Weinzirl et al. 2009) to characterize galaxy structures
(bulges, disks, and bars) with generalized ellipses whose radial light
distributions are described by the  S\'ersic (1968) profile:
\begin{equation}
\rm{ I(r)=I_e \ exp \left( -b_n \left[ \left( r\over{r_\mathrm{e}} \right)  ^{1/n} -1 \right]  \right) },
\end{equation}
where $I_e$ is the surface brightness at the effective radius $r_\mathrm{e}$ 
and $b_n$ \footnote{The precise values of $b_n$
are given from the roots of the equation $\Gamma(2n) - 2\gamma(2n,b_n)=0$, 
where $\Gamma$ is the gamma function and $\gamma$ is the incomplete gamma 
function.} is a constant that depends on S\'ersic index $n$.  

In this paper, we adopt the working assumption that in intermediate and 
high-mass ($M_\star \geq 10^9$ $M_\odot$) galaxies, a low S\'ersic index $n$ 
below a threshold value $n_{\rm disk\_max}$ corresponds to a dynamically cold 
disk-dominated structure. Note we specify ``disk-dominated'' rather than 
``pure disk'' as we refer to barred disks and thick disks.   While this 
assumption is not necessarily waterproof, it is based on multiple lines of 
compelling evidence that are outlined below. 

\begin{enumerate}
\item
Freeman (1970) showed that many large-scale disks of S0 and spiral galaxies are 
characterized by an exponential light  profile (S\'ersic index $n=1$) over 4-6 
disk scalelengths.  Since then, it has become standard practice in studies of 
galaxy structure to model the outer disk of S0s and spirals with an exponential 
profile (e.g., Kormendy 1977; Boroson 1981; Kent 1985; de Jong 1996; 
Baggett et al. 1998; Byun \& Freeman 1995; Allen et al. 2006; Laurikainen 2007; 
Gadotti 2009; Weinzirl et al. 2009). 

\item
On smaller scales, flattened, rotationally supported inner disks with high 
$V/\sigma$ (i.e., disky pseudobulges) have been associated with low S\'ersic 
index $n\lesssim2$ (Kormendy 1993; Kormendy \& Kennicutt 2004; Jogee, 
Scoville, \& Kenney 2005; Athanassoula 2005; Kormendy \& Fisher 2005; 
Fisher \& Drory 2008; Fabricius et al. 2012). This suggests $n_{\rm disk\_max}$ 
should be close to 2. 

Fabricius et al. (2012) explore the major-axis kinematics of 45 S0-Scd galaxies 
with high-resolution spectroscopy. They demonstrate a systematic agreement 
between the shape of the velocity dispersion profile and the bulge type as 
indicated by the S\'ersic index. Low S\'ersic index bulges have both increased 
rotational support (higher $<V^2>/<\sigma^2>$ values) and on average lower 
central velocity dispersions. Classical bulges (disky pseudobulges)
show have centrally peaked (flat) velocity dispersion profiles whether
identified visually or by a high S\'ersic index.

\item
At high ($z\sim2$) redshift, where it is not yet possible to fully resolve 
galaxy substructures, it has become conventional to use the global S\'ersic 
index $n\lesssim2$ in massive galaxies to separate disk-dominated versus 
bulge-dominated galaxies (e.g., Ravindranath et al. 2004; 
van der Wel et al. 2011; Weinzirl et al. 2011). Weinzirl et al. (2011) further 
explore the distributions of ellipticities ($1-b/a$) for the massive $z\sim2$ 
galaxies with low ($n\leq2$) and high ($n>2$) global S\'ersic index. They find 
galaxies with low global S\'ersic index $n\leq2$ have a distribution of 
projected ellipticities more similar to massive $z\sim0$ spirals than to 
massive $z\sim0$ ellipticals.

\end{enumerate}

The above does not allow for low-$n$, dynamically hot structures. 
A low-$n$ dynamically hot structure would be considered in our study as
a pure photometric disk, a low-$n$ bulge, or an unbarred S0 galaxy. The 
error due to misunderstood objects in the first two groups is expected to be 
small or nonexistent.  There is only one pure photometric disk in the sample
(Section~\ref{mdr}) and low-$n$ bulges ($N=20$) only make up 2.2\% of galaxy 
stellar mass (excluding the cDs, Section~\ref{census}).
Furthermore, Figure~15 of Fabricius et al. (2012) shows that no low-$n$ bulge 
turns out to be dynamically hot. 

There are 20 unbarred S0 galaxies in our sample, and these 
account for 18.5\% of the galaxy stellar mass (excluding the cD galaxies). 
About 75\% of these objects have stellar mass and luminosity consistent with 
dwarf spheroidal galaxies (Kormendy et al. 2009).  Even if some of these 
systems are actually dwarf spheroidals, they may not be dynamically hot as
some studies (e.g., Kormendy et al. 2009, Kormendy \& Bender 2012) claim that 
many dwarfs are actually disk systems closely related to 
dIrr, which have been stripped of gas via supernova feedback or environmental 
effects. The remaining 25\% would be misclassified elliptical
galaxies as they are too bright and massive to be dwarfs.  Note,
however, that Figure 33 of Kormendy et al. (2009) shows that elliptical
galaxies with $M_V < -18$ and S\'ersic $n<2$ are very rare. In the
worse-case scenario that all of our unbarred S0 galaxies are dynamically hot 
structures, our measurement of the dynamically hot stellar mass in 
Section~\ref{census} would be too low by $\sim 30\%$.

The second natural  related working assumption in our paper is
 that in intermediate and high-mass 
($M_\star \geq 10^9$ $M_\odot$) galaxies, components with S\'ersic 
$n> n_{\rm disk\_max}$ 
are classical bulge/elliptical components (defined in Section~\ref{intro}).
Such bulges/ellipticals are formed by the redistribution of stars during major 
and minor galaxy collisions. $N$-body simulations
show that minor mergers consistently raise the bulge S\'ersic index
(Aguerri et al. 2001; Eliche-Moral et al. 2006; Naab \& Trujillo 2006).
The effect of successive minor mergers is cumulative (Aguerri et al. 2001; 
Bournaud, Jog, \& Combes 2007; Naab et al. 2009; Hilz et al. 2012).

We empirically set $n_{\rm disk\_max}$ to 1.66 based on looking at the 
S\'ersic $n$ of outer disks in those Coma galaxies that are barred, and by 
definition, must harbor outer 
disk since bars are disk features. Appendix~\ref{2cp} and Appendix~\ref{nuperr} 
discuss the empirical details behind this choice.

\subsection{Overview of Our Structural Decomposition Procedure}\label{special}

For our mass-complete sample of 69 intermediate-to-high mass
($M_\star \geq 10^9$~$M_\odot$) galaxies, we use deep, high-resolution 
($0\farcs1$ or 50~pc), F814W-band images of Coma from $\textit{HST}$/ACS, 
which allow for accurate structural decomposition.
We fit galaxies with one, two, or three S\'ersic profiles, 
plus a nuclear point source, when needed  (see Appendix~\ref{app} 
for details).  We use GALFIT (Peng et al. 2002). 
In a model with one or more S\'ersic profiles, there is expected to be coupling 
between the free parameters, particularly $r_\mathrm{e}$ and $n$, although most 
previous studies have generally ignored this effect. Weinzirl et al. (2009) 
explores the issue of parameter coupling for barred and unbarred spiral 
galaxies.

We take some precautions to ensure accurate decompositions:

\begin{enumerate}
\item  
We fit all structures with a generalized S\'ersic profile where
the S\'ersic index is a free parameter (Section~\ref{scheme}).
This limits the number of a priori assumptions on the physical 
nature or shape of galaxy structures.

\item  
In clusters, the featureless (i.e., no spiral arms delineated by young stars, 
rings of SF, or gas/dust lanes) outer disks of gas-poor S0s are not readily
distinguished from the equally featureless outer stellar components of 
classical ellipticals. We do this in essence by applying $n_{\rm disk\_max}$
to the S\'ersic index $n$ of the outer galaxy structure.

\item Not requiring outer disks to have an exponential $n=1$ profile 
accommodates non-exponential disk structures (e.g., disks with down-bending 
truncations or up-bending anti-truncations Freeman 1970;
van der Kruit 1979; van der Kruit \& Searle 1981a, 1981b;
de Grijs et al. 2001; Pohlen et al. 2002; Matthews \& Gallagher 1997;
Erwin et al. 2005; Pohlen \& Trujillo 2006; Maltby et al. 2012) that are 
rotationally supported. 

\item
Stellar bars, ovals/lenses, and nuclear point sources are modeled when needed, 
which is critical for obtaining a reliable characterization of the bulge 
(e.g., Balcells et al. 2003; Laurikainen et al. 2005, 2007; 
Weinzirl et al. 2009). 
\end{enumerate}

Our structural decomposition scheme and decision sequence are described
in detail in Appendix~\ref{decomp}, illustrated in Figures
\ref{fitflow} and  \ref{fitflow2}, and briefly outlined below:

\begin{itemize}
\item
{\it Stage 1 (Single S\'ersic fit with nuclear point source if needed):}
The single S\'ersic model is adopted if either the galaxy does not show any 
coherent structures (e.g., inner/outer disks, bars, bulges, rings, or spiral 
arms) indicating the need for additional S\'ersic components, or, 
alternatively, if the galaxy has a core - a light profile that deviates 
downward from the inward extrapolation of the S\'ersic profile (see 
Appendix~\ref{core}). Such galaxies are interpreted as photometric ellipticals 
if the single S\'ersic index is above a threshold value $n_{\rm disk\_max}$ 
associated with disks (Section~\ref{approach}, Appendix~\ref{2cp}, and 
Appendix~\ref{nuperr}); otherwise they are considered photometric disks. Three 
galaxies show convincing evidence for being cores, and these are luminous 
objects with high single S\'ersic $n > n_{\rm disk\_max}$ (see 
Appendix~\ref{2cp}, Table~\ref{tabcore}, Appendix~\ref{core}). The results of 
Stage 1 are listed in Table~\ref{1cptab}. See Appendix~\ref{1cp} for 
additional details on the single S\'ersic fits. 

\item
{\it Stage 2 (Double S\'ersic model with nuclear point source if needed):}
Galaxies showing coherent structure in the Stage 1 residuals are subjected to 
a two-component S\'ersic + S\'ersic fit, with nuclear point source if needed 
(see Figure~\ref{fitflow2}). This two-component model is intended to model the 
inner (C1) and outer (C2) galaxy structures. 

There are two possible outcomes.  a) If the outer component C2 is an outer disk 
based on having S\'ersic index $n\leq n_{\rm disk\_max}$, then the galaxy is 
considered a spiral or S0 with an outer disk having a photometric bulge and, 
in some cases, a large-scale bar. b) If the outer component C2 does not meet 
our definition of an outer disk, then the galaxy is considered a photometric 
elliptical having an outer component C2 with $n>n_{\rm disk\_max}$ and an inner
component C1 of any $n$. See Appendix~\ref{2cp} for details.

\item
{\it Stage 3 (Triple S\'ersic model with nuclear point source if needed):}
Case (a) in Stage 2 identifies spiral and S0 galaxies with an outer disk. 
These galaxies are further processed as follows: a) If there is evidence for a 
large-scale bar (see Appendix~\ref{2cp}), then a triple S\'ersic profile is 
fitted in Stage 3 for the photometric bulge, disk, and bar.  b) Otherwise, the 
galaxy is considered as unbarred and the double S\'ersic fit for a photometric 
bulge and disk is adopted.  In both cases (a) and (b), it is important to note 
that the photometric bulge is allowed to have any S\'ersic index $n$, thus 
allowing for structures with $n\leq n_{\rm disk\_max}$ and structures with 
$n>n_{\rm disk\_max}$.
\end{itemize}

\subsection{Overview of  Our Galaxy Classification Scheme}\label{scheme}

The decomposition scheme discussed above and in Figures \ref{fitflow} and 
\ref{fitflow2} leads naturally to the galaxy classification system outlined in 
Figure~\ref{galaxyclass}, where there are five main galaxy types, G1 to G5.  
Systems best fitted by single S\'ersic models (plus a nuclear point source if 
present) represent galaxies of type G1 and G2. Systems best fitted by two or 
three S\'ersic profiles (plus a nuclear point source if present) represent 
galaxies of type G3 to G5.

\begin{enumerate}
\item
G1: Photometric disk with $n\leq n_{\rm disk\_max}$ (plus a nuclear point 
source if present).

\item
G2: Photometric elliptical with $n > n_{\rm disk\_max}$ (plus a nuclear point 
source if present).

\item
G3: Unbarred S0 or spiral having an outer disk with $n\leq n_{\rm disk\_max}$ 
and an inner photometric bulge of any $n$ (plus a nuclear point source if 
present).

\item
G4: Barred S0 or spiral having an outer disk with $n\leq n_{\rm disk\_max}$, 
a bar, and an inner photometric bulge of any $n$ (plus a nuclear point 
source if present).

\item
G5: Photometric elliptical having an outer component with 
$n > n_{\rm disk\_max}$ and an inner component of any $n$.
\end{enumerate}

This galaxy classification scheme has multiple advantages.  Firstly, it 
allows us to identify low-$n$ disk-dominated structures within galaxies, both 
on large scales and in the 
central regions, in the form of outer disks with $n\leq n_{\rm disk\_max}$ 
in spirals and S0s, photometric bulges with $n\leq n_{\rm disk\_max}$ in 
spirals and S0s (representing disky pseudobulges), 
and inner disks within ellipticals represented by a component C1 having 
$n\leq n_{\rm disk\_max}$. Furthermore, it allows a census of galaxy 
components with $n > n_{\rm disk\_max}$ more akin to 
classical bulges/ellipticals. Our scheme does not allow for
low-$n$ dynamically hot components. As discussed in 
Section~\ref{approach}, this is not a problem because in our sample 
such structures are not expected to be present in large numbers.

Table~\ref{tabdecomp} lists 
the distribution of best-fit models for the sample of galaxies with stellar
 mass $M_\star \geq 10^9$~$M_\odot$, and the breakdown of galaxies into classes 
G1 to G5. Table~\ref{multicptab} lists the structural parameters from the best
single or multi-component model.  In summary, we fit 6, 38, and 25 galaxies 
with 1, 2, and 3 S\'ersic profiles, respectively. Our best-fit models have 
reduced $\chi^2$ of order one. In terms of galaxy types G1 
to G5, we assign 1, 5, 24, 25, and 14 objects to classes  G1, G2, G3, G4, and 
G5, respectively. The number of Stage 3 fits implies the bar fraction among 
galaxies with an extended outer disk is $50.0 \pm 7.1\%$, and this is
consistent with the bar fraction in Coma derived by Marinova et al. (2012). 

\section{Empirical Results on Galaxy Structure}\label{results}

\subsection{Galaxy Types and Morphology-Density Relation in the Center of Coma}\label{mdr}

We next map classes G1 to G5 to more familiar Hubble types, namely cD, 
photometric E, S0, and spiral. The Hubble types assigned here depend 
only on the morphology classes (G1 to G5) associated with structural 
decomposition; they are independent of the morphological types from the 
Trentham et al. (in prep.) catalog discussed in Section~\ref{data0}. The 
results are shown in Table~\ref{tabdecomp}, and this process is explained in 
detail below. 

The one object in class G1 (photometric disk) has a single S\'ersic 
index~$n \leq n_{\rm disk\_max}$ and a nuclear point source. This object has 
no visible spiral arms, so it is an S0. Objects assigned to class G2 
(photometric ellipticals) have single S\'ersic index $n > n_{\rm disk\_max}$ 
and include two known central cD galaxies, NGC~4874 and NGC~4889. We label 
these two sources separately as cD galaxies because they contain a 
disproportionately large fraction of the stellar mass.
Classes G3 (unbarred S0, spiral) and G4 (barred S0, spiral) represent S0 or
spiral disk galaxies with a possible large-scale bar. We label the six 
galaxies in either class G3 or G4 showing spiral arms in the data or
residual images as spirals, while the remaining sources are labeled S0.
Class G5 objects are identified as photometric ellipticals 
having an outer component with $n > n_{\rm disk\_max}$ and an inner 
component of any $n$. 

Considering the Hubble types assigned above, we find evidence of a 
strong absence of spiral galaxies. In the projected central 0.5~Mpc of the 
Coma cluster, there are 2 cDs  (NGC~4874 and NGC~4889), spirals are rare, and 
the morphology breakdown of (E+S0):spirals is (25.3\%+65.7\%):9.0\% by numbers 
and (32.0\%+62.2\%):5.8\% by stellar mass. Note that our ratio of 
E-to-S0 galaxies is lower than found elsewhere for Coma (e.g., 
Gavazzi et al. 2003) and for other clusters (e.g., Dressler 1980; 
Fasano et al. 2000; Poggianti et al. 2009), where it is $\sim1-2$. This is 
driven by the effect of cosmic variance on our sample 
(Appendix~\ref{cosmicvar}). Also, the total stellar mass cited here does not 
include the cDs as their stellar mass is quite uncertain (see 
Section~\ref{datas2}).

In contrast to the central parts of Coma, lower-density environments are 
typically dominated by spirals. This is quantitatively illustrated by  
Table~\ref{tabmdr}, 
which compares the results in Coma with the lower-density Virgo cluster and the 
field. We note that Virgo has significantly lower projected galaxy number 
densities and halo mass (Binggeli et al. 1987) than the center of Coma.
McDonald et al. (2009) study a sample of 286  Virgo cluster member 
galaxies that is complete down to $B_T=16$ (Vega mag). At stellar mass 
$M_\star \geq 10^9$~$M_\odot$, if M87 is counted as a giant elliptical,  the 
(E+S0):spirals breakdown is (34.1\%+31.6\%):34.8\%  by numbers and  
(59.2\%+19.3\%):21.4\% by stellar mass. There is evidence  
(Mihos et al. 2005, 2009; Kormendy et al. 2009) that M87 has a cD halo, and 
after excluding M87, the (E+S0):spirals breakdown  changes slightly to 
(33.5\%+31.6\%):34.8\%  by numbers and (57.2\%+20.3\%):22.5\% by stellar mass. 
In the field, the  (E+S0):spiral morphology breakdown is  $\sim$~20\%:80\% by 
number for bright galaxies (Dressler 1980).

\subsection{What Fraction of Total Galactic Stellar Mass is in Disk-Dominated Structures 
Versus Classical Bulges/Ellipticals?}\label{census}
Here and in Section~\ref{discussmass}, we discuss the stellar mass breakdown 
among galaxy components within each galaxy type. Our results are summarized 
in Tables~\ref{masstable2} and \ref{masstable}. 

Recall that in Section~\ref{datas2}, the total stellar masses were 
computed through applying calibrations of $M/L$ to the $HST$ F475W and F814W 
photometry. To calculate the stellar mass in galaxy substructures we assume a 
constant $M/L$ ratio and simply multiply the F814W light ratio of each 
component by the total galaxy stellar mass. 
A more rigorous approach is to also perform the decompositions
in the F475W band and to fold the colors of galaxy substructures into
the calculation. In Appendix~\ref{colorgrad}, we
consider the effect of galaxy color gradients for a subset of galaxies;
the effect of the color gradients on the stellar mass fractions is small 
($\sim5\%$) and does not impact our conclusions.

Table~\ref{masstable2} summarizes our attempt at providing a census of 
the stellar mass among disk-dominated components and classical 
bulges/ellipticals, in the projected central 0.5~Mpc
of Coma, excluding the two cDs. We highlight the main results below.

\begin{enumerate}

\item{\it Stellar mass in low-$n$ flattened disk-dominated structures ($43\%$):}\\
The total stellar mass in small and large-scale disk-dominated components is $\sim36.0\%$. Bars are disk-dominated components in
the sense that they are flattened non-axisymmetric components.  
Bar proportions typically range from 2.5:1 to 5:1 in their equatorial 
plane (Binney \& Tremaine 1987). 
The stellar mass percentage in bars is 6.8\%. 
Thus, the total fraction mass in disk-dominated components is $43\%$.\\

\item{\it Stellar mass in high-$n$ classical bulges/ellipticals ($57\%$):}\\
The remaining stellar mass is in components with $n> n_{\rm disk\_max}$.
These components include the outer components of photometric ellipticals,
the central components with $n> n_{\rm disk\_max}$ in photometric
ellipticals, and the bulges of S0s and spirals with $n> n_{\rm disk\_max}$.
The percent stellar mass in these systems is 57\%.  \\

\item{\it Environmental dependence of disk-dominated structures :}\\
Finally, we discuss how $f_{\rm disk\_dominated}$, the fraction of galactic stellar 
mass in disk-dominated structures, varies with environment.
For the
lower density field-like environments studied by Weinzirl et al.
(2009), this fraction  $f_{\rm disk\_dominated}$  is $\sim89.6\%$  
for galaxies with $M_\star  \geq 10^{10}$~$M_\odot$. Applying the same
mass cut in Coma, the fraction  $f_{\rm disk\_dominated}$ 
is $\sim40.1\%$, which is lower than in the field
as expected.

Due to the effect of cosmic variance on our sample 
(Appendix~\ref{cosmicvar}), our measurement of disk-dominated stellar mass 
is larger by an estimated factor of 1.27, compared to what would be 
obtained from an unbiased sample. This is estimated by weighting the 
fraction of hot and cold stellar mass in elliptical, S0, and spiral 
galaxies (Table~\ref{masstable}) with the morphology-density
distribution from GOLD Mine for the projected central 0.5~Mpc of Coma.

We also note here the results for the Virgo cluster, in which Kormendy
et al. (2009) find that in galaxies with 
$M_\star \gtrsim 5\times 10^9$~$M_\odot$,
more than 2/3 of the stellar mass is in classical
bulges/ellipticals, implying that $f_{\rm disk\_dominated}$
is less than 1/3.  It may seem surprising that our value
of  $f_{\rm disk\_dominated}$ in Coma is higher than the value of 1/3 for Virgo.
However, we believe this apparent discrepancy is due to the fact that
the Virgo study includes the giant elliptical galaxy M87, which is
marginally classified as a cD (Kormendy et al. 2009), while our study excludes
the two cDs in the central part of Coma.  If we include these 2 cDs
and adopt a conservative lower limit for their stellar mass,
then the fraction $f_{\rm disk\_dominated}$  of stellar mass in the low-$n$
component would be less than 27\%,  since the cDs add their mass
to high-$n$ stellar components (see Appendix~\ref{scD}).

\end{enumerate}

\subsection{What Fraction of Stellar Mass within S0, E, Spirals is in 
Disk-Dominated Structures versus Classical Bulges/Ellipticals?}\label{discussmass}

We now discuss how the stellar mass is distributed among E, S0, and spiral 
Hubble types in the projected central 0.5~Mpc of Coma. As above, fractional 
stellar masses are reported without including the cD galaxies. 

\begin{enumerate}
\item {\it Mass distribution among high-$n$ classical bulges/ellipticals versus low-$n$ disky pseudobulges in Coma S0s and spirals:}\\
Bulges account for $\sim30.5\%$ of the stellar mass across E, S0, and spiral 
galaxies. The ratio $R$ of stellar mass in high-$n$ ($n\gtrsim1.7$) classical
bulges to low-$n$  ($n\lesssim1.7$) disky pseudobulges is $28.3\%/2.2\%$ or 
12.9.\\

\item {\it Mass distributions among bulges in Coma S0s versus S0s in lower-density environments:}\\
We next compare the bulges of Coma S0s versus S0s in lower-density
environments (LDEs). The results are summarized in Table~\ref{tabS0}.
We base this comparison on the results of Laurikainen et al. (2010), who
derive structural parameters from $2D$ multi-component decompositions of 
117 S0s in LDEs that include a mix of field and Virgo environments. 
For S0s in these LDEs with $M_\star \geq 7.5\times 10^9$ $M_\odot$,
the ratio $R$ of stellar mass  in high-$n$ ($n\gtrsim1.7$) 
classical bulges to low-$n$  ($n\lesssim1.7$) disky pseudobulges is
30.6\%/4.7\% or 6.5, while it is
41.7\%/2.4\% or 17.4 in the projected central 0.5~Mpc of Coma.
Note the difference in mass stored in high-$n$ and low-$n$ bulges is not 
due to a greater frequency of high-$n$ bulges, which
is similar at this mass range.\\

\item {\it Mass distribution in outer and inner components of photometric ellipticals in Coma:}\\
By definition in Section~\ref{mdr}, photometric ellipticals have no outer disk.
The outer components of these ellipticals have S\'ersic $n$ from 1.72 to 6.95, 
with a median value of 2.1. The total fractional stellar mass of the outer 
structures in ellipticals relative to our sample (minus the cDs) is 
$\sim25.9\%$. Photometric ellipticals may contain an inner component of any 
S\'ersic $n$, and we find a range in $n$ of 0.31 to 5.88 in S\'ersic index, 
with a median of 1.0. Inner components with $n\leq n_{\rm disk\_max}$ represent 
compact inner disks analogous to the disky pseudobulges in S0s and spirals; 
most of these inner components (9/14 or $64.3 \pm 12.8\%$) qualify as inner 
disks.
\end{enumerate}

\subsection{Scaling Relations for Outer Disks and Bulges}\label{discussScaling}
Here, we explore scaling relations for the bulges and outer disks 
in the projected central 0.5~Mpc of the Coma cluster. We assess 
how these structures compare with outer disks and bulges in LDEs, such as 
field, groups, and even low-density clusters similar to the Virgo cluster, 
where environmental processes and merger histories are likely to be different. 

For this comparison, we use the results of Gadotti (2009), who studies face-on
($b/a\geq0.9$) galaxies from the SDSS Data Release~2 in a volume limited sample 
at $0.02\leq z \leq 0.07$. He derives galaxy structure from 2D decompositions
of multi-band $gri$ images that account for bulge, disk, and bar components. 
The Coma sample S0s/spirals have stellar mass 
$10^9 \leq M_\star \leq 6\times 10^{10}$~$M_\odot$, and for this comparison we 
consider only galaxies with stellar mass 
$5\times 10^9 \leq M_\star \leq 6\times 10^{10}$~$M_\odot$. We proceed with the
caveat that the sample from Gadotti (2009) is incomplete in mass for
$M_\star \lesssim 5\times10^{10}$~$M_\odot$.

Figure~\ref{scaling16} compares properties of large-scale disks (size, 
luminosity) with galaxy $M_\star$. Figure~\ref{scaling16}a explores the 
\textit{projected} half-light radius in the $i$-band ($r_\mathrm{e}$) of outer 
disks along the major axis at a given galaxy $M_\star$ in Coma versus LDEs. It 
shows that at a given galaxy $M_\star$, the average disk $r_\mathrm{e}$ is 
smaller in the projected central 0.5~Mpc of Coma compared with LDEs by 
$\sim30-82\%$. 
While the scatter in disk $r_\mathrm{e}$ is large, the separation 
between the two mean values in each mass bin is larger than the sum of the 
errors. The suggestion that outer disks in Coma are more compact is
consistent with the results of previous analyses of disk 
structure in Coma (Guti{\'e}rrez et al. 2004; Aguerri et al. 2004).
Figure~\ref{scaling16}b makes a similar comparison for the outer disk
luminosity between Coma and LDEs. We use here the ACS F814W photometry for
Coma and the SDSS $i$-band photometry from Gadotti (2009). At a given stellar
mass, the average outer disk luminosities are fainter by $\sim40-70\%$,
excluding the lowest mass bin.

We next consider the effect of $M/L$ to test if the difference in outer disk 
luminosity could imply a a difference in outer disk mass.
For Coma, we show the {\it galaxy-wide} $(M/L)_i$ ratio 
estimated, while for the Gadotti (2009) sample we show $i$-band $M/L$ ratios in the outer disks, $(M/L)_{d,i}$.
Figure~\ref{scaling16}c compares the resulting 
$(M/L)$ values against galaxy $M_\star$. The average $(M/L)_i$ in Coma is larger than 
the average $(M/L)_{d,i}$ in LDEs by a factor of $\sim1.3-2$ at a 
given galaxy $M_\star$, excluding 
the lowest mass bin. This difference in $(M/L)_i$ accounts for $\sim48-80\%$ 
of the average offset in disk luminosity. This suggests {\it some} of the difference in 
outer disk luminosity might be driven by a real difference in outer disk mass.
Cappellari (2013), in comparison, concludes that spirals in Coma transformed
into fast rotating early-type galaxies while decreasing in
\textit{global} half-light radius with little mass variation.

Figure~\ref{scaling17} examines how bulge size ($r_\mathrm{e}$), bulge 
luminosity, bulge S\'ersic index,  and bulge-to-disk {\it light ratio} ($B/D$) 
scale with galaxy 
$M_\star$. Figures~\ref{scaling17}a-c show that bulge size, bulge luminosity, 
and bulge S\'ersic index as a function of galaxy $M_\star$ are not
systematically offset in Coma versus LDEs. 
Figure~\ref{scaling17}d shows there is a great scatter in $B/D$ versus 
galaxy $M_\star$. 

Figure~\ref{scaling18}a shows $B/D$ versus bulge S\'ersic index. At a 
given bulge S\'ersic index, galaxies in Coma show a systematically higher 
average $B/D$ ratio than galaxies in LDEs. A linear regression fit reveals 
a clear offset in $B/D$ for a given bulge index. Figure~\ref{scaling18}b 
indicates that at a given bulge S\'ersic index the bulge luminosities in Coma 
and LDEs are very consistent. Figure~\ref{scaling18}c, on the other hand, shows 
a clear offset in disk luminosity ($\sim0.6$ mag),
indicating that differences in $B/D$ are due, at least in part, to outer disk 
size/luminosity.

From this investigation, we have learned of a reduction in the average 
sizes and luminosities in the outer disks of Coma galaxies that may translate 
into a lower mean outer disk stellar mass. This may be explained in part by 
cluster environmental effects. We consider this point further in 
Section~\ref{discuss3}.

\subsection{Environmental Processes in Coma}\label{discuss3}

Many studies provide evidence for the action of environmental processes in Coma.
The predominantly intermediate or old stellar populations in the center
of the cluster (e.g., Poggianti et al. 2001; Trager et al. 2008; Edwards \& Fadda 2011) 
are indirect evidence for the action of starvation.
Furthermore, the properties of Coma S0s display radial cluster trends that 
favor formation processes that are environment-mediated 
(Rawle et al. 2013, Head et al. 2014). Several examples of 
ram-pressure stripping have been directly observed in Coma (Yagi et al. 2007, 
2010; Yoshida et al. 2008; Smith et al. 2010; Fossati et al. 2012). 
There is also much evidence for the violent effects of tidal forces.
The presence of a diffuse intra-cluster medium around Coma central galaxies 
NGC~4874 and NGC~4889 has long been discussed (Kormendy \& Bahcall 1974; 
Melnick et al. 1977; Thuan \& Kormendy 1977; Bernstein et al. 1995; 
Adami et al. 2005; Arnaboldi 2011). At the cluster center, the intra-cluster 
light represents up to $20\%$ of the cluster galaxy luminosity (Adami et al. 
2005).  This central intra-cluster light is not uniform given the presence of 
plumes and tidal tails (Gregg \& West 1998; Adami et al. 2005), and debris 
fields are also found further outside the cluster 
center (Gregg \& West 1998; Trentham \& Mobasher 1998).

Below, we comment on how our results add to this picture.

\begin{enumerate}
\item {\it  Reduced Growth and Truncations of Outer Disks in  Coma S0s/spirals:}\\
In Section~\ref{discussScaling}, we found that at a given galaxy stellar mass, 
the average half-light radius ($r_\mathrm{e}$) of the outer disk in S0s/spirals is $\sim30-82\%$ smaller, and the average disk $i$-band luminosity is 
$\sim40-70\%$ fainter in Coma than in  lower-density environments 
(Figure~\ref{scaling16}).
These observations may be explained in part by cluster environmental effects
(e,g., strangulation, ram-pressure stripping, tidal stripping) that
suppress the growth of large-scale disks.
Hot gas stripping (strangulation) can 
plausibly suppress disk growth by limiting the 
amount of gas that can cool and become part of the outer disk. Tidal stripping 
via galaxy harassment is predicted (e.g., Moore et al. 1999) to be particularly
efficient at removing mass from extended disks. Ram-pressure stripping is
most effective at removing HI gas in the outskirts of a large scale-disk.
The evidence (Yagi et al. 2007, 2010;
Yoshida et al. 2008; Fossati et al. 2012) suggests ram-pressure 
stripping happens quickly, and if so it should be
effective at preventing the growth of large-scale disks after the host galaxy
enters the cluster.\\

\item {\it Low S\'ersic index in S0/spiral outer disks:}\\
Figure~\ref{diskn} demonstrates the majority 
of outer disks have low S\'ersic index ($66.0\pm8.2\%$ with $n<1$ and 
$18.0\pm12.8\%$ with $n<0.5$). 
This effect is not artificially driven by bars because
the low $n<1$ disks  include barred and unbarred galaxies
to similar proportions, and additionally, the disks are fitted
separately from the bars in our work.
Similar examples have been found
in Virgo. Kormendy \& Bender (2012) find several examples of Gaussian 
($n\sim0.5$) disks among both barred and unbarred galaxies, which commonly
occur in barred galaxies (e.g., Kormendy \& Kennicutt 2004). 
Gaussian-like disks among
unbarred galaxies are much more surprising (Kormendy \& Bender 2012).
Figure~\ref{diskn} shows that the large fraction of $n<1$ outer disks in Coma 
is not driven by barred galaxies alone.
It is not easy to compare the fraction of low $n<1$ disks in
Coma versus LDEs because most work to date in LDEs
(e.g., Allen et al. 2006; Laurikainen 2007, 2010; Weinzirl et al. 2009)  fit the outer disk with a 
fixed $n=1$ exponential profile. 

Environmental processes could be creating the Gaussian-like disks. 
Kormendy \& Bender (2012) have suggested this and invoked dynamical heating. We 
could be seeing a stronger and/or different manifestation in Coma.
Ram-pressure stripping and tidal stripping can plausibly reduce the 
S\'ersic $n$ by cutting off the outskirts of the outer stellar/gaseous 
disk. 
\\

\item {\it Bulge-to-disk ratio ($B/D$):}\\
The mean bulge S\'ersic index rises with mean $B/D$ light ratio
in both the central part of Coma and LDEs, consistent with the idea that
the development of high $B/D$ ratio in galaxies is usually associated
with processes, such as major mergers, which naturally results in a
high $n$. Such a correlation was also found previously in field spirals
(e.g., Andredakis et al. 1995; Weinzirl et al. 2009).

We also find that at a given bulge index, the  $B/D$ light
ratio is higher for Coma. This environmental effect appears to be 
due, at least in part, to 
the fact that at a given bulge $n$, the bulge luminosity is similar in
Coma and LDEs, but the outer disks have lower luminosity by a factor
of a few in Coma  (Figure~\ref{scaling18}).  
This reduced disk growth is likely due to cluster environmental effects
suppressing the growth of large-scale outer disks.
This conclusion for Coma nicely parallel studies of 
ram-pressure stripping 
(Cayette et al. 1990, 1994; Kenney et al. 2004, 2008; Chung et al. 2007, 2009)
and dynamical heating (Kormendy \& Bender 2012) in the less extreme Virgo cluster.

\end{enumerate}

\section{Comparison of Empirical Results With Theoretical Predictions}\label{theory}
\subsection{Overview of the Models}\label{theoryoverview}
In this section, we compare our empirical results for Coma with simulations of 
clusters. 
The simulated clusters are derived from a semi-analytical model (SAM) based on 
Neistein \& Weinmann (2010). The SAM is able to produce reasonable matches 
(Wang, Weinmann, \& Neistein 2012) to the galaxy stellar mass function 
determined by (Li \& White 2009) for massive 
$M_\star \gtrsim 5\times 10^8$~$M_\odot$ 
galaxies at low redshift ($0.001<z<0.5$) over all environments 
(including Virgo and Coma) probed in the northern hemisphere component of SDSS
Data Release 7. A brief summary of the SAM formalism is given below. Interested readers should see Neistein \& Weinmann (2010) and Wang, Weinmann, \& 
Neistein (2012) for additional details.

The SAM uses merger trees extracted from the Millennium $N$-body simulation 
(Springel et al. 2005). 
Galaxies are modeled as vectors of stellar mass, cold gas, and hot gas.
Baryonic physics are handled with semi-analytic prescriptions. 
In between merger events, the efficiencies of quiescent evolutionary processes, such as cold and hot gas accretion, gas 
cooling, star formation, and supernovae feedback, are modeled as functions 
of halo mass and redshift only. The star formation rate 
is proportional to the amount of cold gas, and the star formation
efficiency is a function of halo mass and redshift.
In the model, the baryonic mass (i.e., the sum of stellar and cold gas mass)
is used to define major ($M_1/M_2 \geq 1/4$) and minor ($1/10 < M_1/M_2 < 1/4$) mergers. As we will
discuss in Section~\ref{massratio}, the results are highly sensitive to whether
the stellar mass ratio or  baryonic mass ratio are used.

Immediately after a major merger, the remnant's stellar $B/T$ ratio is always 
one. This is because the model assumes any existing stellar disks are 
destroyed, and all stars undergo violent relaxation to form a bulge/elliptical.
After a major merger, an extended stellar disk is rebuilt via gas cooling, 
causing $B/T$ to fall. Any further major mergers will reset $B/T$ to one. 
During a minor merger, the stellar component of the satellite
of baryonic mass $M_1$ is added to the bulge.  

During a major/minor merger, some fraction of cold gas is converted to stars
in a short induced starburst $\sim10$~Myr in duration. The amount of 
merger-induced star formation depends explicitly on the cold gas mass. 
Stars formed in 
{\it major} merger-induced starbursts are considered part of the bulge
(see Section~\ref{datavmodel}). 
This is a reasonable assumption given that a) the
bursts of star formation are much shorter than the overall duration of the
mergers and b) all existing stars {\it from both progenitors} are violently
relaxed during final coalescence. 
It seems less likely the starburst stars induced in minor mergers should be
violently relaxed since minor mergers are not very efficient at violently
relaxing stars in the host galaxy. We consider this issue further in
Section~\ref{datavmodel}.
Therefore, in the model used in this paper, the bulge stellar mass traces 
the mass assembled via major and minor mergers. Galaxies without bulges have 
had no resolvable merger history. The model does not build bulges through the 
coalescence of clumps condensing in violent disk instabilities (Bournaud, 
Elmegreen, \& Elmegreen 2007; Elmegreen et al. 2009).

Galaxy clusters impose additional environmental effects that complicate
modeling with SAMs. The SAM used here accounts for stripping of hot gas 
(i.e., strangulation; Larson et al. 1980) by assuming hot gas is stripped 
exponentially with a timescale of 4 Gyr. Other processes like ram-pressure,
stripping/disruption of stellar mass (Moore et al. 1996, 1998, 1999; 
Gnedin 2003), dynamical friction heating by satellite (El-Zant et al. 2004), 
and gravitational heating by infalling substructures (Khochfar \& Ostriker 
2008) are neglected.
It is not clear how much the inclusion of ram-pressure stripping in the SAM
would affect our results. While hydrodynamical simulations clearly demonstrate
the strong influence of ram-pressure stripping on gas mass, galaxy morphology, 
and star formation (e.g., Quilis et al. 2000; Tonnesen \& Bryan 2008, 2009, 
2010), some SAMs (e.g., Okamoto \& Nagashima 2003; Lanzoni et al. 2005) 
suggest that accounting for ram-pressure stripping has only a small affect. 
Tidal stripping creates a 
population of intra-cluster stars that can contribute between $10-40\%$ of the
optical light in rich clusters (e.g., Bernstein et al. 1995; 
Feldmeier et al. 2004; Zibetti et al. 2005). 
The inclusion of tidal stripping in SAMs is important for addressing
a wide range of systematic effects (e.g., Bullock et al. 2001; 
Weinmann et al. 2006; Henriques et al. 2008, 2010), but tidal stripping is not
present in this SAM.

\subsection{The Mass Function and Cumulative Number Density in Coma}\label{comavagc}

In order to compare galaxies in the simulations with those in the center of 
Coma, we first need to identify model clusters that best represent Coma. 
We do this based on the global properties of Coma, namely the halo mass and 
size, galaxy stellar mass function, and radial profile of cumulative projected 
galaxy number density. As the ACS coverage of Coma encompasses a fraction 
(19.7\%) of the projected central 0.5~Mpc, we calculate these properties in 
Coma with Data Release 7 (DR7) of the NYU Value-Added Galaxy Catalog 
(NYU-VAGC, Blanton et al. 2005), which  provides full spatial coverage of Coma. 
NYU-VAGC DR7 is based on SDSS DR7 data (Abazajian et al. 2009) and provides 
catalogs generated from an independent, and improved, reduction of the public 
data (Padmanabhan et al. 2008).

We select Coma cluster member galaxies from NYU-VAGC 
assuming Coma cluster galaxies have radial velocity in the range 
v$_{\rm min}$ = 4620 km/s to v$_{\rm max}$ = 10,000 km/s, which is the range 
in radial velocity among spectroscopically confirmed members in the ACS 
survey. We also adopt the Coma virial radius and virial mass to be 
2.9$h^{-1}_{70}$~Mpc and $1.4\times10^{15}h^{-1}_{70}$~$M_\odot$, respectively,
measured by Lokas \& Mamon (2003) with a 30\% accuracy, where
$h^{-1}_{70}=H_0/70$.
For our adopted $H_0$ of 73, we scale these numbers by $(73/70)^{-1}$, so that
the virial radius and virial mass are 
2.8$h^{-1}_{73}$~Mpc and $1.3\times10^{15}h^{-1}_{73}$~$M_\odot$, respectively.
We select galaxies with the following criteria:
\begin{enumerate}
\item Radial velocity in range 4620 to 10,000 km/s.
\item Projected radius, ${\rm{R_p}}$ from the cluster center (i.e., NGC~4874) 
less than the virial radius. 
\item Brightness exceeding the SDSS spectroscopic completeness limit of 
$r=17.7$ mag, or $M_r\leq -17.3$ mag at the 100~Mpc distance of Coma. This 
corresponds to a stellar mass of $1.3\times 10^9$~$M_\odot$ assuming a $g-r$ 
color of 0.67, which is the average among Coma galaxies in the NYU-VAGC
selected in this manner.
\end{enumerate}

Panel (b) of Figure~\ref{virp} shows the resulting projected galaxy
density profile for this set of Coma galaxies.  

We next calculate the global galaxy stellar mass function within the virial 
radius. Figure~\ref{virp}c shows the result. This mass function includes
normal massive galaxies  (E, S0, spiral) as well as the two cDs  
(NGC~4874 and NGC~4889). As described in Section~\ref{datas2},
we derive the stellar mass by applying Equations 5 and 6 to SDSS $gr$ 
photometry.

Using the cD galaxy stellar masses as lower limits
at the high mass end of the galaxy stellar mass function in Figure~\ref{virp}c, 
we measure a slope 
$\alpha = -1.16$ and characteristic mass $M^* = 1.25 \times 10^{11}~M_\odot$ 
for the global galaxy stellar mass function of Coma inside the cluster virial 
radius.

\subsection{Global Properties of Model Clusters Versus Coma}\label{globalprop}
Next, we compare the above global properties of the Coma cluster with the 
simulated clusters  in the theoretical model in order to identify the model 
clusters that best represent Coma. We consider all 160 Friend-of-Friend (FOF, 
Davis et al. 1985) groups in the Millennium simulation having a halo mass in 
the range $5\times10^{14} - 10^{16}$~$M_\odot$. We refer to the most massive 
halo, and its gravitationally bound subhaloes, in each FOF group as a 
`cluster'. 

To find potential matching clusters, we identify massive 
($M_\star \geq 10^9$~$M_\odot$) member galaxies in each 
cluster in a way that is consistent with the selection of Coma member galaxies 
in Section~\ref{comavagc}: 

\begin{enumerate}
\item  Radial velocity matching the range in line-of-sight velocities in the 
xy, xz, yz projections of the cluster.
\item Projected radius, ${\rm{R_p}}$, from the cluster center less than the 
cluster virial radius.
\item Luminosity brighter than the SDSS spectroscopic completeness limit of 
$M_r\leq -17.3$ mag at the 100~Mpc distance of Coma. 
\end{enumerate}

To gauge how well the simulated clusters compare with Coma in terms of
global properties, we examine the match in cumulative number density, mass 
function, and halo parameters (virial mass and radius). 

In Figure~\ref{virp},  we gauge how the global properties of Coma compare
with those of  all 160 cluster simulations.
Figure~\ref{virp}a shows the combinations of virial radius and
halo masses of the simulated clusters. The Coma halo parameters (virial mass 
and radius) adopted in Section~\ref{comavagc} are well matched to the largest 
and most massive model clusters.

Figure~\ref{virp}b shows the radial profile of
cumulative galaxy  number density.
The central galaxy number densities in the simulated clusters span three orders
of magnitude from $\sim10^4$ to $\sim4\times 10^5$~Mpc$^{-3}$,
overlapping with the high central density in Coma 
($\sim3\times10^4$~Mpc$^{-3}$). The thick dotted line denotes the 
cluster model with the best-matching halo 
parameters from Figure~\ref{virp}a. This halo model does a 
good job at matching the galaxy number density profile of Coma at projected
radius $R_p>0.7$~Mpc, but not at smaller projected radii. In comparison, the
model with the best matching cumulative number density profile, shown as
the open circle in Figure~\ref{virp}a, is smaller by $\sim60\%$ in halo mass 
than Coma. The next nine best matches to cumulative number density also
differ in halo mass by $\sim30\%$ or more from the halo mass in Coma, which
is estimated to be accurate to within 30\% (Section~\ref{comavagc}).

Figure~\ref{virp}c compares the galaxy stellar mass
function  between Coma and the simulated model clusters.
All the model clusters produce too many extremely massive
($M_\star \gtrsim 5\times 10^{11}$~$M_\odot$) galaxies.
These very high-mass galaxies are not devoid of ongoing star 
formation like ellipticals in Coma are (Section~\ref{discuss3}). Rather,
these galaxies have present day SFR of $\sim10$~$M_\odot$~yr$^{-1}$.
Furthermore, the cluster mass functions show slopes that are marginally too 
steep ($\alpha \sim -1.5$ versus $\alpha = -1.16$) on the low-mass end 
(Section~\ref{comavagc}).

We note that when this SAM model was compared with SDSS observations of 
galaxies averaged over all environments at low redshift (Wang, Weinmann, \& 
Neistein 2012), the model galaxy stellar mass function shows a similar,  but 
less extreme, discrepancy with the galaxy stellar mass function of Li \& White 
(2009)  in terms of producing too many of the most massive galaxies. 
Figure~\ref{virp}c includes the galaxy stellar mass function from Li \& White 
(2009) as a dashed line for comparison.

In Figure~\ref{bestclusters} we make the comparison with three sets of 
model clusters (a total of 30 model clusters) containing the 10 best 
matches to Coma in terms of the cumulative galaxy number density, galaxy 
stellar mass function, and halo parameters.
Matching to one criterion (e.g., cumulative number density) does not ensure a good match to the other two criteria.

We are left with the sobering conclusion that the simulations cannot 
produce a model cluster simultaneously matching multiple global properties 
of Coma, our local benchmark for one of the richest nearby galaxy clusters.
The large discrepancy in the  galaxy stellar mass function between the model 
and Coma could be due to a number of factors. The model currently does not 
include tidal stripping/disruption of stars and ram-pressure stripping
(Section~\ref{theoryoverview}),
which would reduce the stellar mass of galaxies on all mass scales.
The importance of ram pressure stripping is further discussed in 
Section~\ref{coldgasmass}, where we find that the cold gas fraction in the 
model galaxies is much higher in Coma galaxies. 

\subsection{Strong Dependence of Results on Mass Ratio Used to Define Mergers}\label{massratio}

Merger history and galaxy $B/T$ are highly dependent on the mass used 
(stellar mass, baryonic mass, halo mass) to define merger mass ratio $M_1/M_2$.
For a single representative cluster model, Figure~\ref{majordef} highlights 
the key differences that arise when $M_1/M_2$ is defined as the ratio of 
stellar mass (Def 1, left column) versus cold gas plus stars (Def 2, right 
column).  This representative cluster was selected because it is the best 
matching cluster to the cumulative galaxy number density distribution in Coma 
(Figure~\ref{bestclusters}). The first row of Figure~\ref{majordef} shows the 
cumulative percentage of galaxies with a major merger since redshift $z$. 
In the second row of Figure~\ref{majordef}, the histograms show the percentage 
of galaxies with a last major merger at redshift $z$. 
The third row shows the percentage of galaxies with a given $B/T$ value, sorted
by galaxies with and without a major merger. 
Finally, the last row of Figure~\ref{majordef} gives the distribution of 
present-day $B/T$ versus redshift of the last major merger. 

In the following sections, we consider a model where the merger mass ratio 
$M_1/M_2$ depends on stellar mass plus cold gas, as this ratio is understood to 
be the most appropriate definition (Hopkins et al. 2009b). 
Traditionally, observers have tended to  use stellar mass ratios in identifying
mergers (e.g., Lin et al. 2004; Bell et al. 2006; Jogee et al. 2009;
Robaina et al. 2010) as stellar masses are readily measured for a large number
of galaxies. However, with the advent of ALMA, it will be increasingly possible
to incorporate the cold gas mass for a large number of galaxies.

\subsection{Cold Gas Mass in Coma Galaxies Versus Model Galaxies}\label{coldgasmass}

In the SAM used here,  the cold gas fraction  $f_{\rm gas}$ (defined as the 
ratio of cold gas to the baryonic mass made of cold gas, hot gas, and stars)  
and  the ratio ($M_{\rm cold\_gas}/M_\star$) of cold gas to stellar mass are 
both overly high.  The issue of high cold gas fraction in this model was 
highlighted and discussed in Wang, Weinmann, \& Neistein (2012).
Here, we quantify how far off the model values are compared with what is 
expected for a rich cluster like Coma.

Figure~\ref{boselli} illustrates the degree to which the ratio 
($M_{\rm cold\_gas}/M_\star$) is overestimated by comparing with data from 
Boselli et al. (1997), who measure atomic ($M_{HI}$) and molecular gas 
($M_{H_2}$) masses for Coma cluster member galaxies and non-cluster galaxies.  
The top panel shows that the average ratio of cold gas to stellar mass 
($M_{\rm cold\_gas}/M_\star$) 
ranges from $\sim1-12$ for a representative model cluster. The bottom panel
shows that the ratio of $M_{HI+H_2}/M_\star$ for Coma cluster galaxies from 
Boselli et al. (1997) 
is usually $<0.1$; non-cluster galaxies are more gas rich, but the ratio of 
$M_{HI+H_2}/M_\star$ is still $\ll 1$.  At 
$10^{10} \lesssim M_\star \lesssim 10^{11}$~$M_\odot$, the model predicts a 
cold gas to stellar mass ratio that is a factor $\sim25-87$ times 
higher than the median in Coma cluster galaxies. 

\subsection{Data Versus Model Predictions for Stellar Mass in Dynamically Hot and Cold
Components}\label{datavmodel}

We next proceed to compare the observed versus model predictions for the 
distribution of mass in dynamically hot and cold stellar components.
The following comparisons are made in the projected central 0.5~Mpc of Coma 
and the model clusters.

We first start by describing how the model builds bulges and ellipticals.
In the model, the total bulge stellar mass  $M_{\rm \star,Bulge,model}$ 
consists of stellar mass accreted  in major and minor mergers, plus
stellar mass from SF induced in both types of mergers.

Next, we discuss how to compare the model with the data. 
For our sample of  Coma galaxies (excluding the 2 cD systems)
 with $M_\star \geq 10^9$~$M_\odot$, we compute the ratio R1$_{\rm data}$
as the the stellar mass in all components with $n > n_{\rm disk\_max}$
to the sum of galaxy stellar mass.
The reasons for not including the cD systems were discussed in 
Section~\ref{datas2}.
From Section~\ref{census},  R1$_{\rm data}$ is 57\%.

We next compare this ratio to the corresponding quantity
in the model. The comparison is not entirely straightforward as the model
does not give a S\'ersic index. We therefore have to associate components
in the model to the corresponding high 
$n>n_{\rm disk\_max}$ classical bulges/ellipticals in the data.
The most natural step is to assume that the stellar mass built
during major mergers is redistributed into such high-$n$ components.
We call the result R2$_{\rm model}$. We find that for 
$M_\star \geq 10^9$~$M_\odot$, R2$_{\rm model}$ has a wide dispersion: 
$\sim35-79\%$ for the 30 model clusters shown in Figure~\ref{bestclusters}, 
with a median value of $\sim66\%$. The representative cluster discussed in 
Section~\ref{massratio} and Figures~\ref{majordef}-\ref{boselli} 
has a value of $\sim72\%$.

Guidance on the S\'ersic index of structures formed during minor mergers can be 
gleaned from Hopkins et al. (2009b). In the 
general case of an unequal mass merger, the coalescence of the smaller 
progenitor (mass $M_1$) with the center of the primary will destroy (i.e., 
violently relax) the smaller galaxy and also potentially violently relax an 
additional mass $\leq M_1$ in the primary.  The stars that are 
violently relaxed in the minor merger become part of the bulge in the primary 
galaxy. Thus, we define R3$_{\rm model}$ to be R2$_{\rm model}$ plus the 
stellar accretion from minor mergers. For $M_\star \geq 10^9$~$M_\odot$, 
R3$_{\rm model}$ is only slightly higher than
R2$_{\rm model}$ by a few percent. R3$_{\rm model}$ ranges from $\sim35-82\%$,
with a median value of $\sim71\%$,
and the representative cluster (Section~\ref{massratio}, 
Figures~\ref{majordef}-\ref{boselli}) has a value of $\sim71\%$.

The comparison of R1$_{\rm data}$ with R2$_{\rm model}$  and  R3$_{\rm model}$ 
is a global comparison of the total stellar mass fraction within high-$n$ 
components {\it summed over all} the galaxies with 
$M_\star \geq 10^9$~$M_\odot$. Next, we push the data versus model comparison 
one step further by doing it in bins of stellar mass, as shown in 
Figure~\ref{mtcompare}.

The top panel of Figure~\ref{mtcompare} plots the mean ratio of stellar mass
fraction in dynamically hot components ($f_{\rm \star,hot}$) as a function
of total galaxy stellar mass, for data versus model.
For each stellar mass bin shown in Figure~\ref{mtcompare},
the value of $f_{\rm \star,hot}$ is calculated for {\it each galaxy}
as $M_{\rm \star,hot}/M_{\star}$.
In the data, $M_{\rm \star,hot}$ is taken as the stellar mass of any
high $n>n_{\rm disk\_max}$ component in the galaxy.
The model shown here is the best cluster model matched by cumulative galaxy
number density (see Figure~\ref{bestclusters}, column 1).
For this model, two lines are shown: the solid line takes $M_{\rm \star,hot}$
as the stellar mass accreted and formed during major mergers, 
while the dotted line also adds in the stellar mass accreted during minor
mergers. 

In the top panel of Figure~\ref{mtcompare} there is 
significant disagreement between the fractions of $f_{\rm \star,hot}$
for the Coma data and the model. As shown by the second dotted 
model curve in Figure~\ref{mtcompare}, adding in the stellar mass accreted in 
minor mergers to the model only changes the fraction by a few percent.
The values of $f_{\rm \star,hot}$ are chiefly representative
of the contributions from major mergers.

The bottom panel of Figure~\ref{mtcompare} plots the analogous mean ratio of 
stellar mass fraction in dynamically cold components 
($f_{\rm \star,cold} =M_{\rm \star,cold}/M_{\star}$)
as a function of total galaxy stellar mass. In the model, the two lines show 
two  different expressions for $M_{\rm \star,cold}$.
For the solid  line, we  take $M_{\rm \star,cold}$ to be the
mass of the outer disk $M_{\rm \star,Outer\_disk}$, which 
represents the difference between the bulge mass ($M_{ \rm \star,Bulge,model}$) 
and the total stellar mass.
One problem with this approach is that it ignores small-scale  nuclear disks
formed in the bulge region. We tackle this problem by defining a second
dotted model line 
that accounts for stars formed via induced SF during minor mergers.
It is clear in the bottom panel of Figure~\ref{mtcompare} that the model
overpredicts the mass in disks as a function of galaxy stellar mass. 
Note the contribution to $f_{\rm \star,cold}$ from minor merger induced SF is 
$\lesssim17\%$ in a stellar mass given bin.

The main conclusion from Figure~\ref{mtcompare} is that the best-matching
cluster model is underpredicting the mean fraction $f_{\rm \star,hot}$ of 
stellar mass locked in hot components over a wide range in galaxy stellar mass
($10^9 \leq M\star \lesssim 8\times10^{10}$~$M_\odot$).
Similarly this model overpredicts the mean value for $f_{\rm \star,cold}$.
The effect of cosmic variance on our sample (Section~\ref{census} and 
Appendix~\ref{cosmicvar}) means
our measured $f_{\rm \star,hot}$ is lower than the true value 
by an estimated factor of 1.16.
Therefore, the underprediction of $f_{\rm \star,hot}$ in the
model is worse than what we are citing.
While the discussion in this section focused only on a single model cluster,
the results and conclusions would be similar if we had analyzed alternate
simulated clusters, such as those matched to the cluster galaxy stellar mass
function (see Figure~\ref{bestclusters}, column 2) or halo parameters
(see Figure~\ref{bestclusters}, column 3). 

There could be several explanations as to why the models are underproducing the 
fraction of dynamically hot stellar mass ($f_{\rm \star,hot}$) and 
overproducing the fraction of dynamically cold stellar mass 
($f_{\rm \star,cold}$). One possibility 
is that the absence of key cluster processes (especially ram-pressure stripping 
and tidal stripping) in the models is leading to the overproduction of the 
model galaxy's cold gas reservoir (Section~\ref{coldgasmass}), compared to a 
real cluster galaxy, whose outer gas would be removed.  This means that in the 
models, SF in gas that would otherwise be removed from the
galaxy builds additional dynamically cold stellar mass following
the last major merger.  Another possibility is that the models  ignore
the production of bulges via the merging of star forming clumps
(Bournaud, Elmegreen, \& Elmegreen 2007; Elmegreen et al. 2009).
It is still debated whether this mode can efficiently produce
classical bulges, but if it does, then its non-inclusion in the models
could lead to the underprediction of $f_{\rm \star,hot}$.

In summary, our comparison of empirical results to theoretical
predictions underscores the need to include in SAMs
environmental processes, such as ram-pressure
stripping and tidal stripping, which affect the cold gas content of galaxies,
as well as more comprehensive models of bulge assembly.
It is clear that galaxy evolution is a
function of \textit{both} stellar mass and environment.

\section{Summary \& Conclusions}\label{summary}

We present a study of the Coma cluster in which we constrain galaxy assembly
history in the projected central 0.5~Mpc  by performing multi-component 
structural decomposition on  a mass-complete sample of 69 galaxies with 
stellar mass $M_\star \geq 10^9$~$M_\odot$. Some strengths of this study 
include the use of superb high-resolution ($0\farcs1$), F814W images from the 
$\textit{HST}$/ACS Treasury Survey of the Coma cluster, and the adoption of a 
multi-component decomposition strategy where no a priori assumptions are made 
about the  S\'ersic index of bulges, bars  or disks. We use structural 
decomposition to identify the two fundamental kinds of galaxy structure -- 
dynamically cold, disk-dominated components and dynamically hot classical
bulges/ellipticals -- by adopting the working assumption that the S\'ersic 
index $n$ is a reasonable proxy for tracing different structural components.
We define disk-dominated structures as components with a low S\'ersic index
$n$ below an empirically determined threshold value $n_{\rm disk\_max}\sim1.7$
(Section~\ref{approach}). Galaxies with an outer disk are called spirals or
S0s. We explore the effect of environment by performing a census of 
disk-dominated structures versus classical bulges/ellipticals in Coma.
We also compare our empirical results on galaxies in the
center of the Coma cluster  with theoretical predictions 
from a semi-analytical model. Our main results are summarized below.

\begin{enumerate}




\item {\it Breakdown of stellar mass in Coma between
low-$n$ disk-dominated structures and high-$n$ classical
bulges/ellipticals:}\\
We make the first attempt (Section~\ref{census} and 
Tables~\ref{masstable2}--\ref{masstable}) at exploring the distribution of 
stellar mass in Coma in terms of dynamically hot versus dynamically cold 
stellar components. After excluding the 2 cDs because of their uncertain 
stellar masses, we find that in the projected central 0.5 Mpc of the
Coma cluster, galaxies with stellar mass $M_\star \geq 10^9$~$M_\odot$ have
57\% of their  cumulative stellar mass  locked up in
high-$n$  ($n \gtrsim 1.7$)  classical bulges/ellipticals
while the remaining  43\% is in the form of
low-$n$ ($n\lesssim1.7$) disk-dominated structures (outer disks,
inner disks, disky pseudobulges, and bars).
Accounting for the effect of cosmic variance and color gradients in 
calculating these stellar mass fractions would not significantly change
this census (Appendices~\ref{cosmicvar}--\ref{colorgrad}).\\


\item {\it Impact of environment on morphology-density relation:}\\
Using our structural decomposition to assign galaxies the Hubble types E, S0, 
or spiral, we find evidence of a  strong morphology-density relation.
In the projected central 0.5~Mpc of the Coma cluster, spirals are rare, and 
the morphology breakdown of (E+S0):spirals is (91.0\%):9.0\% by numbers and 
(94.2\%):5.8\% by stellar mass (Section~\ref{mdr} and Table~\ref{tabmdr}).\\

\item {\it Impact of environment on outer disks:}\\
In the central parts of Coma, the properties of large scale disks are
likely indicative of environmental processes that suppress disk growth or 
truncate disks (Section~\ref{discuss3}). In particular, 
at a given galaxy stellar mass, outer disks are smaller
by $\sim30-82\%$ and fainter in the $i$-band by $\sim40-70\%$
(Figure~\ref{scaling16}).
The suggestion that outer disks in Coma are more compact is
consistent with the results of previous analyses of disk
structure in Coma (Guti{\'e}rrez et al. 2004; Aguerri et al. 2004).\\

\item {\it Impact of environment on bulges:}\\
The ratio $R$ of stellar mass in high-$n$ ($n\gtrsim1.7$) classical bulges to 
low-$n$ ($n\lesssim1.7$) disky pseudobulges is 17.3 in Coma. We measure $R$ to
be a factor of $\sim2.2-2.7$ higher in Coma compared with various samples from
LDEs (Sections~\ref{census}--\ref{discussmass}, 
Tables~\ref{masstable2}--\ref{masstable}).
We also find that at a given bulge  S\'ersic index $n$, the bulge-to-total
ratio $B/D$, and the $i$-band light ratio are offset to higher values in Coma
compared with LDEs. This effect appears to be due, at least in part, to the 
above-mentioned lower disk luminosity in Coma.\\


\item {\it Comparison of data to theoretical predictions:}\\
We  compare our empirical results on galaxies in the
center of the Coma cluster  with theoretical predictions based on
combining the Millennium cosmological simulations of dark matter
 (Springel et al. 2005) with baryonic physics from a semi-analytical model
(Neistein \& Weinmann 2010; Wang, Weinmann, \& Neistein 2012).

It is striking that no model cluster can simultaneously match the global 
properties
(halo mass/size,  cumulative galaxy number density, galaxy stellar mass 
function) of Coma (Figures~\ref{virp} and \ref{bestclusters}), and the 
cold gas to stellar mass ratios in the model clusters are at least 25 times 
higher than is measured in Coma. 

As suggested by Hopkins et al. (2009b), we find galaxy merger history is 
highly dependent on how the merger mass ratio $M_1/M_2$ is defined.
Specifically, there is a factor of $\sim5$ difference
in merger rate when the merger mass ratio is based on the baryonic mass versus 
the stellar mass (Figure~\ref{majordef}).
Traditionally, observers have tended to use stellar mass ratios in identifying
mergers, but  with the advent of ALMA, it will be increasingly possible  and
important to incorporate the cold gas mass.

For representative ``best-match'' simulated clusters, we compare the empirical 
and theoretically predicted fraction $f_{\rm \star,hot}$ and 
$f_{\rm \star,cold}$ of stellar mass locked, respectively, in high-$n$, 
dynamically hot versus low-$n$, dynamically cold stellar components.
Over a wide range of galaxy stellar mass 
($10^9 \leq M\star \lesssim 8\times10^{10}$),
the model underpredicts the mean fraction $f_{\rm \star,hot}$ of stellar
mass locked in hot components by a factor of $\gtrsim1.5$. Similarly, the model
overpredicts the mean value for  $f_{\rm \star,cold}$ (Section~\ref{datavmodel}
and Figure~\ref{mtcompare}). 

We suggest this disagreement might be 
due to two main factors. Firstly, key cluster processes (especially 
ram-pressure stripping and tidal stripping), which impact the cold gas content 
and disk-dominated components of galaxies, are absent. Secondly, the models 
ignore the production of bulges via the merging of star forming clumps
(Bournaud, Elmegreen, \& Elmegreen 2007; Elmegreen et al. 2009).
These results underscore the need to implement  in theoretical models
environmental processes, such as ram-pressure stripping and tidal stripping,
as well as more comprehensive models of bulge assembly. It is clear that
galaxy evolution is not a solely a function of stellar mass, but it also depends
on environment.


\end{enumerate}

\appendix
\section{Using GALFIT}\label{app}
The proper operation of GALFIT depends on certain critical inputs.  We briefly
describe below how these important inputs are handled:
\begin{enumerate}
\item {\it Point Spread Function (PSF):}\\
Accurate modeling of the PSF is essential in deriving galaxy structural 
properties. GALFIT convolves the provided PSF with the galaxy model in each 
iteration before calculating the $\chi^2$.  Because the PSF varies with 
position across the ACS/WFC chips, it is ideal to separately model the PSF for 
each galaxy position. We use the grid of model ACS PSFs in the F475W and F814W 
filters from Hoyos et al. (2011).  This grid of 
PSFs was created with \texttt{TinyTim} (Krist 1993) and 
\texttt{DrizzlyTim}\footnote{\texttt{DrizzlyTim} is written by Luc Simard}. 

For a given set of multidrizzle parameters, \texttt{DrizzlyTim}
transforms $x-y$ coordinates in the final science frames back to the system of 
individually distorted FLT images.  \texttt{DrizzlyTim} invokes 
\texttt{TinyTim} to create a PSF with the specified parameters (e.g., position 
and filter) and then places the PSF at the appropriate position in blank FLT 
frames.  The FLT frames are passed through \texttt{MULTIDRIZZLE} with the same 
parameters as the science images.  Finally, a Charge Diffusion Kernel is 
applied to the PSFs in the geometrically distorted images.  The grid of ACS 
PSFs from Hoyos et al. (2011) models a PSF for every 150 pixels in the $x$ and 
$y$ directions.  For each galaxy in our sample we select the model PSF closest 
in proximity to the galaxy.

\item {\it Sigma Images:}\\
A sigma image is the 2D map of the $1\sigma$ standard deviations in pixel 
counts of the input image.  GALFIT uses the sigma image as the relative weight 
of pixels for calculating the goodness of fit.  Achieving a reduced 
$\chi^2\sim1$ with a successful model fit requires that the sigma image be 
correct. A sigma image can either be 
provided, or GALFIT can be allowed to calculate one based on the properties of 
the data image (image units of counts or counts/second, effective gain, read 
noise, number of combined exposures).  We choose the latter option and allow 
GALFIT to calculate the sigma images.  

\item {\it Background Subtraction:}\\
While it is possible for GALFIT to freely fit the background sky, this is not
recommended (Peng et al. 2002).  In a multiple-component fit to a galaxy with 
at least two components, freely fitting the sky can exaggerate or suppress the 
wings of the central S\'ersic profile and incorrectly measure the bulge 
half-light radius and S\'ersic index. To avoid this, for each galaxy the 
background sky is measured and held fixed during the fit. The sky background is 
based on ellipse fitting with the \texttt{IRAF/ELLIPSE} task.  Ellipses are fit 
to the galaxy and the surrounding area, with the ellipses in the surrounding 
area being fixed to the shape and orientation of the galaxy. The gradient 
along the semi-major axis is calculated, and the sky is estimated as the mean 
of elliptical annuli over a span in semi-major axis where the gradient reaches 
a prescribed small value.  In each case, the area fitted by the 
ellipses exceeds the area subtended by the galaxy. Visual inspection of the 
ellipse fits shows that the perceived flat gradient corresponds to empty sky 
and not an extended galaxy outer profile with a very small gradient.

\item {\it Image Thumbnails and Masks:}\\
Thumbnail cutouts of the intermediate-mass galaxies are made to lessen the 
computational time for fitting.  Following Hoyos et al. (2011), square image 
thumbnails centered on the target galaxy are made using the output from
\texttt{SExtractor}. Image size in pixels is determined with
\begin{equation}
\rm{ size= 4\times A\_IMAGE \times KRON\_RADIUS}.
\end{equation}

The image units are transformed from counts/second to counts by multiplying by
the exposure time. Image masks are based on the segmentation images provided 
from \texttt{SExtractor}. The segmentation images are modified to unmask the 
background and target galaxy being fitted. Any bright sources that visibly 
overlap with the target galaxy are also unmasked so that overlapping sources 
can be fitted simultaneously. Masks for relatively bright sources that do not 
overlap with the galaxy being fitted are expanded in semi-major axis by a 
factor of 1.5. We visually check by blinking the data image and modified 
segmentation image to verify that the unmasked region encompasses all of the 
target galaxy, including those with large diffuse halos that 
\texttt{SExtractor} does not capture (Hoyos et al. 2011).
\end{enumerate}

\section{Details of Structural Decomposition}\label{decomp}

This appendix contains the full details concerning the structural decomposition
scheme outlined in Section~\ref{special}. 

\subsection{Single S\'ersic Fits}\label{1cp}
We first fit all galaxies with a single S\'ersic profile before attempting the
multi-component fits. This step is useful for measuring the total luminosity of 
a galaxy as well as measuring the centroid (Weinzirl et al. 2009).  The 
S\'ersic profile has seven free parameters: centroid, luminosity, half-light 
radius $r_\mathrm{e}$, S\'ersic index $n$, axis ratio, position angle, and 
diskiness/boxiness.  We fix the diskiness/boxiness so that the fitted 
structures are perfect ellipses.  We estimate the other six parameters based on
the parameters in \texttt{SExtractor} and allow them to optimize in the fit.
The detailed image preparation and inputs for the proper operation of GALFIT 
are described in Appendix \ref{app}.

Figure~\ref{singlesersic} compares our results for the single S\'ersic fits 
(with no point source) with those of Hoyos et al. (2011), who also perform 
single S\'ersic fits with GALFIT and GIM2D using Coma ACS Treasury Survey 
data.  Note that 
the galaxies in our sample requiring one S\'ersic profile are distinguished in 
Figure~\ref{singlesersic}.  With the exception of COMAi125935.698p275733.36 
(NGC~4874), our results for these sources requiring one S\'ersic profile well 
match those derived by Hoyos et al. (2011). For NGC~4874, we measure the 
$r_\mathrm{e}$ and $n$ of NGC~4874 to be 17.3~kpc and 3.05, respectively, while 
Hoyos et al. (2011) measure  $r_\mathrm{e}$ and $n$ to be 3.2~kpc and 1.3.  

For sources requiring more than one S\'ersic profile, our single S\'ersic 
magnitudes agree well in general with those of Hoyos et al. (2011), except for 
one case (COMAi13051.149p28249.90) where Hoyos et al. (2011) underestimate the 
magnitude by $\sim5.5$ mag. 
There are also outliers in both $r_\mathrm{e}$ and $n$. In 10 (5) instances 
(including COMAi13051.149p28249.90), the difference in $r_\mathrm{e}$ ($n$) 
exceeds a factor of 1.5.

There are two key differences in our fitting methodology (see 
Appendix~\ref{app}) compared with Hoyos et al. (2011). Most importantly, we 
entirely unmask the target galaxy and background in the segmentation-based 
masks so that GALFIT fits to pixels beyond what \texttt{SExtractor} associates 
with each galaxy. Hoyos et al. (2011) confine a galaxy to a customized mask 
generated based on the output of \texttt{SExtractor}. This approach misses a 
finite fraction of the flux in the target galaxy. This may explain why in 
Figure~\ref{singlesersic} we measure brighter magnitudes and larger 
$r_\mathrm{e}$ for more extended galaxies, where \texttt{SExtractor} does not 
detect all of the light in the galaxy. Second, we measure and fix the 
background sky while Hoyos et al. (2011) keep the sky as a free parameter.  
Allowing the sky background to freely vary in our fits fails to account for 
most of the scatter between our results and those of Hoyos et al. (2011).  
Rather, the disagreement appears to mainly be the result of differences in 
image masking.

\subsection{Multi-Component Fits}\label{2cp}
For the Stage 2 S\'ersic + S\'ersic fits, we model the `inner' and `outer' 
components (C1 and C2) with S\'ersic profiles that can represent physically 
different components (see Section~\ref{special}). 

Sensible initial guess parameters for Stage 2 are determined from a combination 
of the data image, Stage 1  model, and Stage 1 residuals. Guesses for the inner 
S\'ersic component (C1) are usually based on the Stage 1 model.  The 
centroid of the S\'ersic components (and nuclear point source if present) are 
fixed to the best-fit centroid from the single S\'ersic model. During the 
fits, we allow all other parameters (luminosity, $r_\mathrm{e}$, $n$, axis 
ratio, and position angle) to vary for the inner and outer components without a 
priori fixing the nature of these components.

With one exception, the $\chi^2$ 
in Stage 2 is always lower compared with $\chi^2$ in Stage 1 due to the extra 
S\'ersic component.  While the rare increase in $\chi^2$ from Stage 1 to 
Stage 2 is an indication the latter model is not reliable, the almost universal
decrease in $\chi^2$ is not necessarily a sign that the Stage 2 fit is 
meaningful because, in principle, 
such a decrease in $\chi^2$ could be driven by the extra free model parameters.
We consider a Stage 2 multi-component model to be superior to the Stage 1 fit 
if i) $\chi^2$ drops, ii) the Stage 2 model parameters are well behaved (i.e., 
not unphysically large or small), and iii) the Stage 2 residuals are deemed by 
visual inspection to show a reduction in coherent structure relative to the 
Stage 1 residuals.

Figure~\ref{resid} provides examples where a single S\'ersic model fails to 
model the entire galaxy well and leaves behind coherent structure in the 
residuals.  Such coherent structure is indicative of additional components such 
as compact central structures, rings, annuli and extended components, and 
bars/ovals. We illustrate in Figures~\ref{resid2} and \ref{resid3} how some of 
these examples are best fitted by models with multiple S\'ersic components.

If a galaxy does not require a Stage 2 model, or if the Stage 2 model fails to 
meet the above criteria, then the galaxy is described by a single S\'ersic 
profile + point source, if present. Six galaxies are best represented by Stage 
1. Two (COMAi13017.683p275718.93 and COMAi13018.093p275723.59) cannot be fit 
with multiple S\'ersic models because they are interacting. In the third case, 
(COMAi125931.103p275718.12), the $\chi^2$ increases from Stage 1 to Stage 2. 
The final three cases (NGC~4874, NGC~4889, and COMAi125909.468p28227.35) show 
evidence of a core (see Appendix~\ref{core}).

Galaxies for which the Stage 2 model is deemed an improvement are interpreted 
as follows. Since the outer component C2 could represent a disk, we must 
specify criteria for identifying an outer disk.  The outer component C2 is a 
disk if it satisfies at least one of the following. i) The galaxy is highly 
inclined such that C2 has a low axis ratio $b/a \leq 0.25$ that is below the 
axis ratios found for ellipticals. ii) The galaxy is moderately inclined and 
C2 shows disk signatures (e.g., bars, rings, or spiral arms) in the data images 
and/or Stage 2 residuals. iii) For moderately inclined galaxies without disk 
features that do not satisfy (i) or (ii), we require S\'ersic $n$ be less than 
the threshold value $n_{\rm disk\_max}$. 

Theoretical considerations show that pure disks have $n=1$, suggesting the 
threshold should be $n\sim1$.  However, real galaxy disks are not fitted 
perfectly by S\'ersic profiles. We determine the value empirically from the 
maximum disk S\'ersic index in galaxies satisfying (i) and (ii). Highly 
inclined disks show a range in S\'ersic index of 0.48-0.86. Moderately inclined 
galaxies identified as having spiral arms but no bar have outer disks with 
S\'ersic index 0.63-1.20.  Note that some of the highly inclined 
galaxies could be barred, and this may may account for the small difference in 
average S\'ersic index between the highly inclined and moderately inclined 
barred galaxies.

In order to accurately model the outer disk of moderately inclined barred 
galaxies, a triple S\'ersic profile (see below) is required. After taking this 
extra step, the outer disk S\'ersic index among moderately inclined barred 
galaxies is 0.25-1.66. The maximum S\'ersic index among outer disks in galaxies 
satisfying requirement (i) and (ii) is 1.66, and we therefore set 
$n_{\rm disk\_max}$ to this value. Thus, outer disks span the range 0.25-1.66 
in S\'ersic index and have a median $n$ of 0.84. Figure~\ref{nupgal} shows the 
galaxy (COMAi125950.105p275529.44) on which we base our measurement of 
$n_{\rm disk\_max}$.  Appendix~\ref{nuperr} discusses the uncertainties in the 
adopted value of $n_{\rm disk\_max}$. 

Galaxies that satisfy any of requirements (i), (ii), or (iii) are deemed to 
have an outer disk. Galaxies without an outer disk are considered photometric 
ellipticals. 

We test all galaxies having an outer disk for the presence of a large-scale 
bar/oval in Stage 3 by fitting a triple S\'ersic profile + point source, if 
present. Bars/ovals are modeled with an elongated, low S\'ersic index 
($n\sim0.5$) profile (Peng et al. 2002; Weinzirl et al. 2009). In the text, we 
do not distinguish between bars and ovals, and we use ``bar'' to describe both.

The initial guesses for the three-component models come from the best Stage 2 
model combined with visual inspection. The S\'ersic index for the bar is 
initially guessed to be 0.5, and the shape and position angle of the bar are 
visually estimated using the data image or the residuals of the Stage 2 fit. 
When selecting between the Stage 2 and Stage 3 fits, we applied the same 
constraints described above for the behavior of $\chi^2$.  An additional 
complication is that in galaxies with unbarred outer disks, GALFIT may fit a 
`bar' to any existing spiral arms, rings, or clumpy disk structure. Stage 3 
fits in these cases could be discarded by noting the resulting discrepancies in 
appearance between the galaxy images and the Stage 3 model images.
Figure~\ref{resid3} shows examples of two disk galaxies where adding the third 
S\'ersic component removes the bar signature from the residuals.

\subsection{Nuclear Point Sources}\label{sptsrc}
Nuclear point sources are found in galaxies of all Hubble types. The frequency 
of nuclear point sources is very sample dependent and is particularly sensitive 
to range of galaxy  luminosity.  $\textit{HST}$ studies of early-type galaxies 
(e.g., Ravindranath et al. 2001; C{\^o}t{\'e} et al. 2006) have measured 
nucleation rates of $50\%$ or more.  Ravindranath et al. (2001) find about 
half of early-type (E, S0, S0/a) galaxies have nuclear point sources.  
C{\^o}t{\'e} et al. (2006) show that the frequency of nucleation in ACS images 
of the Virgo cluster is at least 66\% in galaxies with $M_B \leq -15$. 
Graham \& Guzm{\'a}n (2003) discuss 13/15 examples of dwarf ellipticals in the 
Coma cluster showing evidence for nucleation. 
Balcells et al. (2007a) measure a frequency of 58\% for S0 to Sbc galaxies.
B{\"o}ker et al. (2002) measure the frequency of point sources to be 75\% in 
spirals with Hubble types Scd to Sm.  

Although nuclear point sources account for a small percentage ($<1\%$) of a
galaxy's light, it is important to include them during multi-component 
structural decomposition.  Neglecting nuclear point sources can have a 
significant effect on derived parameters of bulges (Balcells et al. 2003; 
Weinzirl et al. 2009). We assess the presence of nuclear point sources with 
visual inspection. If a compact light source is visible by eye in the residuals 
of the single S\'ersic fit, the galaxy is flagged as having a potential point 
source. With this procedure, 49/69 galaxies in sample S2 have a potential 
nuclear point source. 

Galaxies having a potential nuclear point source are fitted with an extra 
nuclear point source component in the best-fit single or multi-component model.
GALFIT models the point source with the user-input PSF. More than half (38/69, 
$55.1 \pm 6.0\%$) of objects in sample S2 have a nuclear point source in the 
final, best-fit structural decomposition. Figure~\ref{ptsrc} shows examples of 
residual galaxy images with point sources.

Figure~\ref{sersic-ptsrc} shows the derived point source luminosities correlate 
with total galaxy magnitude such that more luminous point sources are found in 
brighter galaxies. Similar results been found in earlier work 
(e.g., Graham \& Guzm{\'a}n 2003; Balcells et al. 2007a).

\subsection{cD Galaxies}\label{scD}
cD galaxies are defined by having extra light on cluster-sized ($\sim1$ Mpc) 
scales with respect to the outward extrapolation of the S\'ersic profile fit to 
the inner ($\sim 100$ kpc) portion of the galaxy. Such galaxies are luminous 
and are found in regions of high galaxy number density (Binney \& Merrifield 
1998). Of the three cD galaxies in Coma, two  (NGC~4874 and NGC~4889) lie in 
the projected central 0.5~Mpc and are therefore in our sample. The third cD 
(NGC~4839) lies is in the outer southwest region of Coma and is not part of 
this study.

Definitive proof that NGC~4874 and NGC~4889 are cDs is the detection of 
intra-cluster light in Coma (Kormendy \& Bahcall 1974; Melnick et al. 1977; 
Thuan \& Kormendy 1977; Bernstein et al. 1995; Adami et al. 2005; Arnaboldi 
2011).

The single S\'ersic indices reported in Appendix~\ref{decomp} and 
Table~\ref{1cptab} for the these cD galaxies are $n\sim3-4.4$ because the 
decompositions also include the central core. The central core is a clear 
deviation from the inward extrapolation of the S\'ersic profile that 
characterizes the outer galaxy structure. For this reason, masking the core 
regions (i.e., the central $\sim2\arcsec$) 
is more physically motivated and would yield higher single S\'ersic indices 
$n\gtrsim8$. This is demonstrated in Appendix~\ref{core} and 
Table~\ref{tabcore}. 
We note that both approaches (masking or not masking the core during
the $2D$ decomposition)  lead us to the same conclusion that all of the
cD light is associated with structures  of $n\gg n_{\rm disk\_max}$
(Appendix~\ref{core}). Note in Table~\ref{multicptab} we list the cD
galaxies the structure
parameters from the 2D decomposition where the core is masked. 

The high  $n\gg n_{\rm disk\_max}$ values in the cD galaxies are due to the 
extended wings in the S\'ersic profile resulting from the extended low 
surface brightness envelope of the cD.
This extended envelope is likely made up of intra-cluster  light and the
cumulative debris from galaxies, consistent with the view that cD galaxies 
arise from repeated bouts of galactic  cannibalism and tidal stripping of 
satellite galaxies in a cluster (Ostriker \& Tremaine 1975; 
Aragon-Salamanca et al. 1998; De Lucia \& Blaizot 2007).

\subsection{Cosmic Variance}\label{cosmicvar}
The Coma ACS data only cover 19.7\% of the
projected central 0.5~Mpc radius of Coma.  The relative fractional
numbers of E+S0:spiral, or specifically the ratio of E/S0s,
we derive from this data may not be representative of the
full region in the projected central 0.5 Mpc radius of Coma
due to the incomplete sampling and cosmic variance. In order
to assess the effect of incomplete sampling and cosmic
variance on our results, we perform the following test. 

First, we define the region covered by ACS in the projected central 0.5~Mpc 
radius 
of Coma as R1, and the full area in the projected central 0.5~Mpc radius of 
Coma as R2.  We use the Hubble morphological types (MT) from the GOLD Mine 
database\footnote{\texttt{http://goldmine.mib.infn.it/}} (Gavazzi et al. 2003) 
to compute the fraction of E+S0:spiral galaxies in region R1 and R2 with 
$M_\star \geq 4.4\times10^9$~$M_\odot$, the mass limit of the Coma GOLD Mine 
sample. The MT reported by GOLD Mine are sourced from the literature.
If we take the visual MT from GOLD Mine at face value then we draw
the following conclusions:

\begin{enumerate}
\item The effect cosmic variance causes the ratio of E/S0 within the GOLD Mine
   MT to vary by a factor of 1.11 between region R1 and R2
   for $M_\star \geq 4.4\times10^9$~$M_\odot$.
\item The partial ACS coverage of the projected central 0.5 Mpc and associated
   cosmic variance thus causes our study based on region R1 to
\begin{enumerate}
\item overestimate the ratio of S0/E in the ACS sample 
   for $M_\star \geq 4.4\times10^9$~$M_\odot$ by a factor of 1.4.
\item overestimate the fraction $f_{\rm cold}$ of dynamically cold stellar mass
   (43\%) by a factor of 1.27 (Section~\ref{datavmodel}) for
   $M_\star \geq 10^9$~$M_\odot$. We note that the 
   over-estimation of $f_{\rm cold}$ is not by the same factor as in ii (a) 
   because S0s have a significant fraction of their mass in dynamically hot 
   bulges.
\end{enumerate}
\item Currently, our conclusion in Section~\ref{datavmodel}, based on region R1
   is that the hierarchical models are over-predicting the empirical fraction 
   $f_{\rm cold}$. It is clear from ii (b), that correcting
   for partial ACS coverage and cosmic variance would only strengthen
   this conclusion further.
\end{enumerate}

\subsection{Galaxy Color Gradients}\label{colorgrad}
In Section~\ref{census} we suggest that galaxy color gradients should not bias 
our conclusions concerning the distribution of dynamically hot and cold stellar
mass. Here, we explicitly test this idea.

For a subset of 10 galaxies spanning types G3 to G5 and matching the morphology
distribution of the mass-selected sample (E+S0:spiral = 2+7:1) in 
Table~\ref{tabmdr}, we re-evaluated the fractional mass in hot and cold 
components based on combining structural decompositions of both the F814W and 
F475W images. The new F475W-band decompositions were performed identically to 
the existing F814W decompositions, except that the position angle and axis 
ratio of the galaxy structures were fixed to their values from the F814W-band
decompositions. Stellar masses of the structural components were calculated 
according to Into \& Portinari (2013) after converting the F475W-F814W color
and the F814W luminosity into a $B-I$ color and $I$-band luminosity,
respectively, using the procedure in Section~\ref{datas2}.

In the new F475W decompositions for this subset of galaxies, the half-light 
radii and S\'ersic $n$ are similar to the corresponding values in the F814W band.
The average offset is 5.4\% with a standard deviation of 5.6\%.
Furthermore, the fractional hot stellar mass inferred from a constant global 
F814W $M/L$ ratio is 53.4\%. After calculating the stellar mass of each 
galaxy component from the $B-I$ color, the fractional hot stellar mass is found 
to be 50.5\%. Thus, $M/L$ gradients within a galaxy do not appear to have a 
significant effect on the fractional masses measured in cold versus hot 
components.

\section{Identifying Core Ellipticals}\label{core}

While elliptical galaxies are remarkably well-fit by S\'ersic profiles over 
large dynamic ranges, giant elliptical galaxies contain cores, or ``missing 
light'' at small radii that constitute a downward deviation from from the 
inward extrapolation of the outer S\'ersic profile (Graham et al. 2003; 
Trujillo et al. 2004; Kormendy et al. 2009). Such cores are hypothesized to 
form from scouring induced by binary black holes during dry, dissipationless 
mergers. 

Because cores, which have traditionally been identified with 1D radial light 
profiles, are not an obvious feature of the galaxy's 2D light distribution, 
global S\'ersic fits will encompass any existing core. This is potentially
problematic for at least two reasons.  Including the core in the S\'ersic fit 
will lower the global S\'ersic index. This is of concern in this paper where
the S\'ersic index plays a key role in interpreting galaxy structure 
(Section~\ref{approach}). Secondly, fitting the core region may produce 
features in the residuals that prompt addition of extra nuclear components 
that have no physical justification.

We systematically search for cores in all sample galaxies.
For this task, we use 1D light profiles 
generated from ellipse fitting of deconvolved images. 
The ACS images were deconvolved using a simulated PSF (Appendix~\ref{app} 
for details) and 40 iterations of Lucy-Richardson deconvolution with the IRAF 
task \texttt{LUCY} (Lucy 1974; Richardson 1972). Our approach uses the criteria 
for identifying core galaxies from Trujillo et al. (2004) by fitting S\'ersic 
and core-S\'ersic profiles (Graham et al. 2003) to the 1D light profiles. 

For simplicity, we use the version of the core-S\'ersic profile that assumes 
an infinitely sharp transition between the outer S\'ersic and inner power-law 
regions, namely
\begin{equation}
\rm{I(r)= I_b [(r_b/r)^\gamma u(r_b-r)+ e^{b(r_b/r_\mathrm{e})^{1/n}} e^{-b(r/r_\mathrm{e})^{1/n}}u(r-r_b)]}.
\end{equation}
Here, $r_b$ denotes the division between the outer S\'ersic and inner power-law 
profiles, $I_b$ is the intensity at this radius, $\gamma$ is the inner 
power-law slope, and $u(x-a)$ is the Heaviside step function.  Parameters $n$ 
and $r_\mathrm{e}$ refer to the shape and half-light radius of the outer 
S\'ersic profile.  Additionally, $b$ is a constant that depends on several free 
parameters ($r_b$, $\gamma$, $r_\mathrm{e}$, and $n$).

We require a core galaxy to meet the following criteria:  1) the core-S\'ersic
model provide a better fit than the S\'ersic profile; 2) the cores are 
well-resolved so that the break radius $r_b$ is greater than the second 
innermost data point in the profile; 3) the inner power-law slope $\gamma$ is 
less than the logarithmic slope of the S\'ersic profile ($1/n$) in the core 
region.

Three sample galaxies meet the above criteria for having a core. Two of these 
are the central cD galaxies NGC~4874 and NGC~4889. Table~\ref{tabcore} 
summarizes the $r_b$ and $\gamma$ measured from the core-S\'ersic fit.

We further explore the best way to handle these cored galaxies in the 2D 
luminosity decompositions. Two natural approaches are to fit the whole galaxy, 
including the core, or to mask the galaxy over $r\leq r_b$. 
Masking is more physically motivated because the central core is a clear 
deviation from the inward extrapolation of the S\'ersic profile that 
characterizes the outer galaxy structure. We try both approaches and summarize 
the results in Table~\ref{tabcore}.  Applying a mask versus no mask has a 
nominal effect on COMAi125909.468p28227.35, but there is a 
significant increase in the $r_\mathrm{e}$ and $n$ of the cD galaxies when 
their larger core regions are masked. 

Performing the 2D fit with the core masked is more physically motivated, and we consider these models to represent the best fits for the cD galaxies.
It is worth noting, however, that our result from Appendix~\ref{scD}
that 100\% of the mass in the cDs is
associated with structures of $n\gg n_{\rm disk\_max}$
remains unchanged irrespective of which
approach (mask or no mask) we take.

\section{Systematics of $\MakeLowercase{n_{\rm disk\_max}}$}\label{nuperr}
Our effort in this paper to make a census (Section~\ref{census}) of 
dynamically cold versus dynamically hot stellar mass depends fundamentally on 
the upper limit, $n_{\rm disk\_max}$ (Section~\ref{approach}), measured for the 
S\'ersic index of a disk. In our approach, all structures with S\'ersic index 
$n\leq n_{\rm disk\_max}$ are considered disk dominated, while all other 
structures with higher 
S\'ersic index are considered classical components built in mergers.

The value of $n\leq n_{\rm disk\_max}$ is set by the moderately inclined barred 
galaxy (COMAi125950.105p275529.44) having the highest outer disk S\'ersic 
index. The accuracy of $n_{\rm disk\_max}$ depends on how representative the 
sample is as well as the robustness of the multi-component structural 
decompositions. Figure~\ref{nupgal} shows for this galaxy the data image and 
residuals of the multi-component decompositions.  While this galaxy was 
identified as an ambiguous E/S0 galaxy in Figure~2 of Marinova et al. (2012), 
the barred nature of this galaxy seems clear based on the image residuals 
produced by our improved method (Sections~\ref{special} and \ref{scheme}) 
of structural decomposition.

The value of $n_{\rm disk\_max}$ is subject to sky subtraction errors because 
it is measured from the outermost S\'ersic profile of disk galaxies, and this 
is likely the dominant systematic effect on $n_{\rm disk\_max}$. As described 
in Appendix~\ref{app}, we measure the background sky value with a robust 
method and hold the sky fixed at this value during the fit. To test the 
importance of the sky subtraction, we refitted COMAi125950.105p275529.44 while 
adjusting the mean sky background by $\pm 1\sigma$. This produces a range in 
outer disk $n$ of $n\sim 1.57-1.77$, which spans $\sim0.1$ above and below the 
adopted $n_{\rm disk\_max}$ value of 1.66. Based on the narrow error bars for 
$n_{\rm disk\_max}$, we do not expect the uncertainty to have a significant 
impact on our conclusions.

For completeness, we explore for an alternate value of $n_{\rm disk\_max}$ the 
relative stellar mass fractions that would be interpreted as belonging to cold 
versus hot stellar components. The value $n_{\rm disk\_max}=2$ is in line with 
estimates of the S\'ersic index of small-scale disks (e.g., Fisher \& Drory 
2008; Weinzirl et al. 2009) yet is still above the anticipated range in 
$n_{\rm disk\_max}$ due to sky subtraction errors in this study. 
With this higher $n_{\rm disk\_max}$, we would find that 
$\sim51\%$ stellar mass is in disk-dominated components while 
$\sim49\%$ is still in classical
bulges/ellipticals assembled in major and minor mergers. 
These values are somewhat different from the corresponding values 
(43\% in disk-dominated structures versus 57\% in non-disks) 
derived in Section~\ref{census} excluding the 2 cD galaxies.  
Choosing a higher $n_{\rm disk\_max}$ would increase the 
importance of disk-building processes relative to processes that build 
classical bulges/ellipticals. 

\section*{Acknowledgments}

SJ, TW, acknowledge support from the National Aeronautics and Space 
Administration (NASA) LTSA grant NAG5-13063, NSF grant AST-0607748, 
and \textit{HST} grants GO-11082 and GO-10861 from STScI, which is operated 
by AURA, Inc., for NASA, under NAS5-26555.
SJ and TW  also acknowledge support from the Norman Hackerman
Advanced Research Program (NHARP) ARP-03658-0234-2009,
National Aeronautics and Space Administration (NASA) LTSA grant
NAG5-13063, and NSF grant AST-0607748.
SJ and TW  acknowledge support for this research by
the DFG cluster of excellence "Origin and Structure of
the Universe" (www.universe-cluster.de).
M. Balcells acknowledges support from grant AYA2009-11137 from the Spanish Ministry of Science and Technology.
We thank Roderik Overzier, T.~J. Cox, and 
Jennifer Lotz for stimulating discussions.
This work made use of observations with the NASA/ESA Hubble Space Telescope
obtained at the Space Telescope Science Institute, which is operated by the
Association of Universities for Research in Astronomy, Inc., under NASA
contract NAS 5-26555. These observations are associated with program
GO10861.
We acknowledge the usage of the HyperLeda database (http://leda.univ-lyon1.fr).
Funding for the Sloan Digital Sky Survey (SDSS) has been provided by the 
Alfred P. Sloan Foundation, the Participating Institutions, the National Aeronautics 
and Space Administration, the National Science Foundation, the U.S. Department of 
Energy, the Japanese Monbukagakusho, and the Max Planck Society. The SDSS Web site 
is http://www.sdss.org/.
The SDSS is managed by the Astrophysical Research Consortium (ARC) for the Participating 
Institutions. The Participating Institutions are The University of Chicago, Fermilab, 
the Institute for Advanced Study, the Japan Participation Group, The Johns Hopkins 
University, Los Alamos National Laboratory, the Max-Planck-Institute for Astronomy 
(MPIA), the Max-Planck-Institute for Astrophysics (MPA), New Mexico State University, 
University of Pittsburgh, Princeton University, the United States Naval Observatory, 
and the University of Washington.

\setcounter{figure}{0} \renewcommand{\thefigure}{\arabic{figure}}
\setcounter{table}{0} \renewcommand{\thetable}{\arabic{table}}

 \begin{table}
  \caption{Cross Identifications}
\label{tab:cross_ids}
{\scriptsize
\begin{tabular}{lcccc}
\hline
\multicolumn{1}{c}{Galaxy Name} &  
\multicolumn{1}{c}{SDSS DR8 Name} & 
\multicolumn{1}{c}{2MASS XSC (or PSC Name)} &
\multicolumn{1}{c}{GMP Name} &
\multicolumn{1}{c}{Dressler (1980) Name} \\
\multicolumn{1}{c}{(1)} &  
\multicolumn{1}{c}{(2)} & 
\multicolumn{1}{c}{(3)} &
\multicolumn{1}{c}{(4)} &
\multicolumn{1}{c}{(5)} \\
\hline
COMAi125926.458p275124.81  & 1237667444048658752  &  -  &  GMP3473  &  -  \\
COMAi13007.123p275551.49   & 1237667444048724242  &  2MASSJ13000711+2755511  &  GMP2931  &  -  \\
COMAi125930.270p28115.17   & 1237667324334571563  &  -  &  GMP3406  &  -  \\
COMAi125937.200p275819.97  & 1237667444048658537  &  2MASSJ12593720+2758203  &  GMP3308  &  -  \\
COMAi125953.929p275813.75  & 1237667444048658918  &  -  &  GMP3098  &  -  \\
COMAi13018.351p28333.32    & 1237667324334637348  &  -  &  GMP2787  &  -  \\
COMAi125937.010p28106.95   & 1237667324334571551  &  2MASSJ12593699+2801074  &  GMP3312  &  -  \\
COMAi125946.943p275930.90  & 1237667324334571832  &  2MASSJ12594688+2759308  &  GMP3166  &  -  \\
COMAi13030.954p28630.22    & 1237667324334637213  &  2MASSJ13003091+2806300  &  GMP2626  &  -  \\
COMAi13035.420p275634.06   & 1237667444048724352  &  -  &  GMP2585  &  -  \\
COMAi125950.183p275445.52  & 1237667444048658912  &  -  &  GMP3131  &  -  \\
COMAi125959.476p275626.02  & 1237667444048658878  &  -  &  GMP3034  &  -  \\
COMAi13000.949p275643.85   & 1237667444048658882  &  2MASSJ13000095+2756433  &  GMP3017  &  -  \\
COMAi13034.430p275604.95   & 1237667444048724349  &  2MASSJ13003442+2756047  &  GMP2591  &  -  \\
COMAi125931.893p275140.76  & 1237667444048658763  &  2MASSJ12593186+2751406  &  GMP3383  &  -  \\
COMAi125931.103p275718.12  & 1237667444048658549  &  -  &  GMP3392  &  -  \\
COMAi13041.193p28242.34    & 1237667324334702866  &  2MASSJ13004119+2802424  &  GMP2529  &  -  \\
COMAi125845.533p274513.75  & 1237667323797635368  &  2MASSJ12584558+2745132  &  GMP4035  &  -  \\
COMAi13018.545p28549.62    & 1237667324334637356  &  2MASSJ13001857+2805503  &  GMP2784  &  -  \\
COMAi13021.673p275354.81   & 1237667444048724303  &  2MASXJ13002172+2753545  &  GMP2736  &  -  \\
COMAi13024.823p275535.94   & 1237667444048724320  &  2MASSJ13002482+2755353  &  GMP2692  &  -  \\
COMAi13051.149p28249.90    & 1237667324334702708  &  2MASSJ13005112+2802499  &  GMP2423  &  -  \\
COMAi13011.143p28354.91    & 1237667324334637325  &  2MASSJ13001117+2803551  &  GMP2879  &  -  \\
COMAi125937.990p28003.52   & 1237667324334571647  &  2MASSJ12593798+2800036  &  GMP3292  &  -  \\
COMAi13018.873p28033.38    & 1237667324334637362  &  2MASXJ13001890+2800332  &  GMP2777  &  -  \\
COMAi125911.543p28033.32   & 1237667324334506328  &  2MASSJ12591153+2800334  &  GMP3681  &  -  \\
COMAi125904.797p28301.16   & 1237667324334506316  &  2MASXJ12590475+2803019  &  GMP3780  &  -  \\
COMAi125909.468p28227.35   & 1237667324334506325  &  2MASXJ12590943+2802279  &  GMP3707  &  -  \\
COMAi125935.286p275149.13  & 1237667444048658774  &  2MASXJ12593524+2751488  &  GMP3339  &  -  \\
COMAi13005.405p28128.14    & 1237667324334637091  &  2MASXJ13000538+2801282  &  GMP2960  &  -  \\
COMAi125950.105p275529.44  & 1237667444048658822  &  2MASXJ12595013+2755292  &  GMP3133  &  -  \\
COMAi13018.772p275613.34   & 1237667444048723991  &  2MASXJ13001877+2756135  &  GMP2778  &  -  \\
COMAi125938.321p275913.89  & 1237667444048658535  &  2MASXJ12593827+2759137  &  GMP3291  &  D154  \\
COMAi125940.270p275805.71  & 1237667444048658530  &  2MASSJ12594026+2758058  &  GMP3254  &  D127  \\
COMAi125944.208p275730.38  & 1237667444048658531  &  2MASXJ12594423+2757307  &  GMP3206  &  D126  \\
COMAi125939.659p275714.03  & 1237667444048658528  &  2MASSJ12593965+2757141  &  GMP3269  &  D128  \\
COMAi13044.632p28602.31    & 1237667324334702891  &  2MASXJ13004459+2806026  &  GMP2489  &  D191  \\
COMAi125928.721p28225.92   & 1237667324334571539  &  2MASXJ12592868+2802258  &  GMP3433  &  D177  \\
COMAi125942.301p275529.15  & 1237667444048658653  &  2MASXJ12594234+2755287  &  GMP3222  &  D125  \\
COMAi13017.014p28350.07    & 1237667324334637347  &  2MASXJ13001702+2803502  &  GMP2805  &  D171  \\
COMAi125956.697p275548.71  & 1237667444048658858  &  2MASXJ12595670+2755483  &  GMP3068  &  D123  \\
COMAi13016.534p275803.15   & 1237667444048723984  &  2MASXJ13001655+2758032  &  GMP2815  &  D122  \\
COMAi13006.395p28015.94    & 1237667324334637086  &  2MASXJ13000643+2800142  &  GMP2940  &  D150  \\
COMAi13027.966p275721.56   & 1237667444048724118  &  2MASXJ13002798+2757216  &  GMP2654  &  D119  \\
COMAi13012.868p28431.74    & 1237667324334637140  &  2MASXJ13001286+2804322  &  GMP2861  &  D173  \\
COMAi125943.721p275940.82  & 1237667324334571645  &  2MASSJ12594372+2759409  &  GMP3213  &  D153  \\
COMAi13028.370p275820.64   & 1237667444048724328  &  2MASXJ13002835+2758206  &  GMP2651  &  D147  \\
COMAi13042.832p275746.95   & 1237667444048724176  &  2MASXJ13004285+2757476  &  GMP2510  &  D116  \\
COMAi13038.761p28052.34    & 1237667324334702605  &  2MASXJ13003877+2800516  &  GMP2551  &  D146  \\
COMAi13014.746p28228.69    & 1237667324334637152  &  2MASXJ13001475+2802282  &  GMP2839  &  D172  \\
COMAi13022.170p28249.30    & 1237667324334637189  &  2MASXJ13002215+2802495  &  GMP2727  &  D170  \\
COMAi125931.453p28247.60   & 1237667324334571535  &  2MASXJ12593141+2802478  &  GMP3390  &  D176  \\
COMAi13018.093p275723.59   & 1237667444048723985  &  2MASSJ13001809+2757235  &  GMP2794  &  D120  \\
COMAi13040.838p275947.80   & 1237667324334702869  &  2MASXJ13004081+2759476  &  GMP2535  &  D145  \\
COMAi125852.097p274706.15  & 1237667323797635203  &  2MASXJ12585208+2747059  &  GMP3958  &  D072  \\
COMAi125946.782p275825.99  & 1237667444048658525  &  2MASXJ12594681+2758252  &  GMP3170  &  D152  \\
COMAi13008.003p28442.81    & 1237667324334637131  &  2MASXJ13000803+2804422  &  GMP2922  &  D174  \\
COMAi125929.956p275723.26  & 1237667444048658522  &  2MASSJ12592995+2757231  &  GMP3414  &  D131  \\
COMAi125929.403p275100.46  & 1237667444048658609  &  2MASXJ12592936+2751008  &  GMP3423  &  D088  \\
COMAi125932.771p275901.04  & 1237667444048658523  &  2MASXJ12593276+2759008  &  GMP3367  &  D155  \\
COMAi125944.407p275444.84  & 1237667444048658654  &  2MASXJ12594438+2754447  &  GMP3201  &  D124  \\
COMAi125930.824p275303.05  & 1237667444048658616  &  2MASXJ12593079+2753028  &  GMP3400  &  D103  \\
COMAi13039.767p275526.19   & 1237667444048724135  &  2MASXJ13003975+2755256  &  GMP2541  &  D118  \\
COMAi13042.766p275817.38   & 1237667324334702622  &  2MASXJ13004277+2758166  &  GMP2516  &  D144  \\
COMAi13048.646p28526.69    & 1237667324334702681  &  2MASXJ13004867+2805266  &  GMP2440  &  D168  \\
COMAi13017.683p275718.93   & 1237667444048723981  &  2MASXJ13001768+2757192  &  GMP2798  &  D121  \\
COMAi13051.464p28234.86    & 1237667324334702705  &  2MASXJ13005158+2802341  &  GMP2417  &  D167  \\
NGC 4889  & 1237667444048723983  &  2MASXJ13000809+2758372  &  GMP2921  &  D148  \\
COMAi125935.698p275733.36 (NGC 4874)  & 1237667444048658532  &  2MASXJ12593570+2757338  &  GMP3329  &  D129  \\
\hline
\multicolumn{5}{p{.85\textwidth}}{
Notes. If there is no match in the 2MASS Extended Source catalog (2MASX),
where available, the 2MASS Point Source catalog name (2MASS) is given in column (3). 
GMP name refers to the Godwin, Metcalfe, Peach (1983) catalog.}
\end{tabular}
}
\end{table}

\begin{table}
\caption{Properties Of Cored Ellipticals}
\label{tabcore}
\begin{tabular}{@{}ccccccc}
\hline
Galaxy Name  & Core-S\'ersic ($\gamma$, $r_b$) & 2D S\'ersic Profile w/o & 2D S\'ersic Profile w/ \\
             &                                 & Core Masked ($n$, $r_e$) & Core Masked ($n$, $r_e$)\\
(1) &       (2)                       &   (3)                            &         (4)    \\
\hline
COMAi125909.468p28227.35  & (0.16, 0\farcs13)  &   (2.54, 4\farcs2)      &  (2.54, 4\farcs20)  \\
NGC~4874 (ACS F814W)      & (0.15, 1\farcs40)  &   (2.89, 35\farcs4)     &  (11.4, 875\farcs)  \\
NGC~4874 (SDSS $i$-band)  & (0.15, 2\farcs32)  &   (4.30, 88\farcs3)      &  (4.70, 107\farcs)  \\
NGC~4889 (SDSS $i$-band)  & (0.06, 1\farcs88)  &   (3.90, 42\farcs9)      &  (7.80, 129\farcs)  \\
\hline
\multicolumn{7}{p{.75\textwidth}}{Notes. Galaxies are identified as having a core following the
procedure in Appendix~\ref{core}. Two of the cored galaxies (NGC~4874 and NGC~4889)
are cD galaxies.}
\end{tabular}
\end{table}

\begin{table}
\scriptsize
\caption{Galaxy Properties And Single S\'ersic Profile Structural Parameters}
\label{1cptab}
\begin{tabular}{@{}ccccccc}
\hline
Galaxy Name  & RA & DEC & $M_\star$ & F814W Magnitude & $r_\mathrm{e}$ & $n$ \\
             &    &     &($M_\odot$)&                 &    (kpc)       &     \\
(1)          &(2) & (3) &   (4)     &    (5)          &     (6)        & (7)  \\
\hline
COMAi125926.458p275124.81 & 194.860245 & 27.856893 & 1.03e+09 & 17.43 & 0.97 & 1.88\\
COMAi13007.123p275551.49 & 195.029679 & 27.930971 & 1.03e+09 & 17.42 & 0.62 & 2.76\\
COMAi125930.270p28115.17 & 194.876126 & 28.020883 & 1.05e+09 & 17.46 & 1.04 & 1.90\\
COMAi125937.200p275819.97 & 194.905001 & 27.972214 & 1.12e+09 & 17.68 & 0.29 & 4.51\\
COMAi125953.929p275813.75 & 194.974706 & 27.970489 & 1.17e+09 & 17.41 & 1.10 & 1.30\\
COMAi13018.351p28333.32 & 195.076465 & 28.059258 & 1.17e+09 & 17.13 & 1.32 & 1.23\\
COMAi125937.010p28106.95 & 194.904209 & 28.018598 & 1.23e+09 & 17.34 & 0.69 & 2.08\\
COMAi125946.943p275930.90 & 194.945597 & 27.991917 & 1.40e+09 & 17.02 & 1.49 & 1.65\\
COMAi13030.954p28630.22 & 195.128977 & 28.108396 & 1.46e+09 & 17.07 & 1.41 & 1.94\\
COMAi13035.420p275634.06 & 195.147586 & 27.942797 & 1.72e+09 & 16.83 & 1.92 & 1.81\\
COMAi125950.183p275445.52 & 194.959098 & 27.912647 & 1.81e+09 & 16.95 & 1.57 & 1.41\\
COMAi125959.476p275626.02 & 194.997820 & 27.940564 & 1.83e+09 & 16.53 & 2.41 & 2.30\\
COMAi13000.949p275643.85 & 195.003956 & 27.945514 & 1.92e+09 & 16.40 & 2.00 & 3.38\\
COMAi13034.430p275604.95 & 195.143461 & 27.934709 & 1.94e+09 & 16.65 & 2.11 & 2.10\\
COMAi125931.893p275140.76 & 194.882888 & 27.861324 & 1.96e+09 & 16.78 & 0.94 & 1.88\\
COMAi125931.103p275718.12 & 194.879597 & 27.955035 & 1.99e+09 & 16.87 & 2.47 & 1.61\\
COMAi13041.193p28242.34 & 195.171639 & 28.045097 & 2.15e+09 & 16.83 & 0.86 & 1.66\\
COMAi125845.533p274513.75 & 194.689724 & 27.753820 & 2.21e+09 & 16.56 & 2.14 & 2.09\\
COMAi13018.545p28549.62 & 195.077272 & 28.097119 & 2.25e+09 & 16.64 & 1.62 & 1.73\\
COMAi13021.673p275354.81 & 195.090308 & 27.898559 & 2.71e+09 & 16.37 & 1.46 & 2.57\\
COMAi13024.823p275535.94 & 195.103430 & 27.926652 & 2.73e+09 & 16.41 & 2.68 & 2.15\\
COMAi13051.149p28249.90 & 195.213122 & 28.047197 & 2.78e+09 & 15.52 & 5.35 & 3.17\\
COMAi13011.143p28354.91 & 195.046429 & 28.065253 & 2.83e+09 & 16.38 & 1.42 & 2.12\\
COMAi125937.990p28003.52 & 194.908292 & 28.000979 & 2.97e+09 & 16.53 & 1.16 & 2.00\\
COMAi13018.873p28033.38 & 195.078639 & 28.009273 & 2.98e+09 & 16.58 & 0.38 & 3.29\\
COMAi125911.543p28033.32 & 194.798099 & 28.009258 & 3.26e+09 & 16.40 & 1.22 & 1.84\\
COMAi125904.797p28301.16 & 194.769991 & 28.050323 & 3.78e+09 & 16.09 & 2.16 & 2.28\\
COMAi125909.468p28227.35 & 194.789451 & 28.040932 & 5.14e+09 & 15.94 & 1.88 & 2.54\\
COMAi125935.286p275149.13 & 194.897029 & 27.863650 & 5.27e+09 & 15.96 & 1.06 & 1.54\\
COMAi13005.405p28128.14 & 195.022521 & 28.024486 & 8.40e+09 & 15.11 & 2.61 & 2.58\\
COMAi125950.105p275529.44 & 194.958773 & 27.924845 & 8.88e+09 & 15.62 & 1.35 & 2.32\\
COMAi13018.772p275613.34 & 195.078218 & 27.937041 & 9.56e+09 & 15.09 & 3.23 & 2.57\\
COMAi125938.321p275913.89 & 194.909675 & 27.987192 & 9.96e+09 & 14.93 & 3.67 & 3.50\\
COMAi125940.270p275805.71 & 194.917794 & 27.968254 & 1.01e+10 & 15.04 & 2.52 & 5.94\\
COMAi125944.208p275730.38 & 194.934203 & 27.958439 & 1.20e+10 & 14.60 & 3.55 & 3.89\\
COMAi125939.659p275714.03 & 194.915246 & 27.953900 & 1.21e+10 & 15.05 & 1.33 & 3.65\\
COMAi13044.632p28602.31 & 195.185968 & 28.100644 & 1.38e+10 & 14.86 & 1.82 & 2.84\\
COMAi125928.721p28225.92 & 194.869671 & 28.040534 & 1.39e+10 & 14.95 & 1.67 & 2.97\\
COMAi125942.301p275529.15 & 194.926256 & 27.924765 & 1.61e+10 & 14.56 & 1.16 & 7.49\\
COMAi13017.014p28350.07 & 195.070896 & 28.063911 & 1.79e+10 & 14.75 & 1.35 & 3.91\\
COMAi125956.697p275548.71 & 194.986241 & 27.930200 & 1.82e+10 & 14.35 & 4.71 & 3.80\\
COMAi13016.534p275803.15 & 195.068895 & 27.967542 & 1.84e+10 & 14.24 & 2.98 & 4.58\\
COMAi13006.395p28015.94 & 195.026649 & 28.004430 & 1.87e+10 & 13.82 & 5.01 & 7.52\\
COMAi13027.966p275721.56 & 195.116526 & 27.955989 & 2.11e+10 & 14.51 & 1.55 & 4.68\\
COMAi13012.868p28431.74 & 195.053620 & 28.075485 & 2.11e+10 & 14.52 & 1.40 & 3.21\\
COMAi125943.721p275940.82 & 194.932172 & 27.994675 & 2.13e+10 & 14.48 & 1.38 & 3.81\\
COMAi13028.370p275820.64 & 195.118212 & 27.972400 & 2.17e+10 & 14.24 & 4.29 & 4.04\\
COMAi13042.832p275746.95 & 195.178470 & 27.963042 & 2.49e+10 & 14.17 & 2.70 & 4.47\\
COMAi13038.761p28052.34 & 195.161508 & 28.014541 & 2.51e+10 & 13.87 & 5.34 & 4.31\\
COMAi13014.746p28228.69 & 195.061442 & 28.041304 & 2.88e+10 & 14.06 & 1.58 & 4.70\\
COMAi13022.170p28249.30 & 195.092378 & 28.047029 & 2.88e+10 & 13.84 & 2.97 & 4.08\\
COMAi125931.453p28247.60 & 194.881057 & 28.046557 & 2.88e+10 & 14.22 & 1.82 & 2.84\\
COMAi13018.093p275723.59 & 195.075391 & 27.956554 & 2.89e+10 & 14.41 & 1.27 & 2.41\\
COMAi13040.838p275947.80 & 195.170159 & 27.996612 & 2.97e+10 & 14.00 & 3.15 & 3.27\\
COMAi125852.097p274706.15 & 194.717073 & 27.785042 & 3.05e+10 & 14.02 & 1.90 & 3.29\\
COMAi125946.782p275825.99 & 194.944929 & 27.973886 & 3.44e+10 & 13.83 & 3.38 & 4.33\\
COMAi13008.003p28442.81 & 195.033348 & 28.078560 & 3.51e+10 & 14.10 & 1.13 & 2.59\\
COMAi125929.956p275723.26 & 194.874818 & 27.956462 & 3.92e+10 & 13.32 & 3.85 & 4.89\\
COMAi125929.403p275100.46 & 194.872516 & 27.850130 & 4.26e+10 & 13.87 & 1.68 & 4.07\\
COMAi125932.771p275901.04 & 194.886550 & 27.983624 & 4.49e+10 & 13.20 & 5.50 & 5.86\\
COMAi125944.407p275444.84 & 194.935031 & 27.912457 & 4.62e+10 & 13.67 & 2.26 & 2.96\\
COMAi125930.824p275303.05 & 194.878435 & 27.884182 & 5.02e+10 & 13.61 & 1.54 & 3.77\\
COMAi13039.767p275526.19 & 195.165696 & 27.923943 & 5.02e+10 & 13.44 & 2.93 & 3.64\\
COMAi13042.766p275817.38 & 195.178194 & 27.971495 & 5.73e+10 & 13.43 & 2.37 & 4.00\\
COMAi13048.646p28526.69 & 195.202693 & 28.090749 & 6.69e+10 & 13.30 & 2.21 & 3.02\\
COMAi13017.683p275718.93 & 195.073683 & 27.955259 & 7.06e+10 & 13.28 & 2.13 & 2.85\\
COMAi13051.464p28234.86 & 195.214436 & 28.043019 & 7.48e+10 & 13.07 & 2.85 & 3.92\\
NGC~4889 & 195.033750 & 27.977000 & 5.78e+11 & 10.57 & 22.65 & 4.37\\
COMAi125935.698p275733.36 (NGC~4874) & 194.898743 & 27.959269 & 7.69e+11 & 10.96 & 17.35 & 3.05\\
\hline
\multicolumn{7}{p{.5\textwidth}}{Notes.  Rows are sorted by increasing $M_\star$.
}
\end{tabular}
\end{table}

\begin{table}
\caption{Distribution Of Best-Fit Structural Decompositions For Stellar Mass $M_\star \geq 10^9$ $M_\odot$}
\label{tabdecomp}
\begin{tabular}{@{}cccccccc}
\hline
Morphology & Number Per Bin& Stage 1 & Stage 1 & Stage 2 & Stage 2 &   Stage 3 & Stage 3 \\
& & w/o & w/ & w/o & w/ &  w/o & w/  \\
& & Point Source & Point Source & Point Source &  Point Source & Point Source & Point Source  \\
 (1) &  (2)  &  (3) &  (4) & (5) & (6) & (7) & (8) \\
\hline
       &    &    &    &    &    &    &    \\
All Galaxies & 69 & 3 & 3 & 14 & 24 & 14 & 11 \\
\hline
\multicolumn{8}{c}{\vspace*{2 mm}\textbf{In terms of galaxy types G1 to G5}}\\
G1: photometric disk & 1 & 0 & 1 & 0 & 0 & 0 & 0 \\
G2: photometric E & 5 & 3 & 2 & 0 & 0 & 0 & 0 \\
G3: unbarred S0, spiral & 24 & 0 & 0 & 8 & 16 & 0 & 0 \\
G4: barred S0, spiral & 25 & 0 & 0 & 0 & 0 & 14 & 11 \\
G5: photometric E with & 14 & 0 & 0 & 6 & 8 & 0 & 0 \\
extra inner component&    &   &   &   &   &   &   \\
\hline      
\multicolumn{8}{c}{\vspace*{2 mm}\textbf{In terms of Hubble types cD, E, S0, and spiral}}\\
cD            & 2  & 2 & 0 & 0 & 0  & 0  & 0 \\
Photometric E & 17 & 1 & 2 & 6 & 8  & 0  & 0 \\
S0            & 44 & 0 & 1 & 8 & 12 & 14 & 9 \\
Spiral        & 6  & 0 & 0 & 0 & 4  & 0  & 2 \\
\hline
\multicolumn{8}{p{.9\textwidth}}{Notes. 
This table shows the distribution of best-fit models and the breakdown of 
galaxies into classes G1 to G5 arrived at by applying the structural 
decomposition and galaxy classification schemes described in 
Section~\ref{scheme} and Figures~\ref{fitflow}-\ref{galaxyclass}.
}
\end{tabular}
\end{table}

\begin{table}
\scriptsize
\caption{Structural Parameters For the Best Model}
\label{multicptab}
\begin{tabular}{@{}cccccccccc}
\hline
Galaxy Name  & G$n$ & Hubble Type& Point source$/T$, $C1/T$, $C2/T$, Bar$/T$ & C1 $r_\mathrm{e}$ &  C1 $n$ & C2 $r_\mathrm{e}$ & C2 $n$ & Bar $r_\mathrm{e}$ & Bar $n$ \\
             &      &            & (\%,\%,\%,\%)               &    (kpc)             &            &    (kpc)             &         &  (kpc)             &         \\
(1)  &  (2) &  (3) & (4) & (5)  & (6) & (7) & (8) & (9) & (10)        \\
\hline
COMAi125926.458p275124.81 &  G5 & E &  (0.22, 92.40, 7.36, 0.00)  & 1.05 &  1.72 &  0.42 & 0.85 &  ... & ... \\
COMAi13007.123p275551.49 &  G5 & E &  (0.00, 100.00, 0.00, 0.00)  & 0.63 &  1.94 &  0.32 & 5.88 &  ... & ... \\
COMAi125930.270p28115.17 &  G5 & E &  (0.46, 89.20, 10.30, 0.00)  & 1.13 &  1.83 &  0.62 & 0.70 &  ... & ... \\
COMAi125937.200p275819.97 &  G5 & E &  (0.00, 79.90, 20.10, 0.00)  & 0.31 &  6.20 &  0.20 & 0.89 &  ... & ... \\
COMAi125953.929p275813.75 &  G3 & spiral &  (0.46, 43.00, 56.50, 0.00)  & 0.79 &  1.31 &  1.31 & 0.63 &  ... & ... \\
COMAi13018.351p28333.32 &  G3 & S0 &  (0.05, 42.50, 57.40, 0.00)  & 0.85 &  1.24 &  1.75 & 0.52 &  ... & ... \\
COMAi125937.010p28106.95 &  G5 & E &  (0.51, 91.10, 8.42, 0.00)  & 0.79 &  2.13 &  0.35 & 0.51 &  ... & ... \\
COMAi125946.943p275930.90 &  G3 & S0 &  (0.09, 7.22, 92.70, 0.00)  & 0.31 &  0.98 &  1.47 & 1.01 &  ... & ... \\
COMAi13030.954p28630.22 &  G4 & S0 &  (0.33, 2.13, 75.40, 22.20)  & 0.14 &  1.11 &  1.87 & 1.20 &  0.61 & 0.85 \\
COMAi13035.420p275634.06 &  G3 & S0 &  (0.06, 28.10, 71.80, 0.00)  & 0.73 &  1.05 &  2.51 & 0.76 &  ... & ... \\
COMAi125950.183p275445.52 &  G4 & S0 &  (0.28, 1.85, 86.80, 11.10)  & 0.19 &  0.76 &  1.68 & 0.91 &  0.88 & 0.42 \\
COMAi125959.476p275626.02 &  G3 & S0 &  (0.38, 42.00, 57.60, 0.00)  & 1.07 &  1.67 &  2.67 & 0.72 &  ... & ... \\
COMAi13000.949p275643.85 &  G3 & S0 &  (0.35, 27.70, 72.00, 0.00)  & 0.47 &  1.56 &  2.43 & 1.15 &  ... & ... \\
COMAi13034.430p275604.95 &  G3 & S0 &  (0.00, 40.80, 59.20, 0.00)  & 1.15 &  1.80 &  2.83 & 1.00 &  ... & ... \\
COMAi125931.893p275140.76 &  G5 & E &  (0.19, 89.40, 10.40, 0.00)  & 1.04 &  2.09 &  0.62 & 0.65 &  ... & ... \\
COMAi125931.103p275718.12 &  G1 & S0 &  (0.11, 0.00, 100.00, 0.00)  & ... & ... &  2.38 & 1.52 &  ... & ... \\
COMAi13041.193p28242.34 &  G3 & spiral &  (0.29, 30.60, 69.10, 0.00)  & 0.41 & 1.08 &  1.11 & 0.84 &  ... & ... \\
COMAi125845.533p274513.75 &  G5 & E &  (0.08, 100.00, 0.00, 0.00)  & 2.98 &  2.09 &  1.52 & 1.71 &  ... & ... \\
COMAi13018.545p28549.62 &  G3 & S0 &  (0.14, 34.90, 64.90, 0.00)  & 0.76 &  0.97 &  2.83 & 0.86 &  ... & ... \\
COMAi13021.673p275354.81 &  G3 & S0 &  (0.74, 30.40, 68.90, 0.00)  & 0.40 &  1.14 &  1.71 & 0.56 &  ... & ... \\
COMAi13024.823p275535.94 &  G3 & spiral &  (0.23, 14.20, 85.50, 0.00)  & 0.58 &  1.36 &  3.06 & 1.20 &  ... & ... \\
COMAi13051.149p28249.90 &  G3 & S0 &  (0.05, 16.30, 83.60, 0.00)  & 0.96 &  1.57 &  5.98 & 1.35 &  ... & ... \\
COMAi13011.143p28354.91 &  G3 & S0 &  (0.13, 19.20, 80.70, 0.00)  & 0.66 &  2.64 &  1.48 & 1.40 &  ... & ... \\
COMAi125937.990p28003.52 &  G4 & spiral &  (0.76, 22.60, 67.50, 9.09)  & 0.37 &  2.38 &  1.40 & 0.50 &  0.67 & 0.33 \\
COMAi13018.873p28033.38 &  G4 & S0 &  (0.00, 54.20, 34.70, 11.10)  & 0.18 &  2.95 &  0.86 & 1.04 &  0.51 & 0.56 \\
COMAi125911.543p28033.32 &  G3 & S0 &  (0.19, 23.90, 75.90, 0.00)  & 0.54 &  0.95 &  1.55 & 1.21 &  ... & ... \\
COMAi125904.797p28301.16 &  G4 & S0 &  (0.09, 4.23, 76.80, 18.90)  & 0.21 &  1.22 &  2.72 & 1.12 &  0.83 & 0.85 \\
COMAi125909.468p28227.35 &  G2 & E &  (0.00, 100.00, 0.00, 0.00)  & 1.88 &  2.54 &  ... & ... &  ... & ... \\
COMAi125935.286p275149.13 &  G5 & E &  (0.05, 85.80, 14.10, 0.00)  & 1.37 &  2.08 &  0.70 & 0.31 &  ... & ... \\
COMAi13005.405p28128.14 &  G4 & S0 &  (0.00, 14.20, 72.20, 13.60)  & 0.35 &  1.18 &  3.60 & 0.92 &  1.29 & 0.45 \\
COMAi125950.105p275529.44 &  G4 & S0 &  (0.00, 26.70, 66.00, 7.31)  & 0.80 &  2.37 &  1.99 & 1.66 &  0.51 & 0.28 \\
COMAi13018.772p275613.34 &  G4 & S0 &  (0.00, 9.13, 69.10, 21.80)  & 0.49 &  0.98 &  2.87 & 0.69 &  1.64 & 0.60 \\
COMAi125938.321p275913.89 &  G3 & spiral &  (0.08, 25.30, 74.60, 0.00)  & 0.71 &  2.06 &  3.27 & 0.89 &  ... & ... \\
COMAi125940.270p275805.71 &  G4 & S0 &  (0.00, 33.40, 65.70, 0.94)  & 0.31 &  3.39 &  1.82 & 0.86 &  0.48 & 0.14 \\
COMAi125944.208p275730.38 &  G5 & E &  (0.00, 85.30, 14.70, 0.00)  & 5.43 &  5.82 &  1.95 & 0.56 &  ... & ... \\
COMAi125939.659p275714.03 &  G3 & S0 &  (0.64, 33.70, 65.70, 0.00)  & 0.32 &  1.91 &  1.97 & 1.08 &  ... & ... \\
COMAi13044.632p28602.31 &  G3 & S0 &  (0.00, 26.80, 73.20, 0.00)  & 0.35 &  1.51 &  2.36 & 0.80 &  ... & ... \\
COMAi125928.721p28225.92 &  G4 & S0 &  (0.00, 25.30, 33.70, 41.00)  & 0.40 &  1.65 &  3.28 & 0.57 &  1.54 & 1.05 \\
COMAi125942.301p275529.15 &  G3 & S0 &  (0.00, 25.20, 74.80, 0.00)  & 0.08 &  1.53 &  0.98 & 1.48 &  ... & ... \\
COMAi13017.014p28350.07 &  G4 & S0 &  (0.00, 47.20, 31.60, 21.20)  & 0.70 &  4.67 &  3.57 & 0.58 &  0.80 & 0.64 \\
COMAi125956.697p275548.71 &  G4 & S0 &  (0.18, 49.60, 26.20, 24.00)  & 1.89 &  4.33 &  3.57 & 0.25 &  2.48 & 0.41 \\
COMAi13016.534p275803.15 &  G3 & S0 &  (0.00, 84.10, 15.90, 0.00)  & 3.26 &  6.16 &  3.59 & 0.48 &  ... & ... \\
COMAi13006.395p28015.94 &  G3 & S0 &  (0.00, 66.10, 33.90, 0.00)  & 1.60 &  6.78 &  2.08 & 0.84 &  ... & ... \\
COMAi13027.966p275721.56 &  G4 & S0 &  (0.00, 42.80, 41.30, 15.90)  & 0.42 &  2.67 &  3.32 & 0.32 &  1.01 & 0.98 \\
COMAi13012.868p28431.74 &  G4 & S0 &  (0.00, 67.10, 19.10, 13.80)  & 0.77 &  2.42 &  5.07 & 0.41 &  3.00 & 0.53 \\
COMAi125943.721p275940.82 &  G3 & S0 &  (0.00, 69.20, 30.80, 0.00)  & 0.76 &  3.20 &  1.83 & 0.72 &  ... & ... \\
COMAi13028.370p275820.64 &  G4 & S0 &  (0.05, 33.80, 31.40, 34.80)  & 0.85 &  2.53 &  5.02 & 0.38 &  3.61 & 0.59 \\
COMAi13042.832p275746.95 &  G4 & S0 &  (0.00, 43.90, 46.40, 9.68)  & 0.75 &  3.01 &  3.81 & 0.47 &  1.51 & 0.39 \\
COMAi13038.761p28052.34 &  G4 & S0 &  (0.13, 16.80, 71.90, 11.10)  & 0.46 &  1.68 &  3.81 & 0.85 &  2.86 & 0.61 \\
COMAi13014.746p28228.69 &  G3 & S0 &  (0.43, 78.20, 21.40, 0.00)  & 0.90 &  3.68 &  2.26 & 0.47 &  ... & ... \\
COMAi13022.170p28249.30 &  G4 & S0 &  (0.00, 15.20, 73.40, 11.50)  & 0.31 &  1.35 &  3.49 & 1.24 &  1.31 & 0.48 \\
COMAi125931.453p28247.60 &  G3 & S0 &  (0.00, 78.60, 21.40, 0.00)  & 1.61 &  3.51 &  2.36 & 0.86 &  ... & ... \\
COMAi13018.093p275723.59 &  G2 & E &  (0.17, 100.00, 0.00, 0.00)  & 1.27 &  2.37 &  ... & ... &  ... & ... \\
COMAi13040.838p275947.80 &  G5 & E &  (0.13, 99.91, 0.00, 0.00)  & 3.16 &  2.34 &  0.34 & 1.83 &  ... & ... \\
COMAi125852.097p274706.15 &  G5 & E &  (0.00, 50.00, 50.00, 0.00)  & 2.12 &  6.95 &  1.74 & 1.41 &  ... & ... \\
COMAi125946.782p275825.99 &  G4 & S0 &  (0.00, 15.70, 67.70, 16.60)  & 0.31 &  1.75 &  3.44 & 0.67 &  1.23 & 0.72 \\
COMAi13008.003p28442.81 &  G3 & S0 &  (0.00, 85.00, 15.00, 0.00)  & 0.99 &  3.00 &  1.66 & 0.57 &  ... & ... \\
COMAi125929.956p275723.26 &  G4 & S0 &  (0.37, 24.20, 25.50, 49.90)  & 0.45 &  1.75 &  5.59 & 0.33 &  3.15 & 1.02 \\
COMAi125929.403p275100.46 &  G5 & E &  (0.00, 100.00, 0.00, 0.00)  & 2.37 &  1.86 &  0.28 & 2.19 &  ... & ... \\
COMAi125932.771p275901.04 &  G4 & S0 &  (0.02, 49.00, 46.70, 4.25)  & 2.21 &  6.05 &  3.01 & 0.83 &  0.50 & 0.54 \\
COMAi125944.407p275444.84 &  G4 & S0 &  (0.00, 34.30, 62.10, 3.65)  & 0.75 &  2.59 &  2.92 & 1.09 &  1.38 & 0.23 \\
COMAi125930.824p275303.05 &  G4 & spiral &  (0.36, 41.90, 41.20, 16.50)  & 0.54 &  1.89 &  6.41 & 0.66 &  1.79 & 0.46 \\
COMAi13039.767p275526.19 &  G4 & S0 &  (0.01, 46.30, 53.40, 0.34)  & 1.28 &  3.05 &  3.59 & 1.42 &  0.89 & 0.28 \\
COMAi13042.766p275817.38 &  G4 & S0 &  (0.00, 66.90, 31.30, 1.80)  & 1.27 &  2.99 &  6.23 & 0.35 &  1.42 & 0.17 \\
COMAi13048.646p28526.69 &  G5 & E &  (0.11, 70.20, 29.70, 0.00)  & 3.12 &  5.17 &  1.87 & 1.10 &  ... & ... \\
COMAi13017.683p275718.93 &  G2 & E &  (0.21, 100.00, 0.00, 0.00)  & 2.13 &  2.80 &  ... & ... &  ... & ... \\
COMAi13051.464p28234.86 &  G5 & E &  (0.00, 100.00, 0.00, 0.00)  & 3.90 &  2.01 &  0.61 & 2.21 &  ... & ... \\
NGC~4889 &  G2 & cD &  (0.00, 100.00, 0.00, 0.00)  & 57.7 &  7.8 &  ... & ... &  ... & ... \\
COMAi125935.698p275733.36 (NGC~4874) &  G2 & cD &  (0.00, 100.00, 0.00, 0.00)  & 391.3 &  11.4 &  ... & ... &  ... & ... \\
\hline
\multicolumn{10}{p{1\textwidth}}{Notes. Rows are sorted by increasing $M_\star$. 
In Columns 4-8, the meaning of C1 and C2 depends on Hubble type.
For cD and elliptical (E) 
galaxies, C1 is the outermost structure. For E galaxies, C2 represents the inner
component of any $n$. For S0 and spiral galaxies, C1 is the bulge and C2 is the outer
disk. The bar component represents bars/ovals in S0 and spiral galaxies.
For cored galaxies (NGC~4874, NGC~4889, COMAi125909.468p28227.35), the reported model 
corresponds to the 2D fit where the cored region
of the galaxy has been masked (see Table~\ref{tabcore} and Appendix~\ref{core}). 
}
\end{tabular}
\end{table}


\begin{table}
\caption{Morphology-Density Relation}
\label{tabmdr}
\begin{tabular}{@{}ccccc}
\hline
Region & Mass or Mag Cut& Galaxy Type$^{1,2}$ & \% by Numbers & \% by Stellar Mass \\
 (1)    &  (2)           &  (3)        &  (4)          & (5)        \\
\hline
Central 0.5~Mpc     & $M_\star \geq 10^9$ $M_\odot$ & (E+S0):spiral & (25.3\%+65.7\%):9.0\% &  (32.0\%+62.2\%):5.8\%  \\
of Coma (this work) &   &   &   &    \\
\hline
                    &   &   &   &    \\
                    &   &   &   &    \\
Virgo (McDonald et al. 2009) & $M_\star \geq 10^9$ $M_\odot$ & (E+S0):spiral & (33.8\%+31.3\%):35.0\% & (57.2\%+20.3\%):22.5\% \\
                    &   &   &   &    \\
                    &   &   &   &    \\
Virgo (McDonald et al. 2009) & $M_B \leq -19$ & (E+S0):spiral & (28.2\%+36.9\%):35.0\% & (57.3\% + 20.2\%):22.5\% \\
\hline
                    &   &   &   &    \\
Field (Dressler 1980) & Bright galaxies & (E+S0):spiral & ($\sim20\%$):$\sim80\%$ &  - \\
\hline
\multicolumn{5}{p{.9\textwidth}}{Notes.
$^1$Coma has two cD galaxies in the central 0.5~Mpc.

$^2$M87 in Virgo is considered an elliptical galaxy by McDonald et al. (2009). 
The detection of intra-cluster light around M87 (Mihos et al. 2005, 2009) is 
definitive proof that it is a cD galaxy (see also the discussion in 
Kormendy et al. 2009). Here, we consider M87 a cD galaxy and do not include it 
in the above statistics for ellipticals.
}
\end{tabular}
\end{table}

\begin{table}
\caption{Total Galactic Stellar Mass In Disk-Dominated Structures
  Versus Classical Bulges/Ellipticals}
\label{masstable2} 
\begin{tabular}{@{}lc}
\hline
Structure & \% of Stellar Mass in the \\
          & Projected Central 0.5~Mpc of Coma \\
(1) &  (2)   \\
\hline
\multicolumn{2}{c}{\vspace*{2 mm}\textbf{Disk-dominated components with $n\leq n_{\rm disk\_max}$}}  \\
Outer disks of S0s & 27.7 \\
Outer disks of Spirals & 2.94 \\
Bulges with $n\leq n_{\rm disk\_max}$ (pseudo or disky bulge) in S0s & 2.07 \\
Bulges with $n\leq n_{\rm disk\_max}$ (pseudo or disky bulge) in spirals& 0.13 \\
Inner component with $n\leq n_{\rm disk\_max}$ (inner disks) in photometric E & 3.26 \\

\textbf{Total} & \textbf{36.1} \\

\hline
\multicolumn{2}{c}{\vspace*{2 mm}\textbf{Bars}}  \\
Bars in S0s & 6.11 \\
Bars in spirals & 0.73 \\
\textbf{Total} & \textbf{6.84} \\

\hline
\multicolumn{2}{c}{\vspace*{2 mm}\textbf{Non-disk ``hot'' components with $n> n_{\rm disk\_max}$}}  \\
Outer component with $n> n_{\rm disk\_max}$ in photometric E & 26.0 \\
Inner component with $n> n_{\rm disk\_max}$ in photometric E & 2.74 \\
Inner component with $n> n_{\rm disk\_max}$ in S0s & 26.2 \\
Inner component with $n> n_{\rm disk\_max}$ in Spirals & 2.06 \\
\textbf{Total} & \textbf{57.0} \\

\hline
\multicolumn{2}{p{.9\textwidth}}{Notes.
These numbers apply to the galaxies in the projected central 0.5~Mpc
of Coma, after excluding the 2 cDs. We exclude the 2 cDs due to their
uncertain stellar mass and the reasons  outlined at the end of
Section~\ref{datas2}.}
\end{tabular}
\end{table}

\begin{table}
\caption{Fraction of Stellar Mass In Disk-Dominated Structures Versus
  Classical Bulges/Ellipticals In Different Galaxies} 
\label{masstable} 
\begin{tabular}{@{}lcc}
\hline
Structure & \% of Stellar Mass Within & \% of Stellar Mass in the\\
          &  Each Galaxy Type &  Projected Central 0.5~Mpc of Coma\\
(1)       &     (2)             &  (3)   \\
\hline
\multicolumn{3}{c}{\vspace*{2 mm}\textbf{Photometric E (N=17)}}\\
Outer component with $n> n_{\rm disk\_max}$    &81.2& 26.0\\
Inner component with $n> n_{\rm disk\_max}$    &8.6& 2.74 \\
Inner component with $n\leq n_{\rm disk\_max}$ &10.2&  3.26\\
Point sources                                  &0.09& 0.03 \\
\textbf{Total}                           & \textbf{100}& \textbf{32.0} \\
\hline
\multicolumn{3}{c}{\vspace*{2 mm}\textbf{S0 (N=44)}}\\
Outer disk with $n\leq n_{\rm disk\_max}$                             &44.4& 27.7 \\
Bars                                                                  &9.8& 6.11 \\
Bulges with $n> n_{\rm disk\_max}$      (classical bulge)             &42.2& 26.2 \\
Bulge components with $n\leq n_{\rm disk\_max}$ (disky pseudobulge)  &3.3 & 2.07\\
Point sources                                                         &0.06 & 0.04 \\
\textbf{Total}                           & \textbf{100}& \textbf{62.1} \\
\hline
\multicolumn{3}{c}{\vspace*{2 mm}\textbf{Spiral (N=6)}}\\
Outer disk with $n\leq n_{\rm disk\_max}$                             &50.0& 2.94\\
Bars                                                                  &12.4& 0.73 \\
Bulges with $n> n_{\rm disk\_max}$       (classical bulge)            &35.0& 2.06\\
Bulge components with $n\leq n_{\rm disk\_max}$ (disky pseudobulge)  &2.2 & 0.13 \\
Point sources                                                         &0.3 & 0.02 \\
\textbf{Total}                           & \textbf{100}& \textbf{5.90} \\
\hline

\multicolumn{3}{p{.9\textwidth}}{Notes.
The totals listed in Column 3 correspond to Column 5 of Table~\ref{tabmdr}.
These numbers apply to the galaxies in the projected central 0.5~Mpc
of Coma, after excluding the 2 cDs. We exclude the 2 cDs due to their
uncertain stellar mass and there reasons  outlined at the end of
Section~\ref{datas2}.
}
\end{tabular}
\end{table}

\begin{table}
\caption{Bulge S\'ersic Index In S0s Across Different Environments}
\label{tabS0}
\begin{tabular}{@{}ccccc}
\hline
Bulge S\'ersic Index of S0s &  Environment & Stellar Mass Cut & \% of S0s   & \% of Bulge Stellar Mass in S0s\\
 (1)                        &  (2)         & (3)              & (4)         & (5) \\
\hline
\multicolumn{5}{c}{\vspace*{2 mm}\textbf{This work}}\\
$n \leq n_{\rm disk\_max}$ & Projected Central 0.5~Mpc of Coma, high density &  $M_\star \geq 10^9$ $M_\odot$ & $38.6 \pm 7.3$  &  3.06 \\
$n > n_{\rm disk\_max}$    & Projected Central 0.5~Mpc of Coma, high density &  $M_\star \geq 10^9$ $M_\odot$ & $59.1 \pm 7.4$  &  42.2 \\
\hline
\multicolumn{5}{c}{\vspace*{2 mm}\textbf{This work}}\\
$n \leq n_{\rm disk\_max}$ & Projected Central 0.5~Mpc of Coma, high density& $M_\star \geq 7.5\times 10^9$ $M_\odot$ & $21.4 \pm 7.8$  &  2.4\\
$n > n_{\rm disk\_max}$    & Projected Central 0.5~Mpc of Coma, high density& $M_\star \geq 7.5\times 10^9$ $M_\odot$ & $78.6 \pm 7.8$  &  41.7\\
\hline
\multicolumn{5}{c}{\vspace*{2 mm}\textbf{Laurikainen et al. (2010)}}\\
$n \leq n_{\rm disk\_max}$ &  Lower density & $M_\star \geq 7.5\times 10^9$ $M_\odot$ &  $22.3 \pm 3.9$  &  4.7\\
$n > n_{\rm disk\_max}$    &  Lower density & $M_\star \geq 7.5\times 10^9$ $M_\odot$ &  $77.7 \pm 3.9$  &  30.6\\
\hline
\multicolumn{5}{p{.9\textwidth}}{Notes.
This table shows the fraction of S0 galaxies with bulge
S\'ersic index above and below the value $n_{\rm disk\_max}=1.66$
determined empirically in Appendix~\ref{2cp}.
The first four rows pertain to the Coma cluster and represent
different stellar mass cuts. The bottom two rows are for S0
galaxies studied by Laurikainen et al. (2010) from
much lower density environments than the rich Coma cluster.
Column 5 represents the percent of total bulge stellar mass to 
total galaxy stellar mass calculated over S0s satisfying each 
specific bulge index and stellar mass cut.
}
\end{tabular}
\end{table}


\clearpage
\begin{figure}
\centering
\scalebox{0.75}{\includegraphics{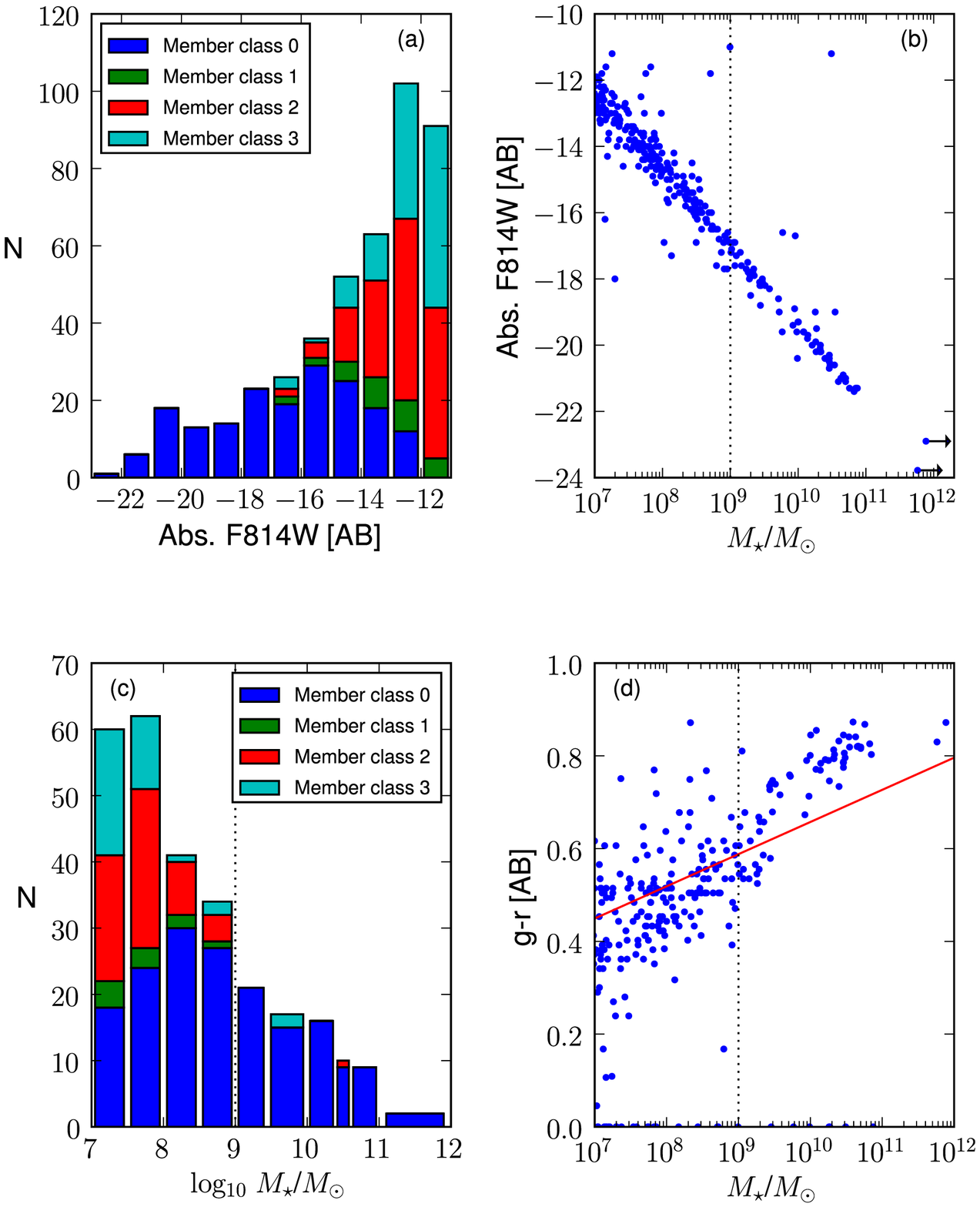}}
\caption{This figure shows in panels (a) to (d) the luminosity, stellar 
mass, and $g-r$ color, respectively, for the 446 galaxies in sample S1 having
F814W$\leq24$, locations within the projected central 0.5~Mpc of the Coma cluster, and 
cluster membership rating 0-3, where rating 0 means spectroscopically confirmed
and rating 1-3 indicate increasingly less likely cluster membership. See 
Section~\ref{datas1} for details. In panel 
(b), the two most massive sources are cD galaxies, and the arrows indicate their 
adopted stellar masses are lower limits (Section~\ref{datas2}). The solid line 
in panel (d) is the color-luminosity break between the red and blue sequence of 
galaxies from Blanton et al. (2005), which we convert from luminosity to stellar 
mass using the relations of Bell et al. (2003). The dotted line in panels (b)-(d)
indicates our main sample of 69 spectroscopically confirmed members with $M_\star \geq 10^9$~$M_\odot$. 
\label{lummasstype}}
\end{figure}

\clearpage
\begin{figure}
\centering
\scalebox{0.55}{\includegraphics{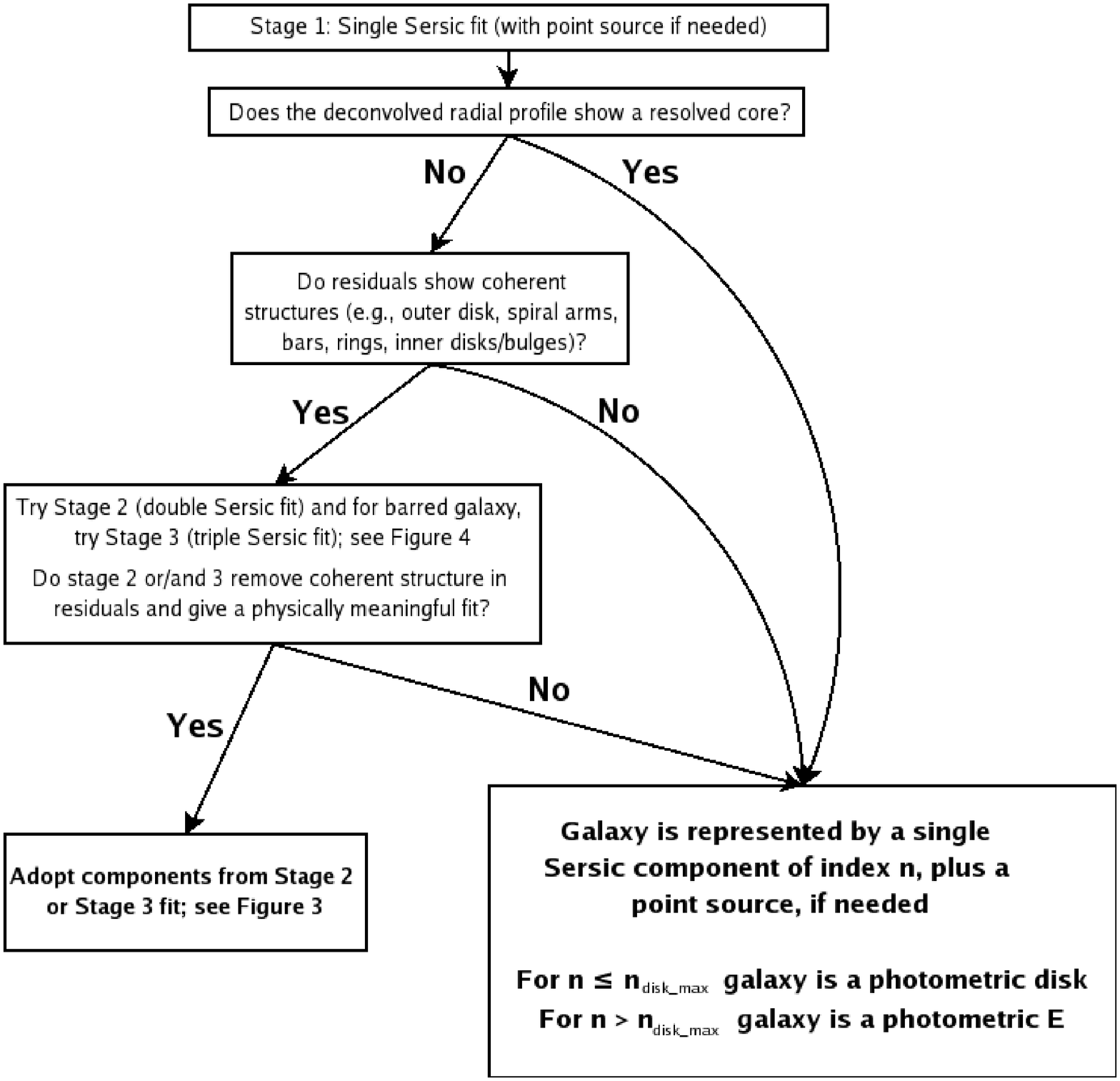}}
\caption{This figure provides an overview of our structural decomposition method.  All galaxies are subjected to
Stage 1, and most are further processed in Stage 2.  A galaxy best fit with a single 
S\'ersic profile plus point source (if needed) is interpreted as a photometric 
elliptical or 
photometric disk. A galaxy with extra 
coherent structure that cannot be described with a single S\'ersic profile is subjected to 
a multiple-component fit in Stage 2 and, if needed, Stage 3. Figure~\ref{fitflow2} describes 
Stage 2 and Stage 3 in more detail. 
\label{fitflow}}
\end{figure}

\clearpage
\begin{figure}
\centering
\scalebox{0.60}{\includegraphics{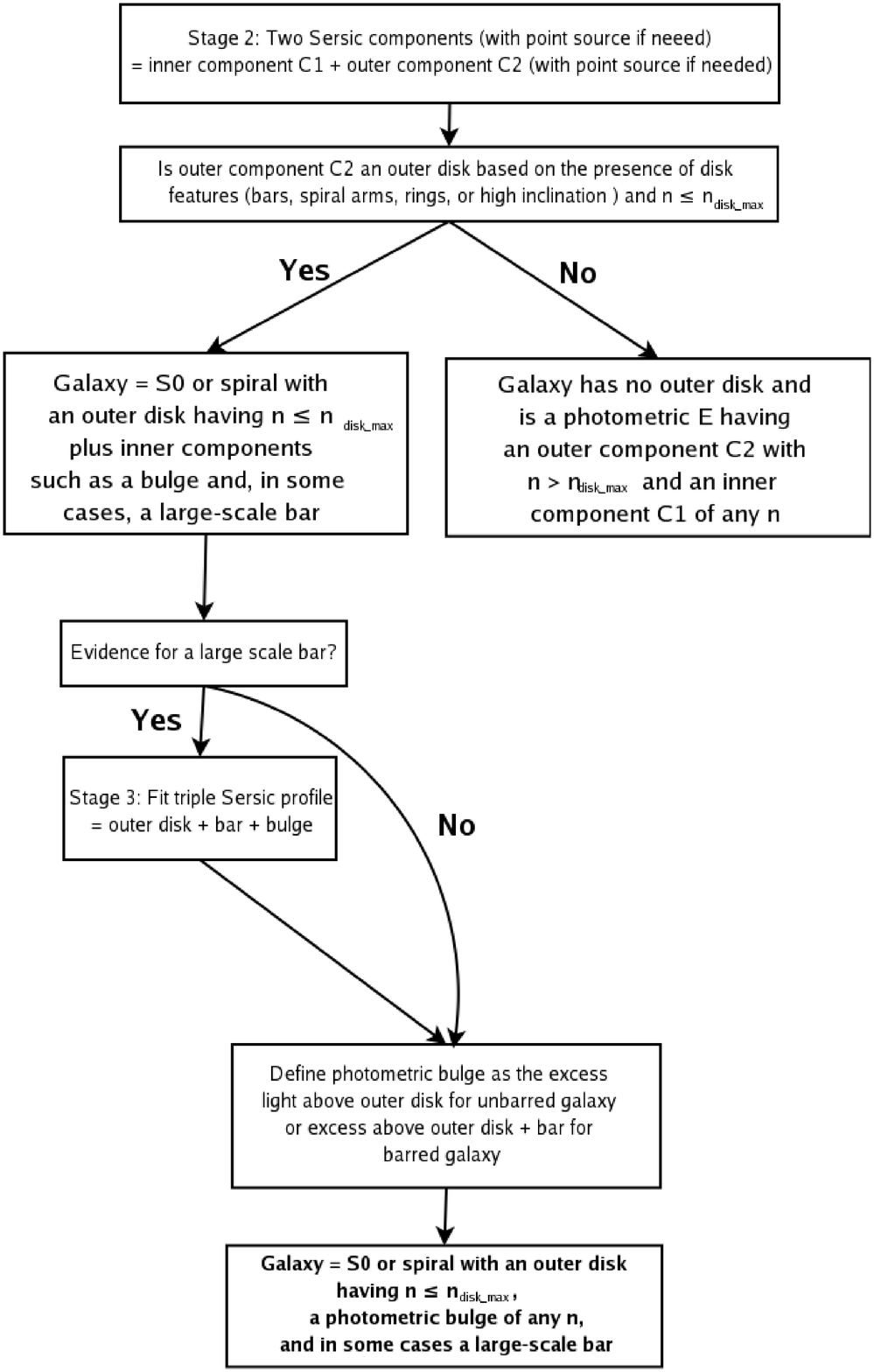}}
\caption{This figure shows the steps following stages 2 and 3 from Figure~\ref{fitflow}.
A galaxy without an extended outer disk is interpreted as a photometric E, while a galaxy 
with such a disk is labeled either an S0 or spiral. When 
evidence for a large scale bar is found in a galaxy with an outer disk, Stage 3 is used to
model the bar component. \label{fitflow2}}
\end{figure}

\clearpage
\begin{figure}
\centering
\scalebox{0.45}{\includegraphics{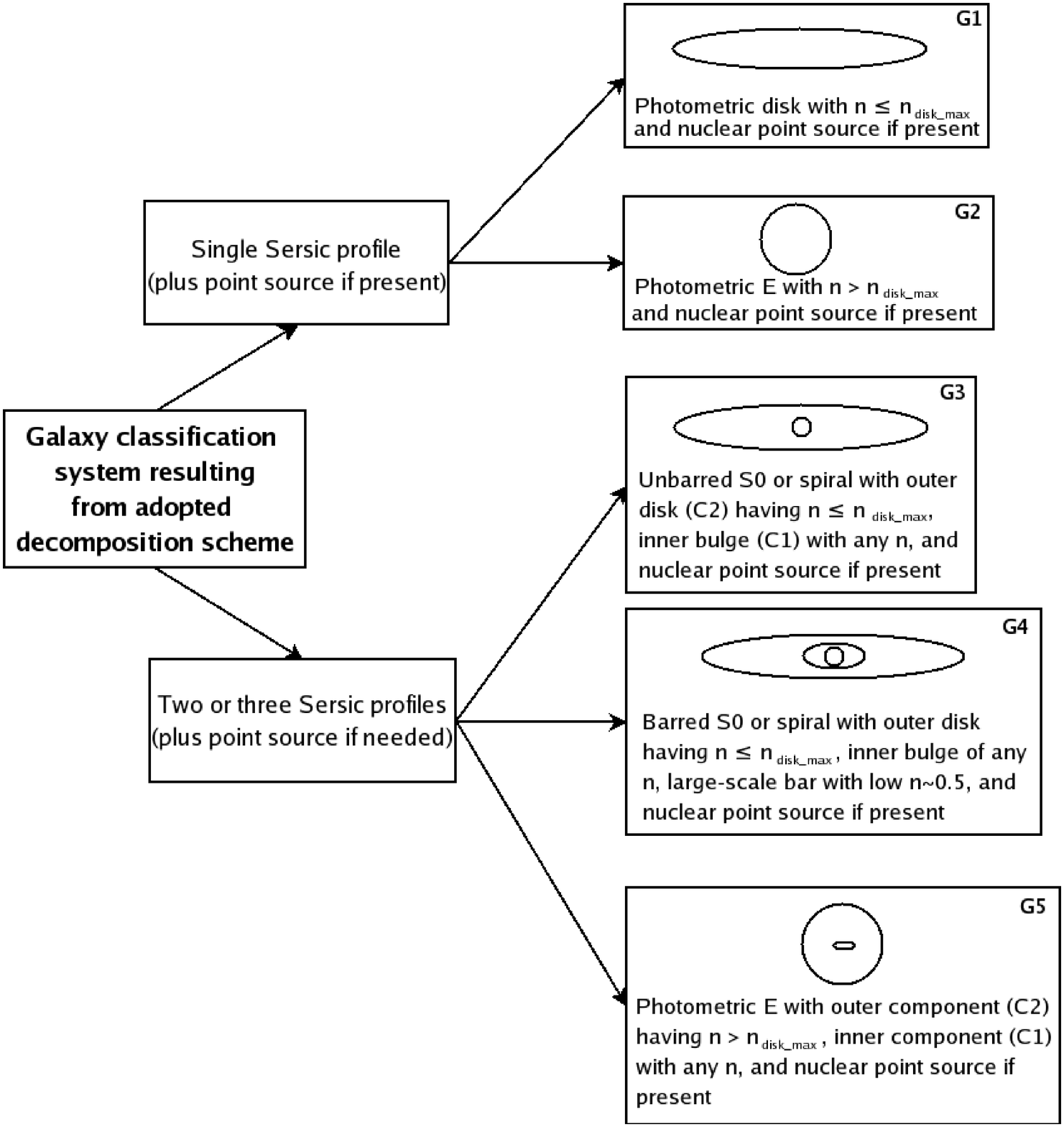}}
\caption{Overview of our galaxy classification system (Section~\ref{scheme}).  Galaxies are deemed to be
best represented by a either a single or multi-component S\'ersic profile 
(plus point source, if needed).  Galaxies fitted with a single S\'ersic profile
are further interpreted as a pure disk (if S\'ersic index $n\leq n_{\rm disk\_max}$) or 
photometric elliptical (if S\'ersic index $n > n_{\rm disk\_max}$). When a multi-component S\'ersic
profile is required, the galaxy is either an unbarred/barred S0 or spiral, or a photometric
E with inner and outer components. S0s and spirals must have an outer component C2 with
S\'ersic index $n\leq n_{\rm disk\_max}$. The inner component C1 can have any $n$. If the outer
component C2 has S\'ersic index $n > n_{\rm disk\_max}$, the galaxy is a photometric 
elliptical with inner
component C1 of any $n$. The value of $n_{\rm disk\_max}$ is set 
to 1.66 based on several considerations (See Appendices~\ref{2cp} and \ref{nuperr}).  We determine $n_{\rm disk\_max}$ to be the maximum 
S\'ersic index of the
outer disk in spiral and S0 galaxies showing clear signs of an outer disk, such as bars, spiral
arms, rings, or high inclination.
\label{galaxyclass}}
\end{figure}

\clearpage
\begin{figure}
\centering
\scalebox{0.75}{\includegraphics{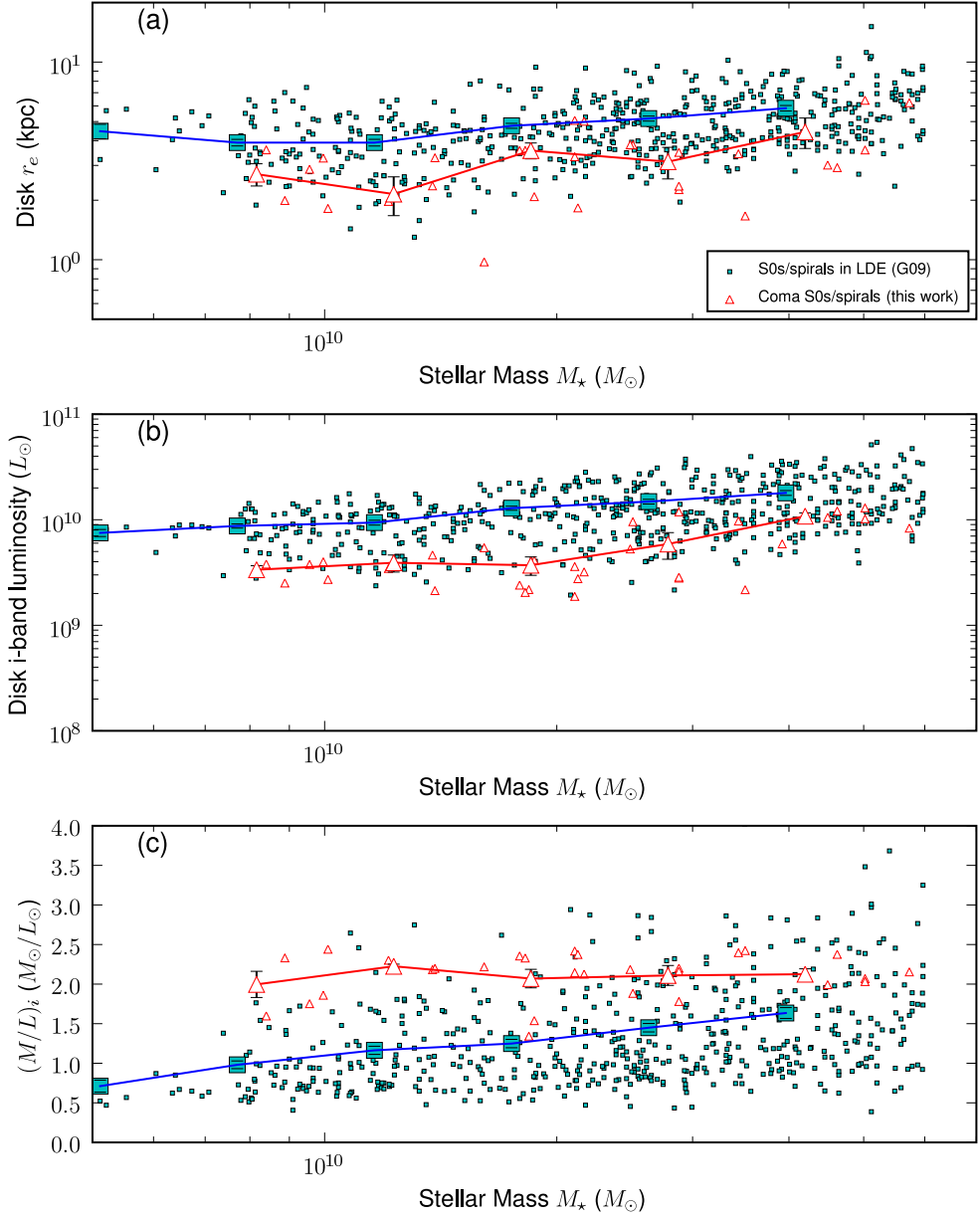}}
\caption{
Panels~(a) and (b) compare the properties of large-scale disks
($r_\mathrm{e}$, luminosity) with galaxy stellar mass $M_\star$. Massive 
($5\times 10^9 \leq M_\star \leq 6\times 10^{10}$~$M_\odot$)
S0/spiral galaxies from the projected central 0.5~Mpc of Coma as well as 
low-density environments (LDEs) are considered. The LDE
sample is from Gadotti (2009) and includes galaxies in SDSS 
Data Release 2 that are face-on ($b/a\geq 0.9$) and have redshift 
$0.02 \leq z \leq 0.07$. In panel~(b), the $i$-band luminosity represents
the ACS F814W photometry for Coma galaxies and the SDSS $i$-band photometry
for the LDE galaxies. Panel~(c) compares $i$-band mass-to-light ratio
$(M/L)_i$ with galaxy $M_\star$. For Coma, the galaxy wide mass-to-light
ratio is plotted, while for LDEs the {\it outer disk} $(M/L)_i$ is shown. 
In all panels, the mean values (large symbols) in galaxy $M_\star$ for Coma and LDEs are 
slightly offset along the x-axis, as shown by the red triangle and blue squares,
in order to avoid the error bars from overlapping. The mean values are calculated
in 0.18 dex bins.
The error bars on the mean values represent the standard error on the mean.
This figure demonstrates that at 
a given galaxy stellar mass, the average disk half-light radius $r_\mathrm{e}$ 
in the $i$-band is smaller in 
the projected central 0.5~Mpc of Coma compared to LDEs.
\label{scaling16}}
\end{figure}

\clearpage
\begin{figure}
\centering
\scalebox{0.80}{\includegraphics{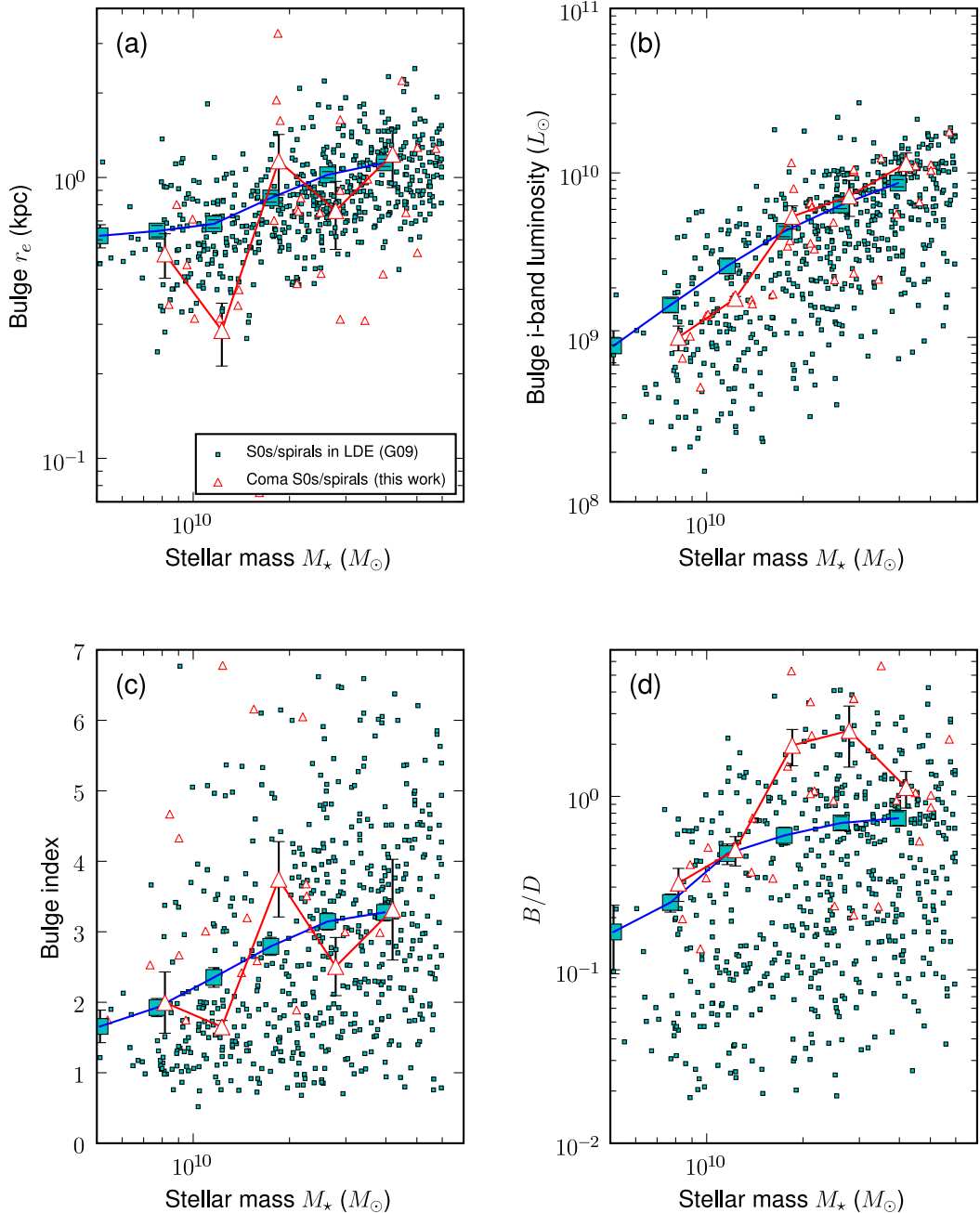}}
\caption{This figure is similar to Figure~\ref{scaling16}, except
that it emphasizes S0/spiral galaxy bulges. See Section~\ref{discussScaling} 
and Figure~\ref{scaling16} for extra details on the sample from Gadotti (2009).
Panels~(a), (b), (c), and~(d) show bulge size ($r_\mathrm{e}$), bulge $i$-band 
luminosity, bulge S\'ersic index, and bulge-to-disk {\it light ratio} ($B/D$), 
respectively, 
versus galaxy $M_\star$. 
In all panels, the error bars on the mean values (large symbols) represent 
the standard error on the mean, and in all panels the mean values in galaxy 
$M_\star$ for Coma and LDEs are slightly offset along the x-axis, as shown by the 
red triangle and blue squares, in order to avoid the error bars from overlapping. 
This figure demonstrates that at a given galaxy stellar mass, there appears to be no 
systematic offset between bulges in Coma and LDEs.
\label{scaling17}}
\end{figure}

\clearpage
\begin{figure}
\centering
\scalebox{0.90}{\includegraphics{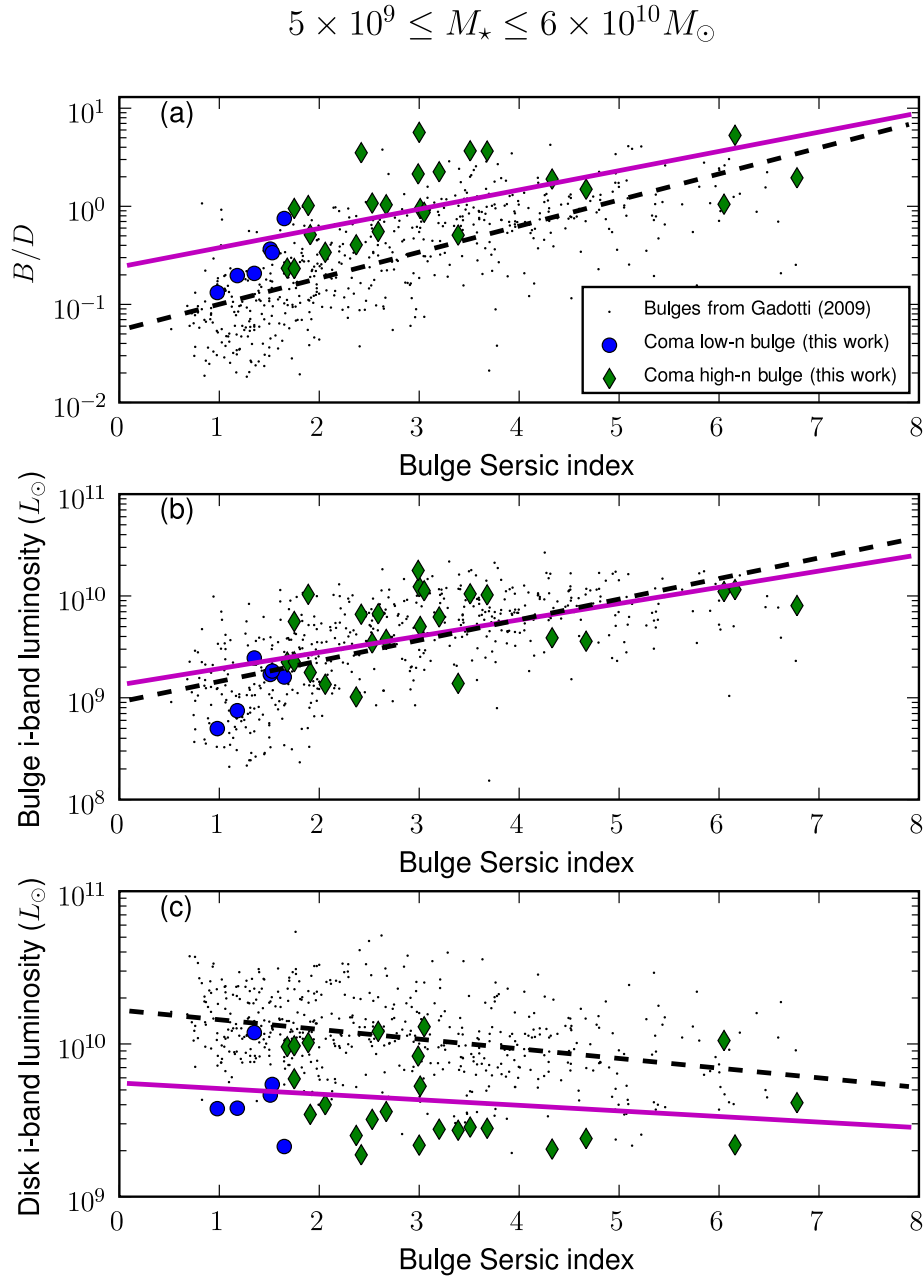}}
\caption{This figure shows a comparison of bulges in massive
($5\times 10^9 \leq M_\star \leq 6\times 10^{10}$~$M_\odot$) 
S0/spiral galaxies in low-density environments (LDEs) from Gadotti (2009) 
versus galaxies in the projected central 0.5~Mpc of Coma. See 
Section~\ref{discussScaling} and 
Figure~\ref{scaling16} for extra details on the sample from Gadotti (2009).
Bulges in Coma are divided into groups of low S\'ersic index 
($n\leq n_{\rm disk\_max}$) and high S\'ersic index ($n > n_{\rm disk\_max}$). 
Panel~(a) 
shows bulge-to-disk {\it light ratio} ($B/D$) versus bulge S\'ersic index.
Panels~(b) and~(c) show bulge and disk $i$-band luminosity, respectively, versus
bulge S\'ersic index. 
In each panel, the {\it solid line} represents
the fit to Coma bulges of all S\'ersic $n$, and the {\it dashed line} is the 
corresponding fit to bulges of all S\'ersic $n$ from LDEs.
The offset in $B/D$ in panel (a) appears to be driven, at least in part, by the offset in
disk luminosity in panel (c).
\label{scaling18}}
\end{figure}

\clearpage
\begin{figure}
\centering
\scalebox{1}{\includegraphics{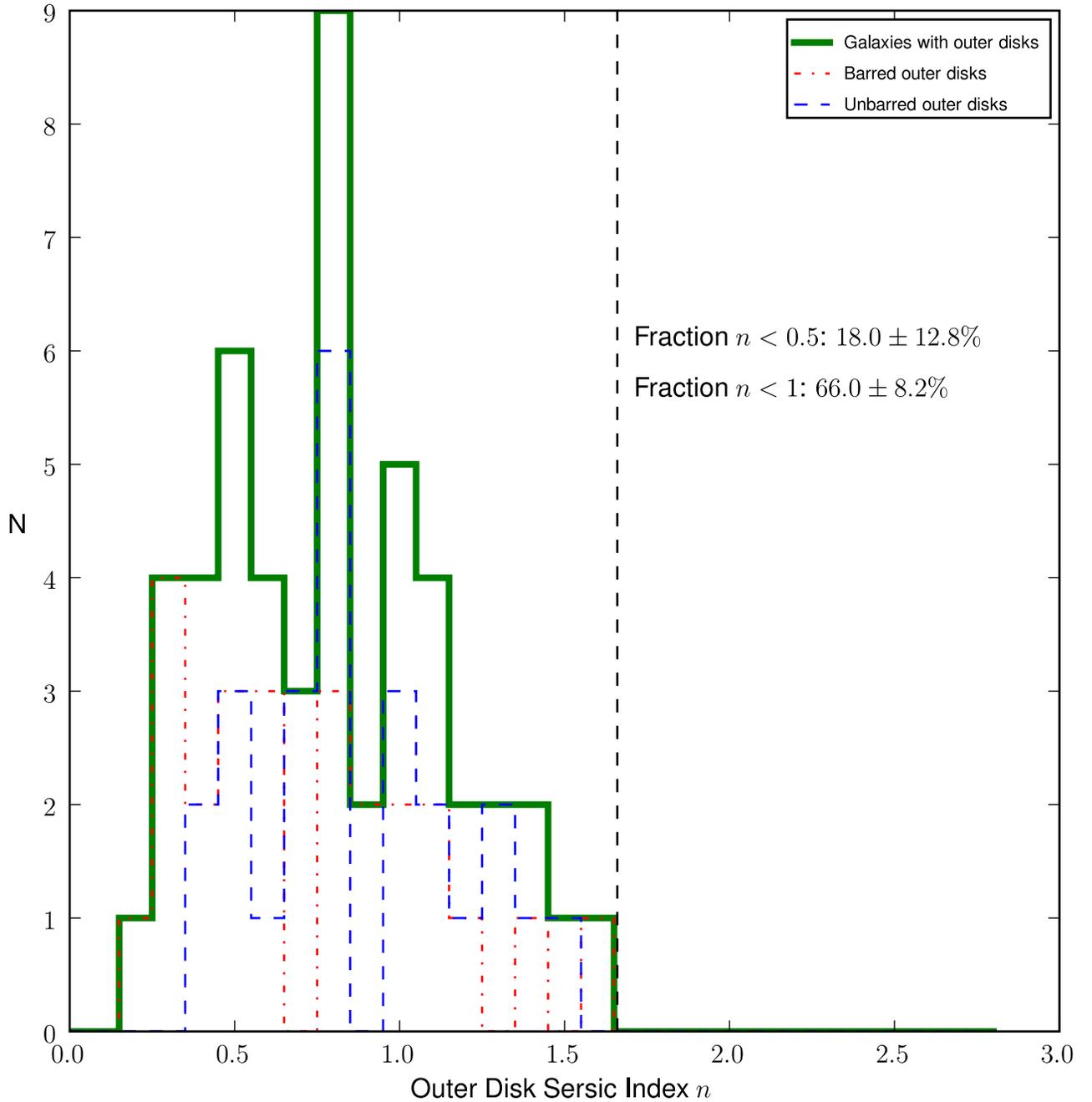}}
\caption{
The distribution of S\'ersic index for outer disks in S0/spiral galaxies. The vertical line 
represents the empirically determined upper limit, $n_{\rm disk\_max}$, in the S\'ersic index 
of outer barred disks, which can be considered unambiguous cases of outer disks due to the presence of a bar. The dash-dotted and dashed lines show the distributions for barred and unbarred
outer disks, respectively. See Section~\ref{scheme} and Appendix~\ref{2cp} for details.
\label{diskn}}
\end{figure}

\clearpage
\begin{figure}
\centering
\scalebox{0.7}{\includegraphics{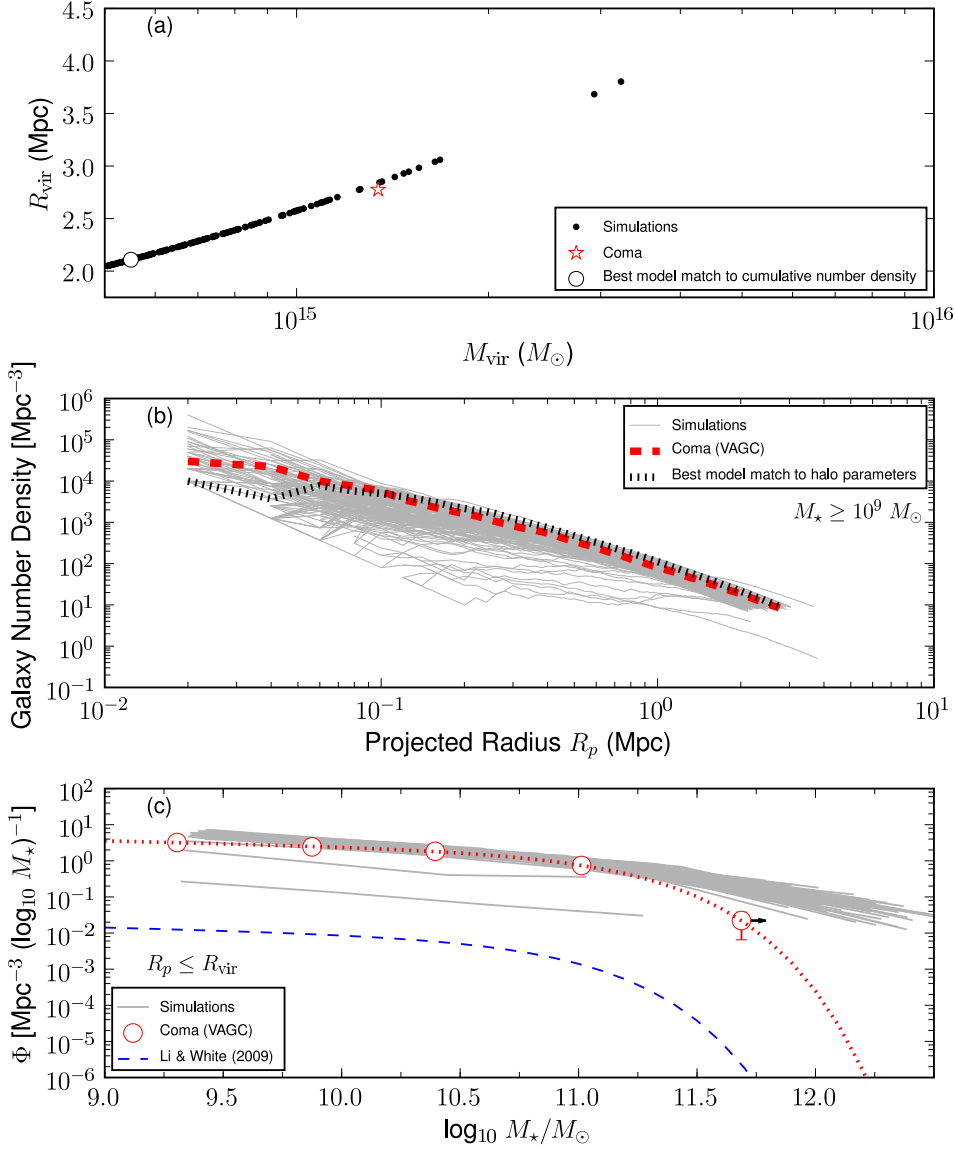}}
\caption{This figure shows how the global parameters of the Coma cluster compare 
with all 160 clusters in the Millennium simulation having a halo mass in the range 
$5\times10^{14} - 10^{16}$~$M_\odot$. The solid lines and black data points 
represent the simulated clusters. In panel (a), the virial mass and virial radius 
adopted for Coma are 2.8$h^{-1}_{73}$~Mpc and 
$1.3\times10^{15}h^{-1}_{73}$~$M_\odot$ (Section~\ref{comavagc}). The open circle 
is the model cluster having the best match
to the projected galaxy number density of Coma. In panels (b) and (c), the 
cumulative 
projected galaxy number density and the galaxy mass stellar function of Coma at 
projected radius $R_{\rm p}\leq R_{\rm vir}$ are based on data from the NYU 
Value-Added Galaxy Catalog (NYU-VAGC, Blanton et al. 2005).
In panel (b), the dotted line represents the cumulative galaxy number density
of the model cluster best matching the Coma halo parameters.
In panel (c), for the Coma galaxy stellar mass function, we measure a slope 
$\alpha = -1.16$ and characteristic mass $M^* = 1.17 \times 10^{11}~M_\odot$. 
The last mass bin in the global mass function for Coma contains the two
cD galaxies, and the arrow on this bin indicates the adopted stellar masses
for the cDs are lower limits.
The simulations are based on a model that produces a reasonable match
to the galaxy stellar mass function of Li \& White (2009) averaged over all 
environments at $0.001<z<0.5$ (dashed line). However,
they cannot  produce a model cluster that simultaneously
matches multiple global properties (halo properties,  galaxy number
density, and galaxy stellar mass function),  of Coma, our local benchmark
for one of the richest nearby galaxy clusters. 
\label{virp}}
\end{figure}

\clearpage
\begin{figure}
\centering
\scalebox{0.8}{\includegraphics{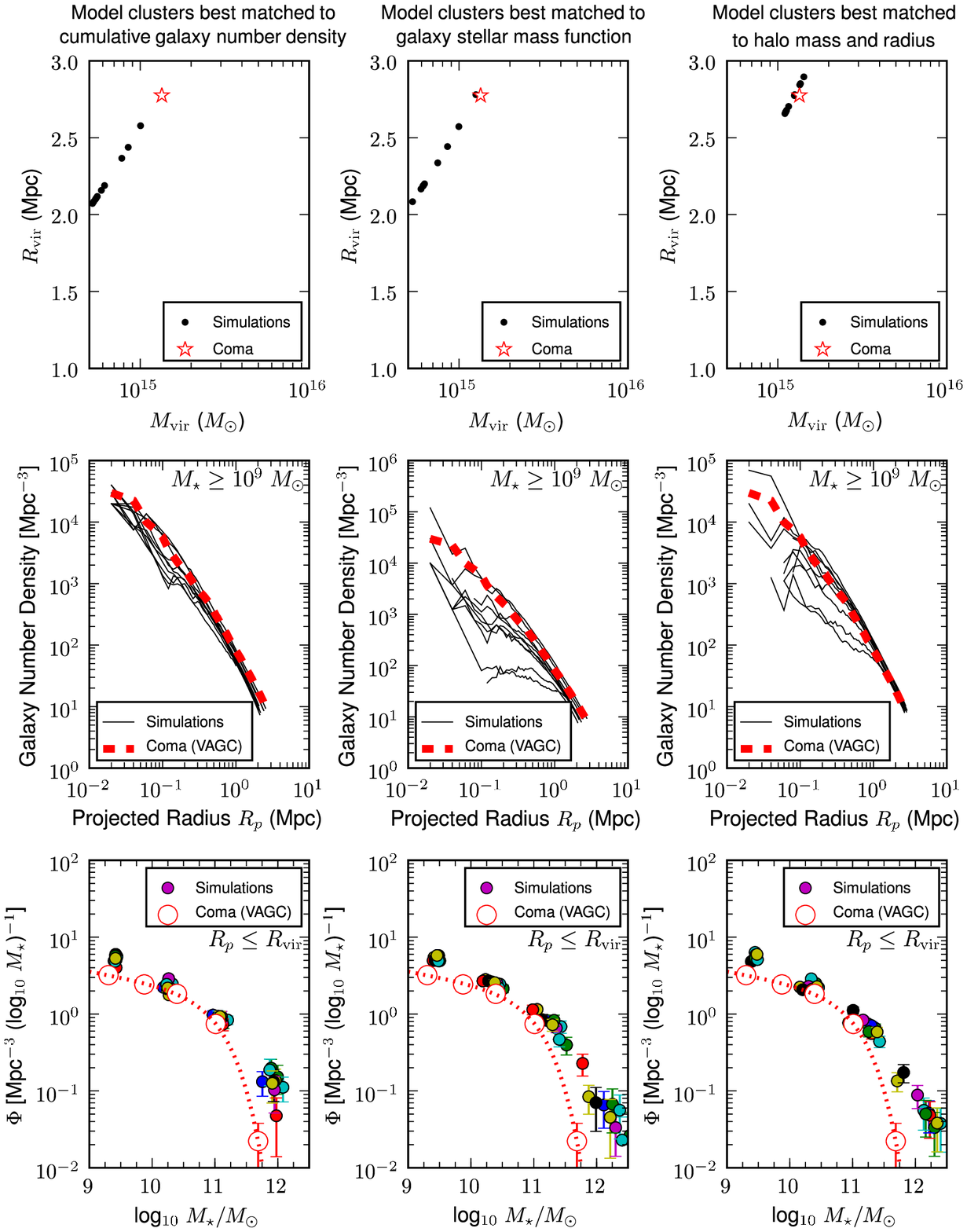}}
\caption{This figure shows the three sets of simulated model clusters 
(30 model clusters in total) chosen to best match, separately, the cumulative 
galaxy number density (Column 1), the galaxy stellar mass function (Column 2),
and halo parameters (halo mass and virial radius, Column 3). The solid 
lines and solid circles in each panel represent the different simulated clusters. 
Rows 1, 2, and 3 show how the different simulated clusters compare with the global 
properties (halo mass and radius, cumulative galaxy number density,
and galaxy stellar mass function) determined with data from the NYU Value-Added 
Galaxy Catalog (NYU-VAGC, Blanton et al. 2005) for Coma in 
Section~\ref{comavagc}. No model cluster simultaneously matches all three
global properties.
\label{bestclusters}}
\end{figure}

\clearpage
\begin{figure}
\centering
\scalebox{0.8}{\includegraphics{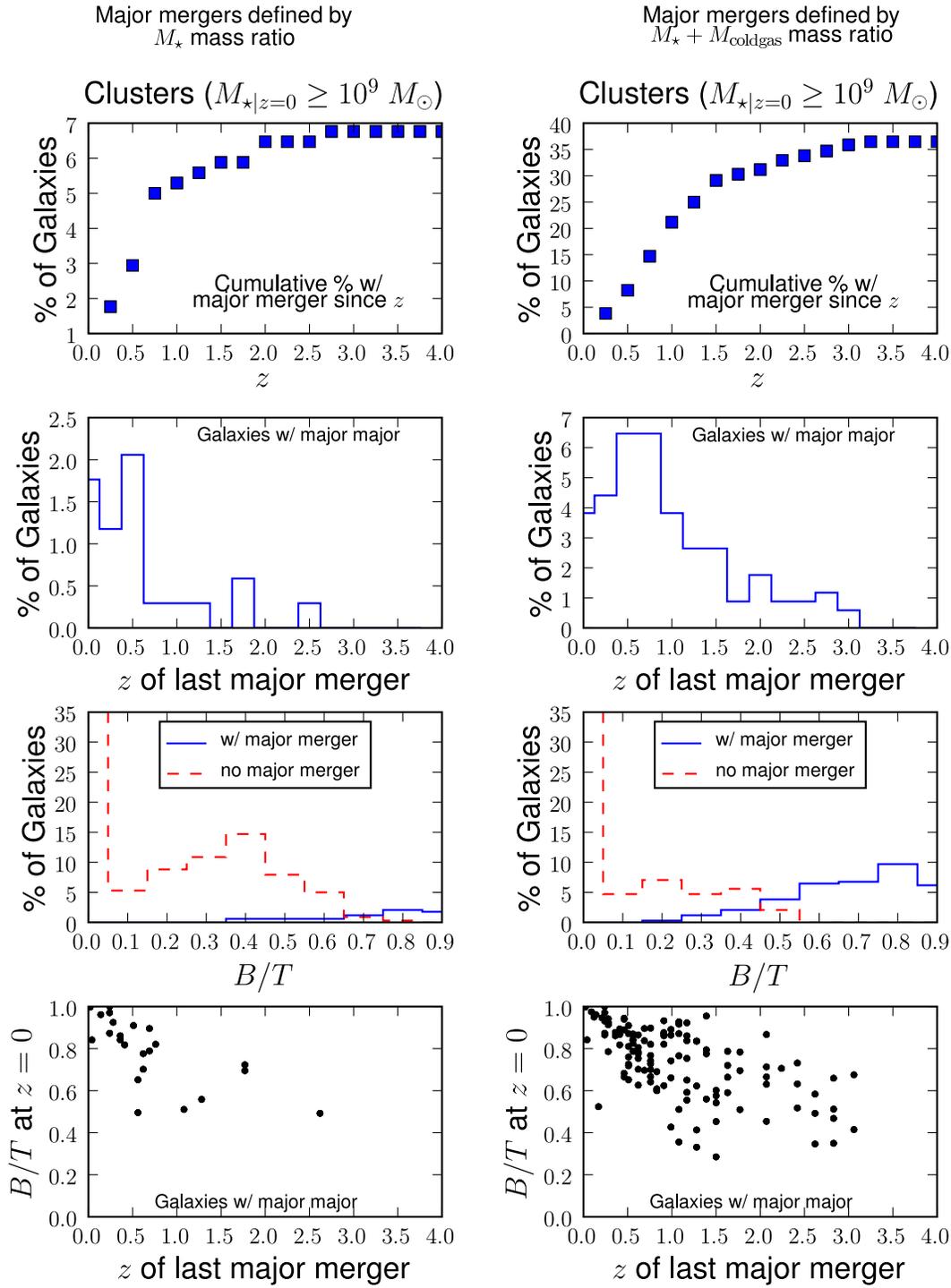}}
\caption{
This figure highlights the effect that the definition of the merger mass ratio 
$M_1/M_2$ has on certain galaxy properties (merger history and $B/T$), for a 
representative model cluster (see Section~\ref{massratio}). Note we require 
a major merger to have $M_1/M_2 \geq 1/4$. The left column is the manifestation 
of the model cluster when $M_1/M_2$ refers to the stellar mass ratio, and in 
the right column the merger mass ratio represents cold gas plus stars. The two
mass ratio definitions lead to vastly different merger histories and significantly
affect the resulting distribution of $B/T$.
\label{majordef}}
\end{figure}

\clearpage
\begin{figure}
\centering
\scalebox{0.8}{\includegraphics{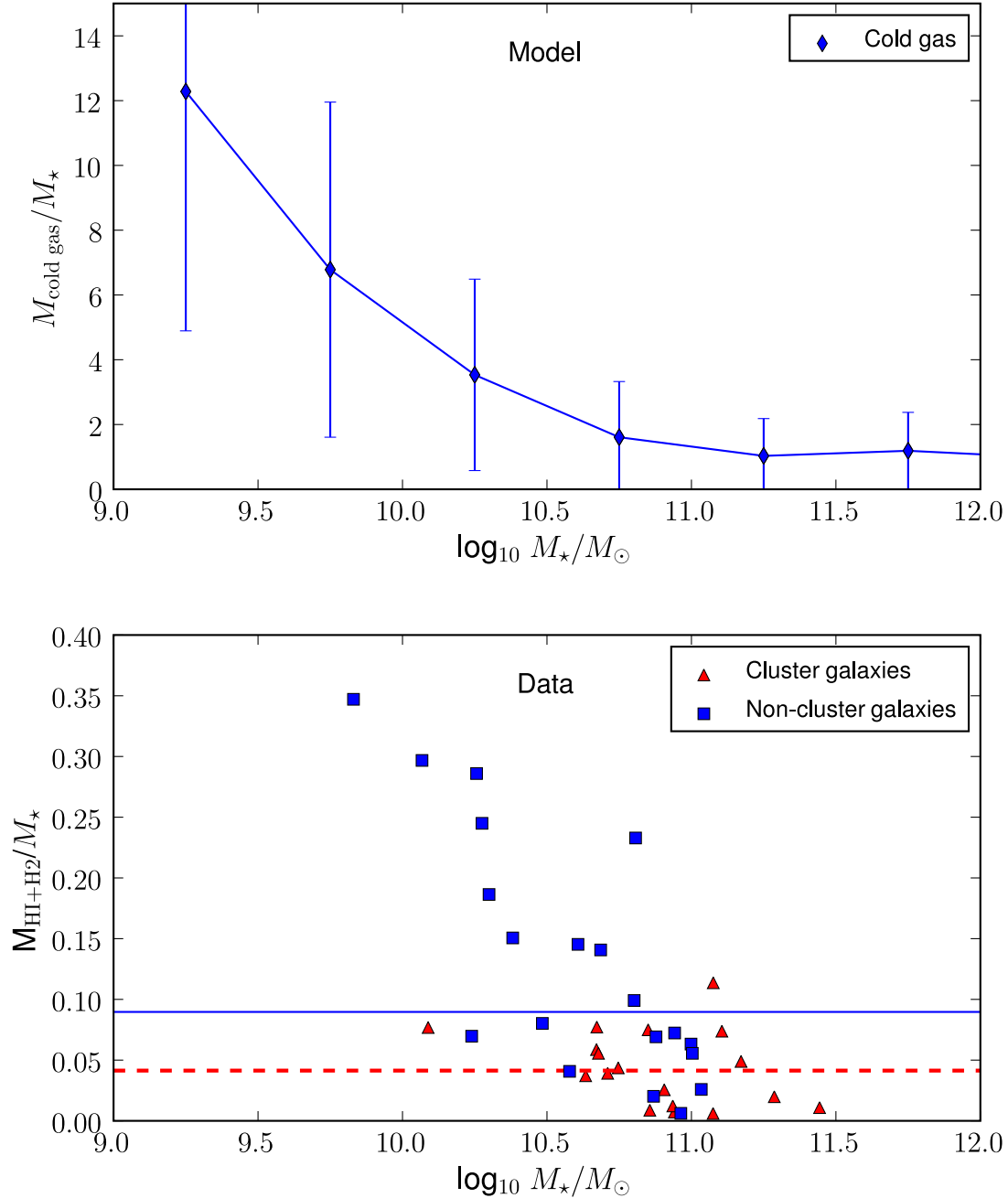}}
\caption{The top panel shows the ratio of cold gas to stellar mass
($M_{\rm cold\_gas}/M_\star$) for 
the best cluster model matched by cumulative galaxy
number density (see Figure~\ref{bestclusters}, column 1).
The error bars represent the $1\sigma$ standard deviation
around the mean value. The bottom panel shows the ratio of observed cold 
gas ($HI+H_2$) to stellar mass ($M_{HI+H_2 }/M_\star$) for galaxies 
studied by Boselli et al. (1997) that are part of 
or near the Coma cluster. The dashed line is the
median ratio (0.04) for Coma cluster galaxies, and the solid line is the median
ratio (0.09) for the non-cluster galaxies.
At $10^{10} \lesssim M_\star \lesssim 10^{11}$~$M_\odot$, the model predicts a cold 
gas to stellar mass ratio that is a factor $\sim25-87$ times higher than the median
value observed in Coma cluster galaxies.
\label{boselli}}
\end{figure}

\clearpage
\begin{figure}
\centering
\scalebox{0.70}{\includegraphics{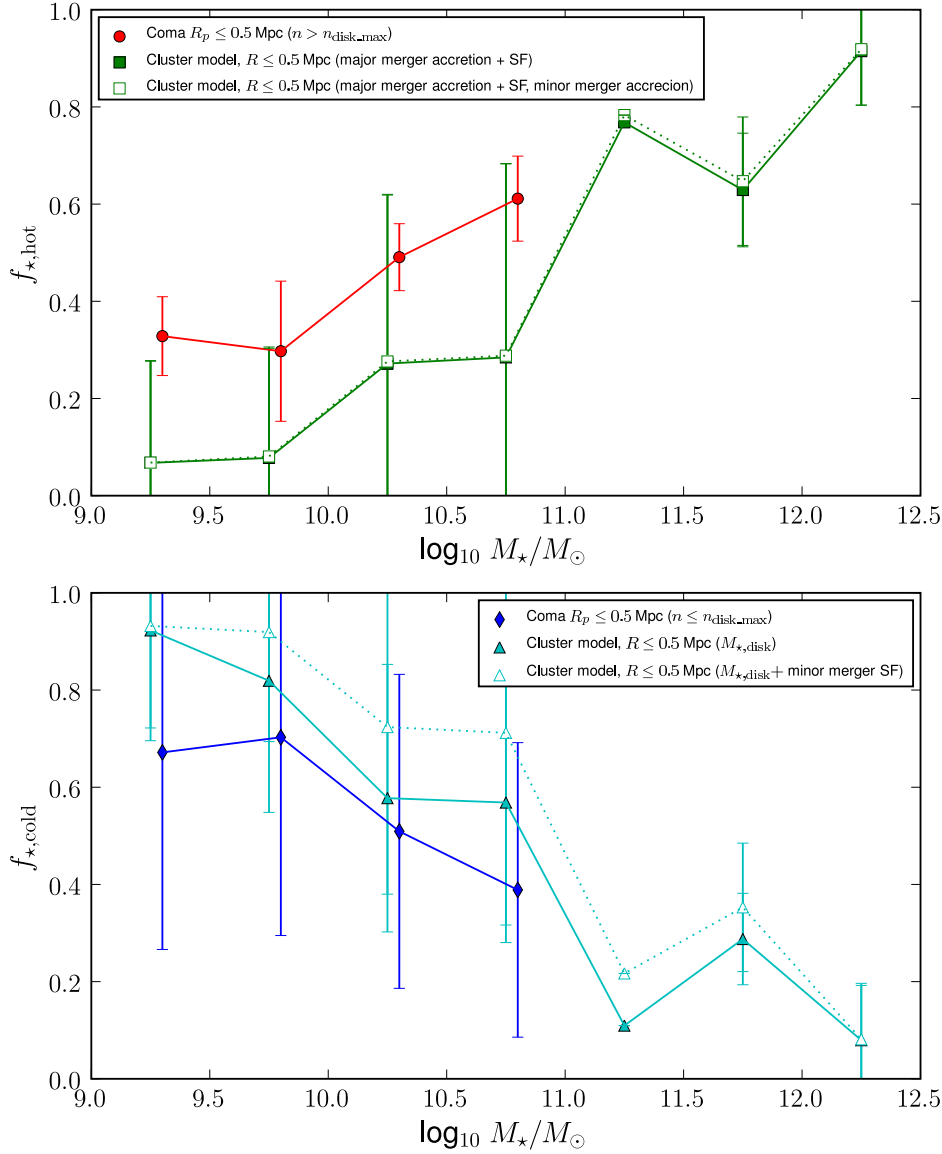}}
\caption{
{\it Top:} The mean ratio of stellar mass fraction in dynamically hot
components ($f_{\rm \star,hot}=M_{\rm \star,hot}/M_\star$) is plotted as a function of total galaxy 
$M_\star$. In the data, $M_{\rm \star,hot}$ is taken as the stellar 
mass of any high $n > n_{\rm disk\_max}$ classical
bulge/elliptical component in the galaxy, excluding the cD galaxies (see 
Section~\ref{datas2}).
The model shown here is the best cluster model matched by cumulative galaxy
number density   (see Figure~\ref{bestclusters}, column 1). For this model,
the solid line takes $M_{\rm \star,hot}$ as the stellar mass built during
major mergers, namely major merger stellar 
accretion plus induced SF, while the dashed line further adds in
minor merger stellar accretion.
{\it Bottom}: The mean stellar mass fraction in dynamically cold
flattened components ($f_{\rm \star, cold} = M_{\rm \star,cold}/M_\star$) is plotted as a function of total
galaxy stellar mass $M_\star$. In the data, $M_{\rm \star,cold}$ is taken as the 
stellar mass of any low $n \leq n_{\rm disk\_max}$  disk-dominated component in 
the galaxy. The model is represented by the solid and dashed lines. With the 
solid line we take $M_{\rm \star,cold}$ to be the
mass of the {\it outer} disk  $M_{\rm \star,Outer\_disk}$,  which is
given by ($M_\star - M_{ \rm \star,Bulge,model}$). For the 
dotted line,  we consider $M_{\rm \star,cold}$ to be the mass
 $M_{\rm \star,all\_disk}$ of inner and outer disks. We compute
the latter mass as $M_{\rm \star,Outer\_disk}$ plus 
the mass of stars formed via induced SF during minor mergers. In both panels, 
only the projected central 0.5~Mpc of the clusters is considered. The error 
bars represent the 
$1\sigma$ standard deviation on the mean. The mean values for Coma are slightly 
offset in $M_\star$ for readability. The main conclusion is that the 
best-matching simulated clusters are underpredicting the mean fraction of 
$f_{\rm \star,hot}$ and overpredicting $f_{\rm \star,cold}$
over a wide range in galaxy stellar mass.
\label{mtcompare}}
\end{figure}

\clearpage
\begin{figure}
\centering
\scalebox{0.75}{\includegraphics{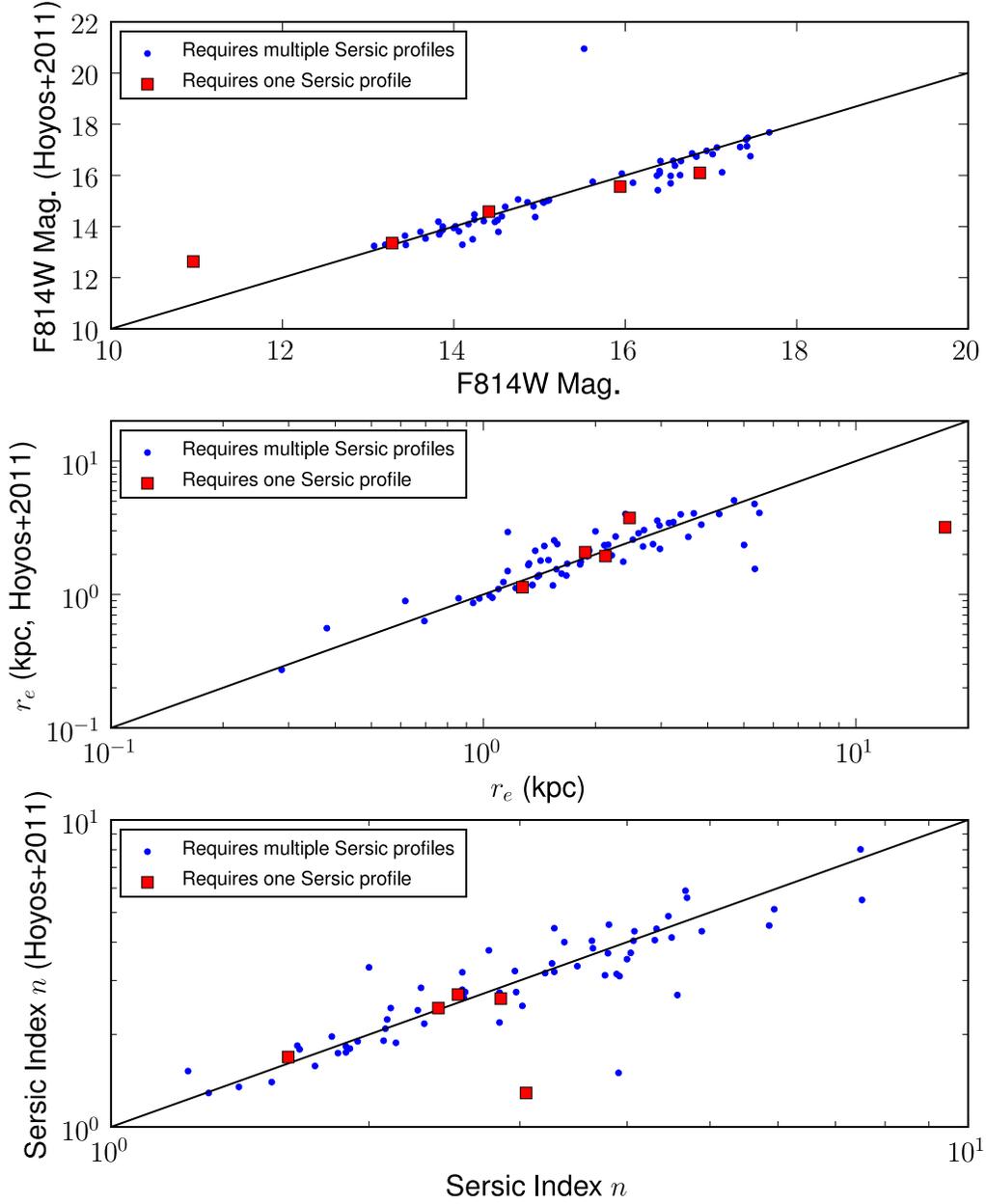}}
\caption{
This figure compares our results for the single S\'ersic fits (with no points source) 
with those obtained by Hoyos et al. (2011) using GALFIT on the same data. The sources in our sample requiring one 
S\'ersic component are labeled separately from sources requiring two or three S\'ersic 
profiles. Our derived magnitudes, $r_\mathrm{e}$, and $n$ for the sources requiring one 
S\'ersic profile agree well with the parameters from Hoyos et al. (2011), with the 
one exception being cD galaxy NGC~4874 (COMAi125935.698p275733.36) with $n\sim3$. 
Note that cD galaxy NGC~4889 requires only one S\'ersic profile but it is not 
included here as it is not in  the Hoyos et al. (2011) sample. See Appendix~\ref{1cp} 
for additional details.
\label{singlesersic}}
\end{figure}

\clearpage
\begin{figure}
\centering
\scalebox{1}{\includegraphics{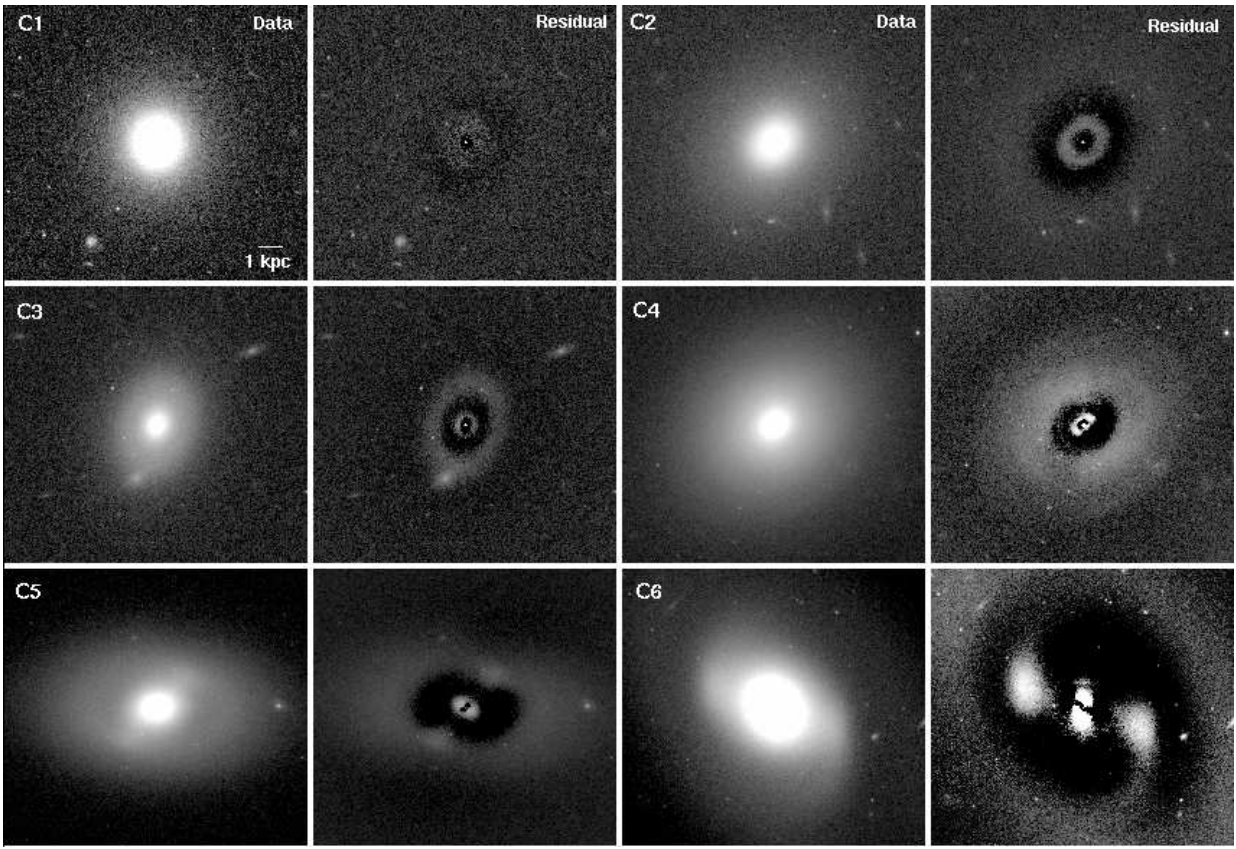}}
\caption{
This figure shows examples C1 to C6 where a single S\'ersic model (plus point source if 
needed) does not provide a good fit to coherent galaxy structure that is best modeled 
with one or more additional S\'ersic profiles. Such residual structure includes 
central compact structures (C2, C3, C4, C5, C6), rings (C3, C4), annuli and extended 
components (C1, C4), and bars/ovals (C5, C6).
These systems are better fitted by models with multiple S\'ersic components 
(see Figures \ref{resid2} and \ref{resid3}).
Columns 1 and 3 show the input $I$-band images. Columns 2 and 4 show the
residuals after subtracting the best single S\'ersic fit.
Note C1$=$COMAi125931.893p275140.76, C2$=$COMAi125935.286p275149.13, C3$=$COMAi13021.673p275354.81, C4$=$COMAi13014.746p28228.69, C5$=$COMAi13027.966p275721.56, and C6$=$COMAi125930.824p275303.05.
\label{resid}}
\end{figure}

\clearpage
\begin{figure}
\centering
\scalebox{1}{\includegraphics{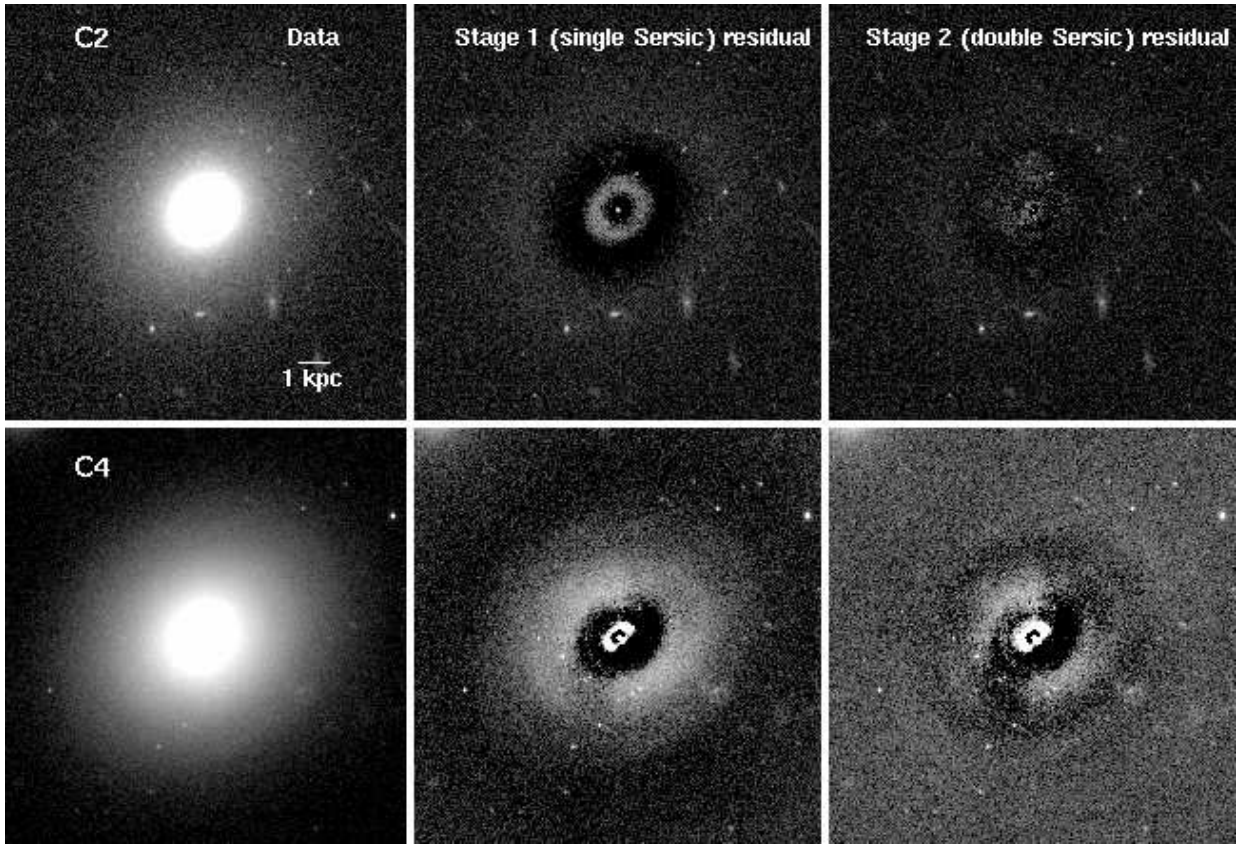}}
\caption{
This figure shows how some of the galaxies (C2$=$COMAi125935.286p275149.13 and 
C4$=$COMAi13014.746p28228.69) poorly fitted by a single S\'ersic model 
(plus point source if needed) in Stage 1 can be better fitted by two S\'ersic models 
(plus point source if needed) in Stage 2. Each row shows the data, residual after Stage 1,
and the residual after Stage 2. Galaxy C2 is best-fit as having an inner disk ($n=0.31$)
and an outer elliptical structure ($n=2.08$). Galaxy C4 is best fit with an inner bulge 
$n=3.68$ and an outer disk ($n=0.47$).
\label{resid2}}
\end{figure}

\clearpage
\begin{figure}
\centering
\scalebox{1}{\includegraphics{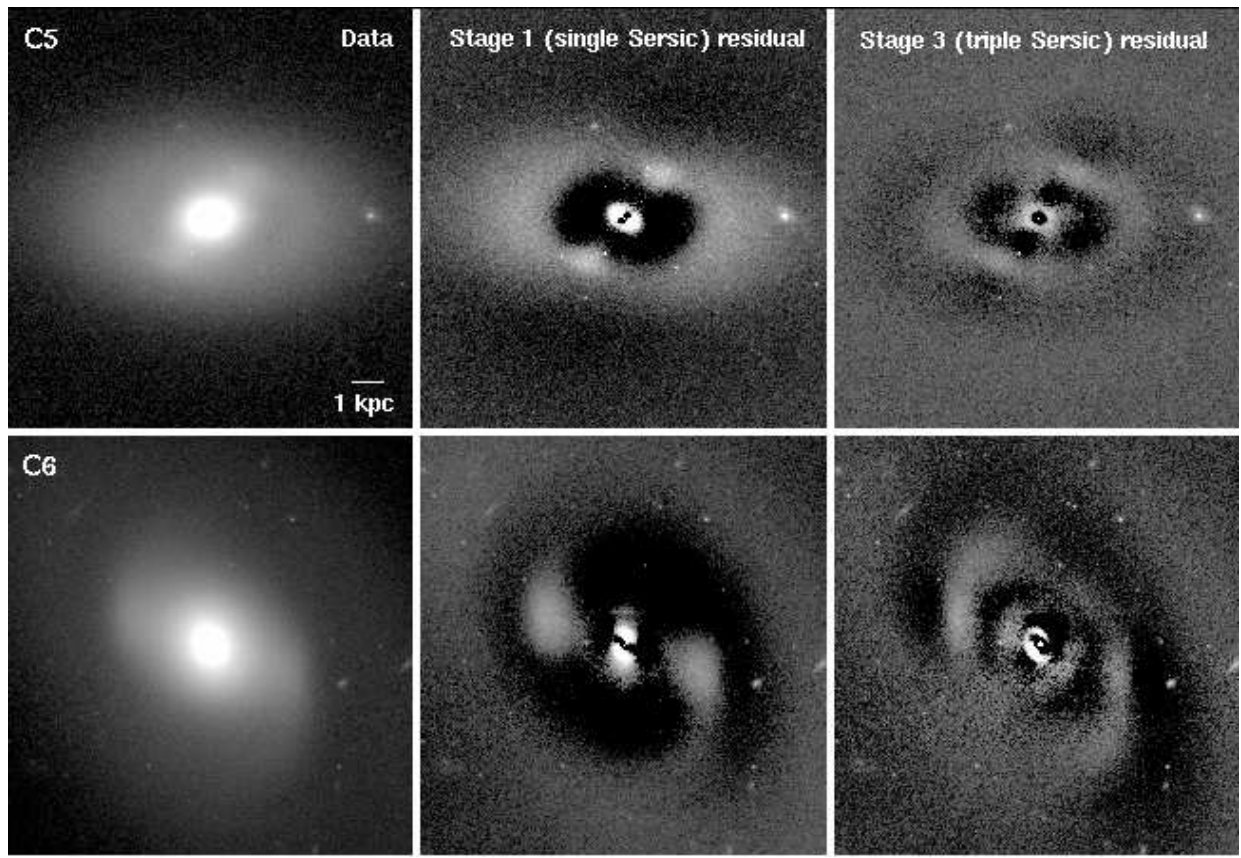}}
\caption{
This figure shows two examples of barred galaxies (C5$=$COMAi13027.966p275721.56 and 
C6$=$COMAi125930.824p275303.05) poorly fitted by a Stage 1 single S\'ersic model that are better
fitted by a Stage 3 triple S\'ersic (plus point source if needed) model. Column 1 shows
the data images while columns 2 and 3 show the residuals after the Stage 1 and Stage 3 
model, respectively.
\label{resid3}}
\end{figure}

\clearpage
\begin{figure}
\centering
\scalebox{1}{\includegraphics{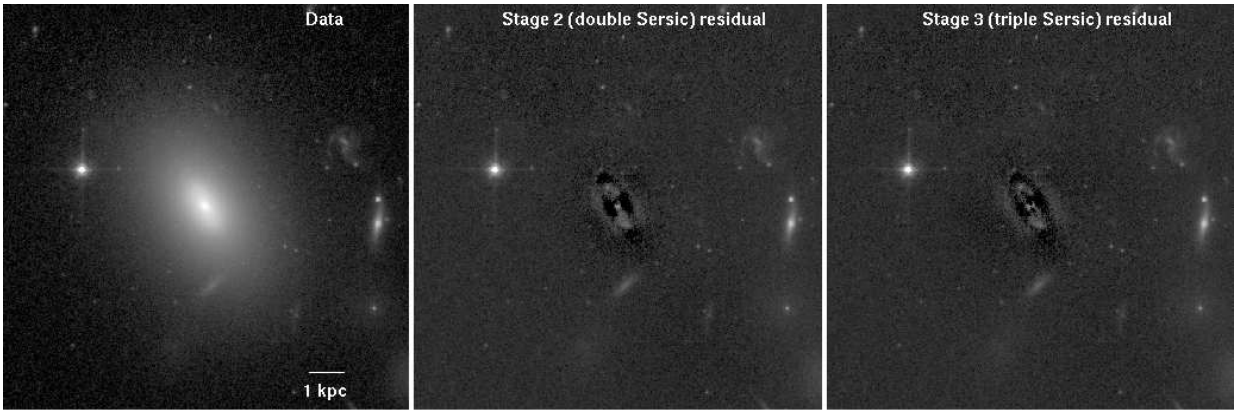}}
\caption{
This figure shows the decomposition of moderately inclined, barred galaxy 
COMAi125950.105p275529.44, in which we measure the highest outer disk S\'ersic
index $n=1.66$. Thus, this galaxy sets the 
empirically determined upper limit on disk S\'ersic index, $n_{\rm disk\_max}=1.66$.
Column 1 shows the data images while columns 2 and 3 show the residuals for the Stage 2 
and Stage 3 model, respectively. The bar signature is clearly present in the residuals.
\label{nupgal}}
\end{figure}

\clearpage
\begin{figure}
\centering
\scalebox{1}{\includegraphics{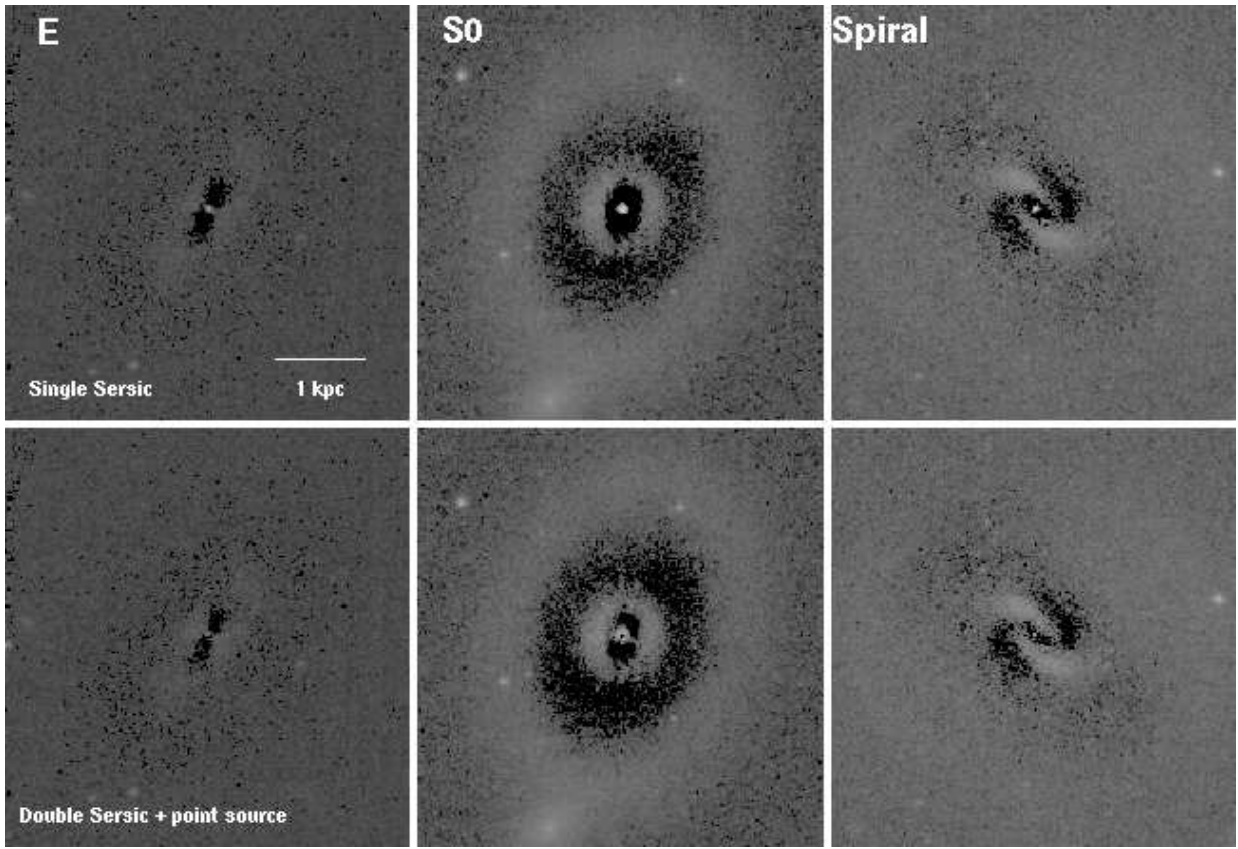}}
\caption{
This figure compares residuals after fitting a single S\'ersic model (top row) versus
the best fit double S\'ersic + nuclear point source model (bottom row) for an elliptical
(COMAi13030.954p28630.22), S0 (COMAi13021.673p275354.81), and spiral 
(COMAi13041.193p28242.34). The nuclear point source is visible in the residuals in the
top row.
\label{ptsrc}}
\end{figure}

\clearpage
\begin{figure}
\centering
\scalebox{1}{\includegraphics{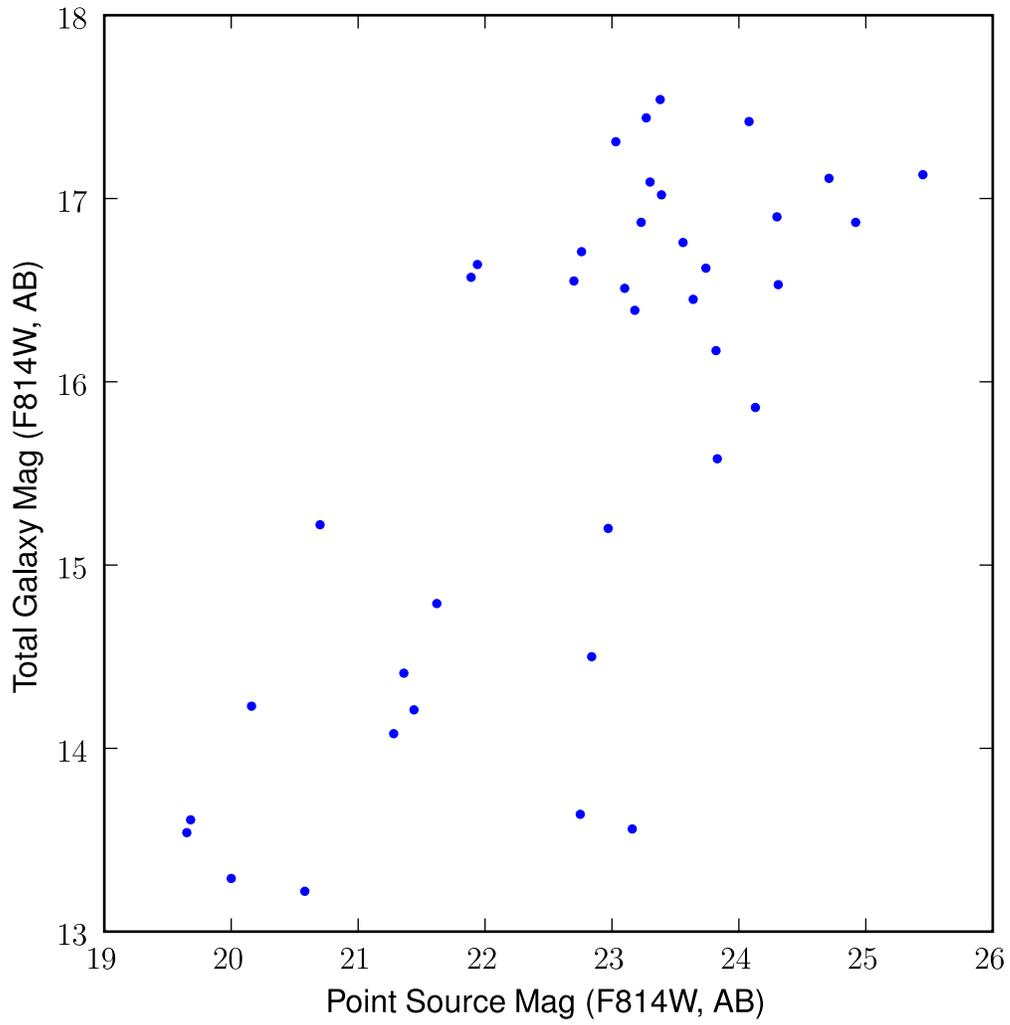}}
\caption{
This panel shows the relation between total galaxy luminosity and 
point source luminosity for objects having a nuclear point source in the 
final, best structural decomposition. 
\label{sersic-ptsrc}}
\end{figure}

\end{document}